\documentclass[11pt,a4paper,preprintnumbers]{article}

\usepackage{jheppub}
\usepackage{graphicx}
\usepackage{amsmath}
\usepackage{xspace}
\usepackage{xcolor}
\usepackage{float}
\usepackage{comment}
\usepackage{subcaption}
\usepackage{textcomp}
\usepackage{placeins}
\usepackage{bm}
\usepackage{pstricks}
\usepackage{color}
\usepackage{braket}
\usepackage{slashed}
\usepackage{soul}
\newcommand{\eq}[1]{eq.~\eqref{eq:#1}}

\newcommand{\eqs}[2]{eqs.~\eqref{eq:#1} and \eqref{eq:#2}}
\newcommand{\eqscs}[2]{eqs.~\eqref{eq:#1}, \eqref{eq:#2}}
\renewcommand{\sec}[1]{sec.~\ref{sec:#1}}

\newcommand{\fig}[1]{fig.~\ref{fig:#1}}

\newcommand{\figs}[2]{figs.~\ref{fig:#1} and \ref{fig:#2}}
\newcommand{\app}[1]{App.~\ref{app:#1}}

\newcommand{\tab}[1]{table~\ref{tab:#1}}
\newcommand{\refcite}[1]{ref.~\cite{#1}}
\newcommand{\refscite}[1]{refs.~\cite{#1}}
\newcommand{\ord}[1]{\mathcal{O}(#1)}

\newcommand{\ORd}[1]{\mathcal{O}\Bigl(#1\Bigr)}

\newcommand{\rescaletwoplots}{0.49\textwidth}

\newcommand{\df}{\mathrm{d}}

\newcommand{\as}{\alpha_{s}}

\newcommand{\Tau}{\mathcal{T}}

\newcommand{\GeV}{\ \mathrm{GeV}}
\newcommand{\TeV}{\ \mathrm{TeV}}

\newcommand{\nn}{\nonumber}

\newcommand{\id}{\mathbf{1}_{ij}}

\newcommand{\cP}{\mathcal{P}}
\newcommand{\cI}{\mathcal{I}}
\newcommand{\cL}{\mathcal{L}}

\newcommand{\cut}{\mathrm{cut}}

\newcommand{\FO}{\mathrm{FO}}

\newcommand{\NLO}{\mathrm{NLO}}
\newcommand{\NNLO}{\mathrm{NNLO}}

\newcommand{\NS}{\mathrm{NS}}

\newcommand{\one}{{(1)}}

\newcommand{\lqcd}{\Lambda_\mathrm{QCD}}
\newcommand{\dsigMC}{\df\sigma^\textsc{mc}}

\newcommand{\de}{\mathrm{d}}

\newcommand{\geneva}{\textsc{Geneva}\xspace}

\newcommand{\pythiaEight}{\textsc{Pythia8}\xspace}

\newcommand{\Matrix}{\textsc{Matrix}\xspace}
\newcommand{\scetlib}{\texttt{scetlib}\xspace}
\newcommand{\radish}{\textsc{RadISH}\xspace}

\renewcommand{\thesection}{\arabic{section}}

\allowdisplaybreaks[2]

\newcommand{\openloopsTwo}{\textsc{OpenLoops2}\xspace}

\makeatletter
\g@addto@macro\bfseries{\boldmath}
\makeatother

\begin{document}

%%%%%%%%%%%%%%%%%%%%%%%%%%%%%%%%%%%%%%%%%%%%%%%%%%%%%%%%%%%%%%%%%%%%%%%%%%%%%%%%
% Title page
%%%%%%%%%%%%%%%%%%%%%%%%%%%%%%%%%%%%%%%%%%%%%%%%%%%%%%%%%%%%%%%%%%%%%%%%%%%%%%%%

\title{Refining the GENEVA method for Higgs boson production via gluon fusion}
\preprint{\vbox{\hbox{DESY 23-009}\hbox{UWThPh-2023-2}}}

\author[a]{Simone Alioli,}

\author[a]{Georgios Billis,}

\author[a,b]{Alessandro Broggio,}

\author[a,c]{Alessandro Gavardi,}

\author[a]{Stefan Kallweit,}

\author[c,d]{Matthew A.~Lim,}

\author[a]{Giulia Marinelli,}

\author[a]{Riccardo Nagar}

\author[a]{and Davide Napoletano}

\affiliation[a]{Universit\`{a} degli Studi di Milano-Bicocca \& INFN, Piazza della Scienza 3, Milano 20126, Italy\vspace{0.5ex}}
\affiliation[b]{Faculty of Physics, University of Vienna, Boltzmanngasse 5, A-1090 Wien, Austria\vspace{0.5ex}}
\affiliation[c]{Deutsches Elektronen-Synchrotron DESY, Notkestr. 85, 22607 Hamburg, Germany\vspace{0.5ex}}
\affiliation[d]{Department of Physics and Astronomy, University of Sussex, Sussex House, Brighton, BN1 9RH, UK\vspace{0.5ex}}
\emailAdd{simone.alioli@unimib.it}
\emailAdd{georgios.billis@unimib.it}
\emailAdd{alessandro.broggio@univie.ac.at}
\emailAdd{alessandro.gavardi@desy.de}
\emailAdd{stefan.kallweit@unimib.it}
\emailAdd{m.a.lim@sussex.ac.uk}
\emailAdd{g.marinelli10@campus.unimib.it}
\emailAdd{riccardo.nagar@unimib.it}
\emailAdd{davide.napoletano@unimib.it}

\date{\today}

%%%%%%%%%%%%%%%%%%%%%%%%%%%%%%%%%%%%%%%%%%%%%%%%%%%%%%%%%%%%%%%%%%%%%%%%%%%%%%%%
\abstract{
We describe a number of improvements to the \geneva method for matching NNLO calculations to parton shower programs. In particular, we detail changes to the resummed calculation used in the matching procedure, including disentangling the cross section dependence on factorisation and beam scales, and an improved treatment of timelike logarithms. We also discuss modifications in the implementation of the splitting functions which serve to make the resummed calculation differential in the higher multiplicity phase space. These changes improve the stability of the numerical cancellation of the nonsingular term at small values of the resolution parameter.
As a case study, we consider the gluon-initiated Higgs boson
production process $gg\to H$. We validate the NNLO accuracy of our
predictions against independent calculations, and compare our showered and hadronised results
with recent data taken at the ATLAS and CMS experiments in the diphoton decay channel, finding good agreement. }
%%%%%%%%%%%%%%%%%%%%%%%%%%%%%%%%%%%%%%%%%%%%%%%%%%%%%%%%%%%%%%%%%%%%%%%%%%%%%%%%

\maketitle
\flushbottom
%% optional: table of contents
% \tableofcontents

%%%%%%%%%%%%%%%%%%%%%%%%%%%%%%%%%%%%%%%%%%%%%%%%%%%%%%%%%%%%%%%%%%%%%%%%%%%%%%%%
\section{Introduction}
\label{sec:intro}
%%%%%%%%%%%%%%%%%%%%%%%%%%%%%%%%%%%%%%%%%%%%%%%%%%%%%%%%%%%%%%%%%%%%%%%%%%%%%%%%
In recent years, the quest for precision at the Large Hadron Collider (LHC) has seen many impressive milestones in the development of theoretical tools used
to describe hadronic collisions.
Many processes are currently known at next-to-next-to-leading order~(NNLO) in perturbative QCD, and several $2 \to 1$ processes even at one order higher~(N$^3$LO).
One particular direction in which much fruitful progress has been made is in the matching of higher order perturbative calculations to parton shower~(PS) programs, resulting in Monte Carlo event generators which combine the advantages of fixed-order calculations with the flexibility of parton shower tools.
 This paradigm generally goes under the name of NNLO+PS.

Several different methods which reach NNLO+PS accuracy have been proposed~\cite{Hamilton:2012rf,Alioli:2013hqa,Hoche:2014uhw,Alioli:2015toa,Monni:2019whf,Monni:2020nks,Campbell:2021svd}, with most applications using a resummed calculation -- either directly or via the Sudakov factor in a shower Monte Carlo -- in a suitable resolution variable alongside the fixed order to achieve the matching. Of these methods, the \geneva approach~\cite{Alioli:2012fc,Alioli:2015toa} has the advantage of being particularly flexible with regard to the framework used for the resummed calculation and the choice of the resolution variable, while also exploiting the possibility of reaching higher logarithmic accuracies in both direct QCD and soft-collinear effective theory (SCET) formalisms. This has resulted in the application of the method to a number of colour singlet processes~\cite{Alioli:2015toa,Alioli:2019qzz,Alioli:2020fzf,Alioli:2020qrd,Alioli:2021qbf,Cridge:2021hfr,Alioli:2021egp,Alioli:2022dkj}, as well as first steps towards implementations involving coloured final states~\cite{Alioli:2021ggd}.

In this work, we describe a number of improvements to the \geneva event generator, which both extend the capabilities of the program and improve its numerical performance.
Specifically, we detail a new treatment of the splitting functions,
which were first introduced in the original \geneva implementation~\cite{Alioli:2012fc}
and serve to make the resummed calculation used in the matching procedure differential in the higher multiplicity phase space. The new approach significantly increases the performance of the code in extreme soft and collinear regions, where the cancellation of large logarithmic terms is extremely delicate. We also implement a more rigorous treatment of the theoretical uncertainties by disentangling the  factorisation and renormalisation scale dependences in the cross section and allowing their independent variations. This puts our uncertainty estimation on a robust and more conservative theoretical footing, and will also prove important for the implementation of processes featuring perturbatively generated heavy flavours in the initial state. Finally, we also discuss the treatment of timelike logarithms in our calculation, the inclusion of which has been shown to improve the perturbative convergence of colour singlet production processes such as $gg\to H$~\cite{Bakulev:2000uh,Ahrens:2008qu,Ebert:2017uel}.

In order to study the various improvements to the \geneva program, we have implemented the gluon-initiated Higgs boson production process ($gg \to H$). We use a resummed calculation in the zero-jettiness resolution variable $\Tau_0$ obtained via SCET up to N$^3$LL accuracy. The process is interesting from both an experimental and a theoretical perspective for a number of reasons.

Experimentally, the gluon-fusion production channel was of utmost importance for the discovery of the Higgs boson~\cite{ATLAS:2012yve, CMS:2012qbp}. Nowadays Higgs physics remains a crucial aspect of the LHC programme~\cite{ATLAS:2013dos,ATLAS:2014yga,ATLAS:2014xzb,ATLAS:2016vlf,ATLAS:2017qey,ATLAS:2018hxb,ATLAS:2020wny,CMS:2015qgt,CMS:2015zpx,CMS:2016ipg,CMS:2018ctp,CMS:2018gwt,CMS:2020dvg}, and constraining the scalar boson's properties and couplings to probe the nature of the Higgs sector is a priority for Run 3 of LHC and beyond~\cite{Cepeda:2019klc}.

From the theory side, many calculations work in the limit in which the top-quark mass is considered to be large compared to other scales present in the process, the so-called heavy-top limit~(HTL).
This significantly simplifies the computational complexity since the top-quark loop coupling the Higgs boson to gluons is integrated out, resulting in an effective $ggH$ vertex and further effective vertices with more gluons and Higgs bosons. Consequently, calculations including QCD corrections up to N$^3$LO are now available in this limit~\cite{Harlander:2002wh,Anastasiou:2002yz,Ravindran:2003um,Anastasiou:2015vya, Anastasiou:2016cez,Mistlberger:2018etf,Chen:2021isd, Billis:2021ecs}, including matching to resummed calculations up to N$^3$LL$'$ accuracy in transverse momentum~\cite{Bozzi:2005wk,Becher:2012yn,Neill:2015roa,Bizon:2017rah,Chen:2018pzu,Bizon:2018foh,Gutierrez-Reyes:2019rug,Becher:2020ugp,Billis:2021ecs,Re:2021con} and in jet veto observables~\cite{Berger:2010xi,Becher:2012qa,Stewart:2013faa,Gangal:2020qik,Campbell:2023cha}. There has also been a considerable amount of work on improving calculations beyond the HTL by including quark mass effects~\cite{Pak:2009dg,Harlander:2009mq,Harlander:2009my,Harlander:2012hf,Ball:2013bra};
  this has culminated in a calculation of the exact top-quark mass dependence at NNLO in QCD~\cite{Czakon:2020vql,Czakon:2021yub}. Additionally, the fact that the perturbative series is known to be poorly convergent has motivated the study of alternative scale choices which include $\pi^2$ terms arising from kinematic logarithms at all orders.
Finally, the simplicity of this process in terms of its kinematics and matrix elements
makes it a particularly suitable testing ground for the improvements which we will detail in this work.

The rest of the paper is organised as follows. In \sec{theory}, we
provide a brief recap of the \geneva method and its application to gluon-induced Higgs production, before discussing the new
features which have been implemented in the program in
\sec{features}. In \sec{validation_intro} we validate the NNLO
accuracy of our calculation for the $gg \to H$ process and discuss the matching to the parton shower
provided by \pythiaEight
~\cite{Sjostrand:2007gs}. Finally, we compare our results with
the $pp \to H \to \gamma \gamma$ data collected at the
ATLAS and CMS experiments \cite{ATLAS:2022fnp,CMS:2022wpo} in \sec{data}. We give our conclusions in \sec{conclusions}.
In \app{splitting-dy} we show an application of the novel splitting function implementation to the Drell-Yan process.

%%%%%%%%%%%%%%%%%%%%%%%%%%%%%%%%%%%%%%%%%%%%%%%%%%%%%%%%%%%%%%%%%%%%%%%%%%%%%%%%
\section{Theoretical framework}
\label{sec:theory}
%%%%%%%%%%%%%%%%%%%%%%%%%%%%%%%%%%%%%%%%%%%%%%%%%%%%%%%%%%%%%%%%%%%%%%%%%%%%%%%%

In the following we lay out the theoretical framework we work in.
We start by giving a summary of the \geneva event generator formalism, which includes the matching procedure of the fixed-order calculation to the resummed prediction.
We then focus on the definition of the process under study, i.e.~Higgs boson production via gluon fusion, and on its zero-jettiness resummation.

%%%%%%%%%%%%%%%%%%%%%%%%%%%%%%%%%%%%%%%%%%%%%%%%%%%%%%%%%%%%%%%%%%%%%%%%%%%%%%%%
\subsection{The G{\scriptsize ENEVA} method}
\label{sec:geneva}
%%%%%%%%%%%%%%%%%%%%%%%%%%%%%%%%%%%%%%%%%%%%%%%%%%%%%%%%%%%%%%%%%%%%%%%%%%%%%%%%

The complete derivation of the \geneva method has been presented extensively in several publications, e.g.~in \refscite{Alioli:2015toa,Alioli:2019qzz}. Here, we explicitly refrain from entering into
the finer details of the method, and we only briefly recall the general formulae,
highlighting some key features that are important for this
process.

We use $N$-jettiness~\cite{Stewart:2010tn} to resolve the QCD emissions that can be associated with each event produced by \geneva:
$\Tau_0$ as the zero-jet resolution parameter, and $\Tau_1$ to separate between one or more emissions.
The partonic event space is then divided into three regions: $\Phi_0$ for events with no extra emissions, $\Phi_1$ for one-jet events, and $\Phi_2$ for the remaining events with two jets in the final state.
These phase space regions are defined via two thresholds, $\Tau_0^\cut$ and $\Tau_1^\cut$.

The differential cross section for the production of events
with no extra emissions is given by
%%%
\begin{align}
\frac{\dsigMC_0}{\df\Phi_0}(\Tau_0^\cut) &= \frac{\df\sigma^{\rm NNLL'}}{\df\Phi_0}(\Tau_0^\cut) - \frac{\df\sigma^{\rm NNLL'}}{\df\Phi_{0}}(\Tau_0^\cut)\bigg\vert_{\NNLO_0} \nn \label{eq:0full}\\
&\qquad +(B_0+V_0+W_0)(\Phi_0)\, +\,  \int \frac{\mathrm{d} \Phi_1}{\mathrm{d} \Phi_0} (B_1 + V_1)(\Phi_1)\,\theta\big( \Tau_0(\Phi_1)< \Tau_0^{\mathrm{cut}}\big) \nn \\
&\qquad+  \int \frac{\mathrm{d} \Phi_2}{\mathrm{d} \Phi_0} \,B_2 (\Phi_2) \, \theta\big( \Tau_0(\Phi_2)< \Tau_0^{\mathrm{cut}}\big)\,.
\end{align}
Here we use the primed counting for the resummation order as in e.g.~\refcite{Berger:2010xi}.
For the case of a single extra emission we have two contributions: that above $\Tau_0^\cut$
\begin{align}
\frac{\dsigMC_{1}}{\df\Phi_{1}} (\Tau_0 > \Tau_0^\cut; \Tau_{1}^\cut) &= \Bigg\{\Bigg[ \frac{\df\sigma^{\rm NNLL'}}{\df\Phi_0\df\Tau_0}-\frac{\df\sigma^{\rm NNLL'}}{\df\Phi_0\df\Tau_0}\bigg\vert_{\NLO_1}\,\Bigg]\, \cP(\Phi_1)\,+ (B_1 + V_1^C)(\Phi_1)  \Bigg\} \nn \label{eq:1masterfull}\\
&\qquad\times\, U_1(\Phi_1, \Tau_1^\cut)\, \theta(\Tau_0 > \Tau_0^\cut) \nn \\
&\qquad+\int\ \bigg[\frac{\df\Phi_{2}}{\df\Phi^\Tau_1}\,B_{2}(\Phi_2) \, \theta\!\left(\Tau_0(\Phi_2) > \Tau_0^\cut\right)\,\theta(\Tau_{1} < \Tau_1^\cut) \nn \\
&\qquad\quad - \frac{\df\Phi_2}{\df \Phi^C_1}\, C_{2}(\Phi_{2}) \, \theta(\Tau_0 > \Tau_0^\cut) \bigg] \nn \\
&\qquad- B_1(\Phi_1)\, U_1^\one(\Phi_1, \Tau_1^\cut)\, \theta(\Tau_0 > \Tau_0^\cut)\,,
\end{align}
and the nonsingular below $\Tau_0^\cut$, arising from non-projectable configurations,
\begin{align}
\frac{\dsigMC_{1}}{\df\Phi_{1}} (\Tau_0 \le \Tau_0^\cut; \Tau_{1}^\cut) &= (B_1+V_1)(\Phi_1)\, \overline{\Theta}^{\mathrm{FKS}}_{\mathrm{map}}(\Phi_1) \, \theta(\Tau_0<\Tau^{\mathrm{cut}}_0)\,. \label{eq:1belowtau0}
\end{align}
Similarly the case of two extra emissions also receives two contributions,
\begin{align}
\frac{\dsigMC_{\geq 2}}{\df\Phi_{2}} (\Tau_0 > \Tau_0^\cut, \Tau_{1}>\Tau_{1}^\cut) &= \Bigg\{ \bigg[ \frac{\df\sigma^{\rm NNLL'}}{\df\Phi_0\df\Tau_0} - \frac{\df\sigma^{\rm NNLL'}}{\df\Phi_0\df\Tau_0}\bigg|_{\NLO_1}\bigg]\, \cP(\widetilde{\Phi}_1) \nn \\
&\qquad + (B_1 + V_1^C)(\widetilde{\Phi}_1)\Bigg\} \,  U_1'(\widetilde{\Phi}_1, \Tau_1)\, \theta(\Tau_0 > \Tau_0^\cut) \Big\vert_{\widetilde{\Phi}_1 = \Phi_1^\Tau\!(\Phi_2)} \nn\\
&\qquad\quad\times\, \cP(\Phi_2) \, \theta(\Tau_1 > \Tau_1^\cut)
\nn \\
&\qquad + \Big\{ B_2(\Phi_2)\, \theta(\Tau_{1}>\Tau^{\mathrm{cut}}_{1})- B_1(\Phi_1^\Tau)\,U_1^{\one\prime}\!\big(\widetilde{\Phi}_1, \Tau_1\big)\nn\\
&\qquad\quad\times\,\cP(\Phi_2)\, \Theta(\Tau_1 > \Tau_1^\cut)
\Big\}\, \theta\left(\Tau_0(\Phi_2) > \Tau_0^\cut\right)\,, \label{eq:2masterfull}
\end{align}
and
\begin{align}
\frac{\dsigMC_{\geq 2}}{\df\Phi_{2}} (\Tau_0 > \Tau_0^\cut, \Tau_{1} \le \Tau_{1}^\cut) &= B_2(\Phi_2)\, \overline{\Theta}_{\mathrm{map}}^\Tau(\Phi_2) \, \theta(\Tau_1 < \Tau_1^\cut)\, \theta\left(\Tau_0(\Phi_2) > \Tau_0^\cut\right)\,,  \label{eq:2belowtau1}
\end{align}
%%%
above and below $\Tau_1^\cut$, respectively.

In the formulae above, $B_n$, $V_n$ and $W_n$ are the $0$-, $1$- and $2$-loop
matrix elements for $n$ QCD partons in the final state (including parton densities); analogously, we denote by $\mathrm{N}^k\mathrm{LO}_n$ a quantity with $n$ additional partons in the final state
computed at $\mathrm{N}^k\mathrm{LO}$ accuracy.
Since it is necessary to evaluate the resummed and resummed-expanded terms on phase space points resulting
from a projection from a higher to a lower multiplicity, we introduce a shorthand for such projected phase space points, $\widetilde{\Phi}_{N}$.
We use the abbreviation
%%%
\begin{align}
\label{eq:dPhiRatio}
 \frac{\df \Phi_{M}}{\df \Phi_N^{\cal O}}  = \df \Phi_{M} \, \delta[ \widetilde{\Phi}_N - \Phi^{\cal O}_N(\Phi_M) ] \,\Theta^{\cal O}(\Phi_M)
\end{align}
%%%
to indicate an integration over the portion of the $\Phi_M$ phase space which can be reached from a $\Phi_N$ point while keeping some observable $\cal O$ also fixed, with $N < M$.
The $\Theta^{\cal O}(\Phi_M)$ term additionally limits the integration to
the phase space points belonging to the singular contribution for the
given observable $\cal O$.  For example, when generating $1$-body
events we use
%%%
\begin{equation} \label{eq:Phi1TauProj}
\frac{\df\Phi_2}{\df\Phi_1^\Tau} \equiv \df\Phi_2\,\delta[\widetilde{\Phi}_1 - \Phi^\Tau_1(\Phi_2)]\,\Theta^\Tau(\Phi_2)
\,,\end{equation}
%%%
where the $1 \to 2$ mapping has been constructed
to preserve $\Tau_0$, i.e.
%%%
\begin{equation} \label{eq:Tau0map}
\Tau_0(\Phi_1^\Tau(\Phi_2)) = \Tau_0(\Phi_2)
\,,\end{equation}
%%%
and $\Theta^\Tau(\Phi_2)$ guarantees that the $\Phi_2$ point is reached from a genuine QCD splitting of the $\Phi_1$ point.
The use of a $\Tau_0$-preserving mapping is necessary to ensure that the point-wise singular $\Tau_0$ dependence is alike among all terms in
\eqs{1masterfull}{2masterfull} and that the cancellation of said singular
terms is guaranteed on an event-by-event basis.

The non-projectable regions of $\Phi_1$ and $\Phi_2$, on the other hand, are assigned to the cross sections in \eqs{1belowtau0}{2belowtau1}.
These events are entirely nonsingular in nature. We denote the constraints due to the choice of map by $\Theta_{\mathrm{map}}$, using the FKS
map~\cite{Frixione:2007vw} for the $\Phi_1 \to\widetilde{\Phi}_0$ projection and, as mentioned above, a $\Tau_0$-preserving map for the
$\Phi_2 \to\widetilde{\Phi}_1$ projection. Their complements are denoted by $\overline{\Theta}_{\mathrm{map}}$.

The term $V_1^C$ denotes the contributions of soft and collinear origins in a standard NLO local subtraction,
%%%
\begin{align} \label{eq:FOFKS}
  V_1^C(\Phi_1) = V_1(\Phi_1) + \int \frac{\df\Phi_2}{\df \Phi_1^C} \, C_2(\Phi_2)\,,
\end{align}
%%%
with $C_2$ a singular approximant of $B_2$; in practice we use the
subtraction counterterms which we integrate over the radiation variables
$\df\Phi_2 / \df \Phi_1^C$  using the singular limit $C$ of the phase space
mapping.

In the formulae involving one or two extra emissions, $U_1$ is a next-to-leading-logarithmic (NLL) Sudakov factor which resums large logarithms of $\Tau_1$, and $U_1'$ its derivative with respect to $\Tau_1$; the $\ord{\alpha_s}$
expansions of these quantities are denoted by $U_1^\one$ and $U_1^{\one\prime}$ respectively.

We extend the differential dependence of the resummed terms from the $N$-jet to the $(N\!+\!1)$-jet phase space using a normalised splitting probability
$\mathcal{P}(\Phi_{N+1})$ which satisfies
%%%
\begin{align}
\label{eq:cPnorm}
\int \! \frac{\df\Phi_{N+1}}{\df \Phi_{N} \df \Tau_N} \, \cP(\Phi_{N+1}) = 1
\,.\end{align}
%%%
The two extra variables are chosen to be an energy ratio $z$ and an
azimuthal angle $\phi$.
The functional forms of the $\cP(\Phi_{N+1})$ are in principle only constrained
by \eq{cPnorm}. However, in order to correctly model the soft-collinear limit behaviour,
we find it useful to write them in terms of the Altarelli-Parisi splitting kernels, weighted by parton
distribution functions (PDFs).

In previous implementations of the \geneva method,
the splitting functions $\mathcal{P}(\Phi_{N+1})$
were computed using a
``hit-or-miss'' method based on precomputed upper bounds, which did
not require knowledge of an analytic expression for the
integration limits of $z$ and $\phi$
(see section II.B.4 of \refcite{Alioli:2015toa} for the definition of the splitting function
and Appendices C and D of \refcite{Alioli:2010xd} for the computation of the upper bounds in a similar situation). 
At the same time, however, this introduced some numerical instabilities.
In this work, we improve on this situation by including the exact integration
limits and evaluate the splitting
functions directly for each phase space point, as detailed in \sec{splitting}.

%%%%%%%%%%%%%%%%%%%%%%%%%%%%%%%%%%%%%%%%%%%%%%%%%%%%%%%%%%%%%%%%%%%%%%%%%%%%%%%%
\subsection{Higgs boson production via gluon fusion}
\label{sec:ggh_definition}
%%%%%%%%%%%%%%%%%%%%%%%%%%%%%%%%%%%%%%%%%%%%%%%%%%%%%%%%%%%%%%%%%%%%%%%%%%%%%%%%

We consider the production of a stable Higgs boson   via the
gluon fusion channel in proton-proton scattering, $pp \to
H + X$, where $X$ denotes any additional hadronic radiation in the final state.
At leading order (LO) in the strong coupling this results in a single contribution $gg \to H$ at partonic level~\cite{Georgi:1977gs}, while at next-to-leading order~(NLO) (anti)quark-initiated channels also start to contribute~\cite{Dawson:1990zj,Djouadi:1991tka,Spira:1995rr}.

For a stable Higgs boson it is phenomenologically reasonable to work
in the HTL effective field theory (EFT), in which the contributions
from the top-quark loops coupling the Higgs boson to gluons have been
integrated out. This EFT supplements the Standard Model~(SM) vertices with additional, effective couplings between gluons and Higgs bosons.
Introducing these effective vertices has the advantage of reducing the complexity of the matrix element computations.
The cross section dependence on the top-quark mass $m_t$ can be partially restored  by
rescaling the HTL result by a factor equal to the ratio between
the LO $m_t$-exact result and that obtained in pure EFT. This is later referred to as rescaled EFT~(rEFT), and reproduces the
exact $m_t$ dependence of the LO cross section by construction.
It is known to be a good approximation, for inclusive quantities, at least up to
NNLO~\cite{Czakon:2021yub}. The resulting
approximation can instead be problematic for differential
distributions, for instance the transverse momentum of the Higgs boson when the accompanying radiation resolves the
top-quark loop, i.e.~when its transverse momentum is
larger than $m_t$.
For the case of a finite top-quark mass, the NNLO corrections have been recently calculated for the inclusive cross section~\cite{Czakon:2020vql,Czakon:2021yub}, and those at NLO for Higgs boson production in association with up to two hard jets~\cite{Jones:2018hbb,Chen:2021azt}. At this level of precision, however, one also needs to take into account the interference between contributions including both massive top and bottom quarks, which is known at NLO for the Higgs plus jet case~\cite{Lindert:2017pky,Bonciani:2022jmb}.
Since the problem of including the quark mass effects for precise phenomenological studies is largely independent of the matching of fixed-order and resummed calculations to parton showers in the \geneva method, which is the topic of the present study, we leave the investigation of these effects to future work.

In this work, the Higgs boson is always produced on shell with a mass $m_H\!=\!125.09 \GeV$.
When comparing with data in the fiducial regions of the ATLAS or CMS experiments, we will consider Higgs boson decays.
In this case we work in the narrow-width approximation, which for the Higgs boson is particularly accurate since $\Gamma_H/m_H \sim \ord{10^{-5}} $.
The Higgs decay products can always be added a posteriori due to the scalar nature of the boson, which implies that they are isotropically distributed without spin correlations with the initial state.
For the rest of this work we will consider a collider energy of $\sqrt{S} = 13 \TeV$ and assume the following values for the SM parameters affecting our calculations:
\begin{equation}
\label{eq:input_parameters}
  G_F = 1.16639 \times 10^{-5}, \quad
  m_t  = 173.1 \GeV\,.
\end{equation}

For the matrix elements in the HTL approximation
we use the \texttt{heftpphj} and \texttt{heftpphjj} libraries of \openloopsTwo
\cite{Buccioni:2019sur,Cascioli:2011va,Buccioni:2017yxi}, which we then rescale by the rEFT factor $r_\text{EFT} = 1.06545$.

%%%%%%%%%%%%%%%%%%%%%%%%%%%%%%%%%%%%%%%%%%%%%%%%%%%%%%%%%%%%%%%%%%%%%%%%%%%%%%%%
\subsection{$\Tau_0$ resummation}
\label{sec:tau0}
%%%%%%%%%%%%%%%%%%%%%%%%%%%%%%%%%%%%%%%%%%%%%%%%%%%%%%%%%%%%%%%%%%%%%%%%%%%%%%%%

The formulae presented in \sec{geneva} require the evaluation of the resummed spectrum and cumulant in the resolution variable $\Tau_0$ up to at least NNLL$'$ accuracy. Although the \geneva method does not depend on any particular resummation formalism, in practice we often find it convenient to exploit results derived via SCET~\cite{Bauer:2000ew,Bauer:2000yr,Bauer:2001ct,Bauer:2001yt,Bauer:2002nz,Beneke:2002ni,Beneke:2002ph}. Within this framework, a factorisation theorem for the zero-jettiness was first derived in~\refscite{Stewart:2009yx,Stewart:2010pd} for colour singlet production.
In the case of the gluon-fusion channel for Higgs production it reads
\begin{align}\label{eq:tau0_factorisation_def}
\frac{\textrm{d} \sigma^{\text{SCET}}}{\textrm{d} \Phi_0 \textrm{d} \Tau_0}
= H_{gg \to H} (Q^2,\mu) \! \int\! \df t_a \df t_b \, B_g(t_a, x_a, \mu) \, B_g(t_b, x_b, \mu) \, S_{gg}\left(\Tau_0-\frac{t_a + t_b}{Q},\mu \right)
\,,
\end{align}
where $H_{gg \to H}$, $S_{gg}$, and $B_g$ are the hard, soft and beam functions, respectively.

The process-specific hard function $H_{gg \to H}(Q^2,\mu)$ is defined as the square of the Wilson coefficient that results from matching the QCD Hamiltonian to the SCET operators, and encodes information about the Born and virtual squared matrix elements. It depends only on the Higgs boson virtuality $Q^2$. In this section and whenever we consider Higgs boson production specifically, we set $Q = m_H$; elsewhere, we consider $Q$ to be a generic hard scale.

Given that we work in the HTL approximation, we perform a two-step matching procedure: we first integrate out the hard degrees of freedom above the top-quark mass, and subsequently match the resulting EFT onto SCET. The final hard function then arises from the product of two Wilson coefficients, the first from the HTL approximation and the second from the matching to SCET; we evaluate both at the same scale $\mu$. In principle, within this approach one could resum $\ln(m_t/m_H)$ contributions by renormalisation group equation~(RGE) evolution. However, given the values of the top quark and Higgs boson masses, these logarithms are never large and, consequently, we include them only at fixed order in the hard function.
Alternatively, if one wants to include the full top-quark mass effects, a single-step matching can be performed as in e.g.~\refcite{Berger:2010xi} at NNLL.
Extending this to NNLL$'$  accuracy requires the three-loop hard function with the exact top-quark mass dependence~\cite{Czakon:2020vql,Czakon:2021yub}.

The beam functions $B_g(t,x,\mu)$ are the inclusive gluon beam functions~\cite{Stewart:2009yx}, which depend on the transverse virtualities $t_{a,b}$ of the initial-state partons that participate in the hard interaction
and on their momentum fractions $x_{a,b}$. While they are nonperturbative objects, for $t \gg \lqcd$ they admit an operator product expansion (OPE),
\begin{align} \label{eq:beam_func_ope_def}
B_i(t, x, \mu)
&= \sum_j \int_{x}^{1} \frac{\df \xi}{\xi}\, \mathcal{I}_{ij}\bigg(t,\frac{x}{\xi},\mu\bigg) \, f_j(\xi,\mu)\, \Bigl[1 + \ORd{\frac{\lqcd^2}{t}} \Bigr]
\nn \, \\
&\equiv \sum_j \left[\cI_{ij} \otimes_x f_j\right]\left(t, x, \mu\right) \, \Bigl[1 + \ORd{\frac{\lqcd^2}{t}} \Bigr]
\,,\end{align}
where the $\mathcal{I}_{ij}(t,z,\mu)$ are matching coefficients that describe the collinear virtual and real initial-state radiation (ISR)
and the $f_j(\xi,\mu)$ are the usual PDFs.
For later use, we denote the Mellin convolution via the symbol $\otimes_x$.

Finally, $S_{gg}(k,\mu)$ is the gluon hemisphere soft function for beam thrust.
Like the beam functions, $S_{gg}(k, \mu)$ is a nonperturbative object and for $k\gg \lqcd$ it also satisfies
an OPE, where the LO matching coefficient is calculable in perturbation theory.
Its perturbative component depends only on the colour representation of the hard partons, and therefore the gluon case can be derived from
that of the quark channel via Casimir scaling.
In our calculation we neglect the nonperturbative part of the soft function.
We then rely on the hadronisation model of the parton shower to provide the missing contribution.

The functions in \eq{tau0_factorisation_def} are all evaluated at a common scale $\mu$ and satisfy RGEs.
The scale dependence in each of these functions involves potentially large logarithms of ratios of disparate scales, which may impact their perturbative convergence.
In order to reduce the effect of these large logarithms, we evaluate each function at its characteristic (canonical) scale, i.e.~$\mu_S = \Tau_0$, $\mu_H = m_H$, and $\mu_B = \sqrt{\mu_S \mu_H}$.
Since the cross section needs to be evaluated at a common scale $\mu$, we use the RGEs to evolve each function to $\mu$. In doing so, we resum said logarithms at all orders in perturbation theory.
The resummed formula for the $\Tau_0$ spectrum is then given by (see e.g.~\refcite{Alioli:2019qzz} for more details)
\begin{align} \label{eq:standardresum}
\frac{\de \sigma^{\rm resum}}{\de \Phi_0 \de \Tau_0} &=
  H_{gg \to H}(Q^2,\mu_H)\, U_H(\mu_H, \mu)
                                                           \nonumber \\
 & \times \int \de t_a \,\de t_b  \left[ B_g(t_a,x_a,\mu_B) \otimes U_B(\mu_B, \mu) \right]
    \left[B_g(t_b,x_b,\mu_B) \otimes U_B(\mu_B,\mu) \right]\nonumber \\
 & \qquad \times \left[S_{gg}( \Tau_0 - \frac{t_a+t_b}{Q},\mu_S)\otimes U_S(\mu_S, \mu)\right],
\end{align}
where we denote the standard convolutions between the different functions and the RGE evolution factors via the $\otimes$ symbol.

In order to achieve NNLL$'$ accuracy in the $\Tau_0$ resummation, each of the hard, soft and beam function boundary terms must be known at 2-loop order.
For the beam function they were calculated at 2-loops in~\refcite{Gaunt:2014cfa}, and in fact they are known up to 3-loop order~\cite{Ebert:2020unb}. Our implementation of the gluon beam function relies on an interface to \scetlib~\cite{Billis:2019vxg,Billis:2021ecs,scetlib}, a library which provides ingredients for resummed calculations in SCET.
The soft function has been known at 2-loops for some time~\cite{Kelley:2011ng,Monni:2011gb}, and recent work has aimed to push this calculation to the 3-loop order~\cite{Baranowski:2021gxe,Baranowski:2022khd,Chen:2020dpk}.
The hard function has appeared several times in the literature, see e.g.~\refscite{Idilbi:2006dg, Berger:2010xi}, and is known analytically with full top-quark mass dependence at NNLO~\cite{Czakon:2020vql}.
In addition, the anomalous dimensions and the beta function which enter the evolution factors and the fixed-order
expansion of \eq{standardresum} must be known at 2-loop (noncusp~\cite{Berger:2010xi}) and 3-loop (cusp~\cite{Moch:2004pa, Vogt:2004mw, Korchemsky:1987wg},
$\beta(\alpha_s)$~\cite{Tarasov:1980au, Larin:1993tp}) order.
By including them at one order higher \cite{Berger:2010xi,vanRitbergen:1997va,vonManteuffel:2020vjv}, one can achieve resummation at N$^3$LL.

The resummation of $\Tau_0$ for the case of Higgs boson production via gluon fusion has already been studied in \refcite{Berger:2010xi} up to NNLL accuracy. In the present work, we extend this calculation to NNLL$^\prime$ and N$^3$LL.
For the determination of the canonical scales we employ the $\Tau_0$-dependent profile functions
described e.g.~in sec.~3 of \refcite{Alioli:2019qzz} with $ \{x_0,
x_1,x_2,x_3\} =  \{1.5 \GeV/m_H, 0.2, 0.35, 0.5 \}$.
The use of such $\Tau_0$ dependent scales is known to cause a difference between the integrated spectrum and the cumulant, which is formally of higher order. This is a result of the noncommutativity of the scale setting and the integration steps.
In previous \geneva implementations, this problem has been alleviated by explicitly adding higher order terms to restore the cumulant cross section (see eq.~(45) of \refcite{Alioli:2019qzz}).
This can be done either by using a `brute-force' approach, in which the integrated spectrum is simply replaced by the cumulant, or by smoothly transitioning from one to the other as a function of $\Tau_0$. In all \geneva implementations thus far we have followed the latter approach, which has the advantage of preserving the $\Tau_0$ spectrum in its peak region.

In the case of $gg \to H$ production at $13 \TeV$, the difference between the integrated spectrum and the cumulant amounts to $\sim 18 \%$ of the total cross section.
Given the size of these corrections, we found the previously adopted solution to be insufficient to completely solve the mismatch. In particular, our smooth fix modifies the $\Tau_0$ spectrum in the region between $\sim 10$ and $\sim 30 \GeV$ by too large an amount, moving the central value of the first outside the uncertainty bands of the second.
We therefore revert to the brute-force approach, and only require the preservation of the resummed cumulant cross section by fixing $\kappa(\Tau_0) =1$ (see eq.~(45) of \refcite{Alioli:2019qzz}) such that the spectrum is exactly equal to the derivative of the cumulant.

%%%%%%%%%%%%%%%%%%%%%%%%%%%%%%%%%%%%%%%%%%%%%%%%%%%%%%%%%%%%%%%%%%%%%%%%%%%%%%%%
\section{Novel features of the G{\scriptsize ENEVA} method}
\label{sec:features}
%%%%%%%%%%%%%%%%%%%%%%%%%%%%%%%%%%%%%%%%%%%%%%%%%%%%%%%%%%%%%%%%%%%%%%%%%%%%%%%%

In this section we discuss the new improvements that have been incorporated in the \geneva method.
Here we focus on their impact on the $gg \to H$ process,
however we note that they can be straightforwardly generalised to several other processes (and indeed have already been tested for Drell-Yan, double Higgs \cite{Alioli:2022dkj}, and $t\bar{t}$ production \cite{Alioli:2021ggd}).

%%%%%%%%%%%%%%%%%%%%%%%%%%%%%%%%%%%%%%%%%%%%%%%%%%%%%%%%%%%%%%%%%%%%%%%%%%%%%%%%
\subsection{Improved treatment of splitting functions}
\label{sec:splitting}
%%%%%%%%%%%%%%%%%%%%%%%%%%%%%%%%%%%%%%%%%%%%%%%%%%%%%%%%%%%%%%%%%%%%%%%%%%%%%%%%

\newcommand{\sss}{\mathchoice %
{\displaystyle} %
{\displaystyle} %
{\scriptscriptstyle} %
{\scriptscriptstyle} %
}
\newcommand{\biggwhite}{{\color{white} \bigg|}}
\newcommand{\DS}{\displaystyle} % fix formulae in arrays and fractions
\newcommand{\EI}{\<\<\<\<} % fix spacing in eqnarrays
\renewcommand{\L}{\left}
\newcommand{\R}{\right}
\renewcommand{\>}{\,} % +3mu
\newcommand{\<}{\!} % -3mu
\renewcommand{\b}{\bar} % bar
\newcommand{\h}{\hat} % hat
\renewcommand{\t}{\tilde} % tilde
\newcommand{\CF}{C_F}
\newcommand{\CA}{C_A}
\newcommand{\TF}{T_F}
\newcommand{\alphaS}{\alpha_s}
\newcommand{\fac}{{\sss\rm F}}
\newcommand{\ren}{{\sss\rm R}}
\newcommand{\muF}{\mu_F}
\newcommand{\muR}{\mu_R}
\newcommand{\real}{{\sss\rm real}} % real
\newcommand{\AP}{{\sss\rm AP}} % Altarelli-Parisi
\newcommand{\CS}{{\sss\rm CS}} % colour singlet
\newcommand{\MAX}{{\sss\rm max}} % maximum
\newcommand{\MIN}{{\sss\rm min}} % minimum

The $N$-jettiness spectra computed through resummation techniques
cannot be directly used for generating events with $N+1$ final-state
partons, since they do not carry a dependence on the full $\Phi_{N+1}$
configurations, but only on $\Tau_N$ and the projected $\Phi_N$
configurations with $N$ final-state partons.
For this reason, a splitting function $\mathcal{P}(\Phi_{N+1})$
was introduced in \refcite{Alioli:2012fc} in order to make
the resummed calculation fully differential in the higher order phase space.

In general, the $N \to N+1$ splitting function $\mathcal{P}\<\L(\Phi_{N+1}\R)$ is defined such that for every integrable function
$g\<\L(\Phi_N,\Tau_N\R)$
\begin{equation}
  \label{eq:splitting_function_condition}
  \int \df\Phi_{N+1} \> \mathcal{P}\<\L(\Phi_{N+1}\R) \, g\<\L(\Phi_N,\Tau_N\R)= \int \df\Phi_N \, \df\Tau_N \>
  g\<\L(\Phi_N,\Tau_N\R)\,.
\end{equation}
If the function $g\<\L(\Phi_N,\Tau_N\R)$ is the $\Tau_N$ spectrum, then multiplying it by the $\mathcal{P}$
functions makes it differential over the $\df\Phi_{N+1}$ phase space
without affecting the distributions of observables that only
depend on $\Phi_N$ and $\Tau_N$.

In order to provide an explicit expression for $\mathcal{P}$, we
write the phase space of the $\Phi_{N+1}$ configurations
with a valid $\Phi_N$ projection as the product of $\df\Phi_N$,
$\df\Tau_N$ and the phase space parametrised by two additional radiation
variables $z$ and $\phi$. In this way the integral over
the projectable $\Phi_{N+1}$ configurations at fixed $\Phi_N$ and
$\Tau_N$ can be expressed as
\begin{equation}
  \int \frac{\df\Phi_{N+1}}{\df\Phi_N \> \df\Tau_N} = \sum_{k=1}^{N+2}
  \int_{z^\MIN_k\<\L(\Phi_N, \Tau_N\R)}^{z^\MAX_k\<\L(\Phi_N,
    \Tau_N\R)} \df z \> J_k\<\L(\Phi_N, \Tau_N, z\R)
  \int_{\phi^\MIN_k\<\L(\Phi_N, \Tau_N, z\R)}^{\phi^\MAX_k\<\L(\Phi_N,
    \Tau_N, z\R)} \df\phi,
\end{equation}
where the index $k$ runs over the $N+2$ possible emitter partons (mothers) of the $\Phi_N$ configurations.
For each mother $k$ and its associated mapping, we assume that the Jacobian
\begin{equation}
  J_k\<\L(\Phi_N, \Tau_N, z\R) = \L.\frac{\df\Phi_{N+1}}{\df\Phi_N \>
    \df\Tau_N \> \df z \> \df\phi}\R|_k
\end{equation}
does not depend on $\phi$. This is true for all the mappings
considered in this paper.
The
integral over the $\Phi_{N+1}$ configurations summed over the
$n_\real$ partonic subprocesses with $N+1$ final-state partons for a generic function $g_\beta (\Phi_{N+1})$ can now be written
as
\begin{eqnarray}
  && \sum_{\beta=1}^{n_\real} \int \df\Phi_{N+1} \, g_\beta (\Phi_{N+1}) =
  \sum_{\beta=1}^{n_\real} \int_{{\rm unproj.~} \Phi_{N+1}}
  \df\Phi_{N+1} \, g_\beta (\Phi_{N+1}) + \\
  && \quad \sum_{\alpha=1}^{n_{\rm Born}} \int \df\Phi_N \, \df\Tau_N
  \sum_{k=1}^{N+2} \int_{z^\MIN_k}^{z^\MAX_k} \< \df z \>
  J_k\<\L(\Phi_N, \Tau_N, z\R) \int_{\phi^\MIN_k}^{\phi^\MAX_k} \< \df\phi
  \sum_{j=1}^{n^{\rm split}_k}  \, g_\alpha^{k \to i+j} (\Phi_{N}, \Tau_N, z, \phi)\,, \nonumber
\end{eqnarray}
where $n_{\rm Born}$ is the number of subprocesses with $N$ final-state
partons, and $n^{\rm split}_k$ the number of possible QCD splittings
$k \to i+ j$, with $i$ the emitted parton and $j$ the sister.
The function $g_\alpha^{k \to i+j} (\Phi_{N}, \Tau_N, z, \phi)$ on the right-hand side is equal to $g_\beta (\Phi_{N+1})$ expressed in terms of the underlying Born process index $\alpha$ and the splitting indices $k$ and $j$.
For ease of notation, the full dependence of the $z$ and $\phi$ integration limits on the phase space variables is not shown.
The unprojectable $\Phi_{N+1}$ configurations are those for
which either the two closest partons do not represent a valid QCD splitting,
the $\Phi_N$ configuration obtained from the projection is not
kinematically allowed, or the flavour configuration of the $\Phi_N$ is invalid.

In order to fulfil the condition presented in
\eq{splitting_function_condition}, we choose splitting
functions $\mathcal{P} \<\L(\Phi_{N+1}\R)$ that depend on the mother and sister indices and vanish in the unprojectable $\Phi_{N+1}$ configurations:
\begin{equation}
\mathcal{P} (\Phi_{N+1}) = \Bigg\{
  \begin{array}{l l}
  0 & \mbox{if $\Phi_{N+1}$ is unprojectable,}
  \\
  \mathcal{P}_{kj} (\Phi_N, \Tau_N, z, \phi) \quad & \mbox{if $\Phi_N \to \Phi_{N+1}$ via the $k \to i + j$ splitting.}
  \end{array}
\end{equation}

The $\mathcal{P}_{kj}$ must then satisfy the equation
\begin{equation}
  \sum_{k=1}^{N+2} \int_{z^\MIN_k\<\L(\Phi_N,
    \Tau_N\R)}^{z^\MAX_k\<\L(\Phi_N, \Tau_N\R)} \< \df z \>
  J_k\<\L(\Phi_N,\Tau_N,z\R) \int_{\phi^\MIN_k\<\L(\Phi_N, \Tau_N,
    z\R)}^{\phi^\MAX_k\<\L(\Phi_N, \Tau_N, z\R)} \< \df\phi
  \sum_{j=1}^{n^{\rm split}_k} \mathcal{P}_{kj} \<\L(\Phi_N, \Tau_N, z, \phi\R) = 1
\end{equation}
for all values of $\Phi_N$ and $\Tau_N$. Without loss of generality, in
the projectable $\Phi_{N+1}$ configurations we can express them as
\begin{eqnarray}
  \label{eq:splitting_function_phi}
  &&  \mathcal{P}_{kj} \<\L(\Phi_N, \Tau_N, z, \phi\R) =
  \\
  && \quad \frac{\DS f_{kj}\<\L(\Phi_N, \Tau_N, z, \phi\R)}{\DS
    \sum_{k'=1}^{N+2} \int_{z^\MIN_{k'}\<\L(\Phi_N,
      \Tau_N\R)}^{z^\MAX_{k'}\<\L(\Phi_N, \Tau_N\R)} \<\< \df z'
    J_{k'}\<\L(\Phi_N, \Tau_N, z'\R) \int_{\phi^\MIN_{k'}\<\L(\Phi_N,
      \Tau_N, z'\R)}^{\phi^\MAX_{k'}\<\L(\Phi_N, \Tau_N, z'\R)} \<\<
    \df\phi' \< \sum_{j'=1}^{n^{\rm split}_{k'}} f_{k'j'}\<\L(\Phi_N,
    \Tau_N, z' \< , \phi'\R)}\,,
  \nonumber
\end{eqnarray}
where $f_{kj}$ is a generic function that we specify later. If we choose it to be independent of $\phi$,
the above expression simplifies to
\begin{eqnarray}
  \label{eq:splitting_function}
  &&  \mathcal{P}_{kj} \<\L(\Phi_N, \Tau_N, z \R) = \\
  && \quad \frac{\DS f_{kj}\<\L(\Phi_N, \Tau_N, z\R)}{\DS
    \sum_{k'=1}^{N+2} \int_{z^\MIN_{k'}\<\L(\Phi_N,
      \Tau_N\R)}^{z^\MAX_{k'}\<\L(\Phi_N, \Tau_N\R)} \df z' \>
    J_{k'}\<\L(\Phi_N, \Tau_N, z'\R) \Delta \phi_{k'}\<\L(\Phi_N, \Tau_N,
    z'\R) \sum_{j'=1}^{n^{\rm split}_{k'}} f_{k'j'}\<\L(\Phi_N,
    \Tau_N, z'\R)}\, , \nonumber
\end{eqnarray}
where $\Delta \phi_k\<\L(\Phi_N,\Tau_N,z\R) = \phi^\MAX_k\<\L(\Phi_N,\Tau_N,z\R) - \phi^\MIN_k\<\L(\Phi_N,\Tau_N,z\R)$.

In order to perform the integral in the denominator of
\eq{splitting_function}, we compute the integration limits on $z$ and $\phi$ and the Jacobian $J_k$
both for the $0 \to 1$ and $1 \to 2$ splitting mappings for each $\Phi_{N+1}$ configuration.
In the previous \geneva implementation of the splitting functions the computation of the integration limits was avoided
by precomputing their upper bounds and then using a
``hit-or-miss'' integration method.
We highlight that, whenever the constraints on $z$ and $\phi$
are in the form of an inequality involving both the variables,
we only compute an overestimate of the true integration limits on $z$ analytically.
We then determine the true limits numerically by imposing the condition $\Delta \phi_k\<\L(\Phi_N,\Tau_N,z\R) > 0$.

\subsubsection{Infrared limits}

For this section we introduce the acronyms ISRA
(initial-state radiation A), ISRB (initial-state radiation
B), and FSR (final-state radiation) to indicate the $N+2$
possible mothers we have to deal with: the parton from the first (A)
and second (B) beam, respectively, and the final-state partons. We furthermore label
ISRA and ISRB collectively as ISR.

The exact form of the function $f_{kj}$ in \eq{splitting_function_phi} can significantly affect the efficiency of the
Monte Carlo event generator in the region of small $\Tau_N >
\Tau_N^\cut$. In this region, the logarithmically enhanced terms appearing in
the fixed-order calculation have to cancel those coming from the
resummed-expanded contributions. For this reason the main criterion we
follow in the choice of $f_{kj}$ is to achieve a good approximation of the behaviour of the associated matrix element in the infrared
limit when $\Tau_N \to 0$.

For simplicity, in practical applications we choose not to include the azimuthal dependence in the form of the $f_{kj}$ functions, using \eq{splitting_function}.
We define
\begin{equation}
  \label{eq:Altarelli-Parisi_splitting_functions}
  f_{kj}\<\L(\Phi_N, \Tau_N, z\R) = \left\{
  \begin{array}{l l}
  \DS \alphaS\<\L(\muR\R) f_a^A\<\L(x_a,\muF\R)
  f_{b}^B\<\L(x_b,\muF\R) z \> \hat{P}_{jk}\<\L(z\R) \quad & \mbox{if
    $k$ is ISR,}
  \\
  \DS \alphaS\<\L(\muR\R) f_{a}^A\<\L(x_a,\muF\R)
  f_{b}^B\<\L(x_b,\muF\R) \hat{P}_{kj}\<\L(z\R) & \mbox{if $k$ is
    FSR,}
  \end{array}
  \right.
\end{equation}
where $a$ and $b$ are the initial-state partons, $\alphaS\<\L(\muR\R)$ is the strong coupling
evaluated at the renormalisation scale $\muR$, and $f_i^H\<\L(x_i,\muF\R)$
is the PDF of the parton $i$ in the hadron $H$ evaluated at longitudinal
momentum fraction $x_i$ and factorisation scale $\muF$.
The renormalisation and factorisation scales are fixed to $\muR = \muF = Q$,
where $Q$ is the virtuality of the colour singlet system.
The
$\hat{P}_{kj}$ are the unregulated Altarelli-Parisi splitting functions
\begin{alignat}{2}
\label{eq:unregulated_dglap_P}
&\hat{P}_{qq}\<\L(z\R) = \CF \> \frac{1+z^2}{1-z}\,, \quad
&&\hat{P}_{qg}\<\L(z\R) = \TF \L[z^2 + \L(1-z\R)^2\R]\,, \nonumber \\
&\hat{P}_{gq}\<\L(z\R) = \CF \> \frac{1+\L(1-z\R)^2}{z}\,, \qquad
&&\hat{P}_{gg}\<\L(z\R) = 2\CA \L[\frac{z}{1-z} + \frac{1-z}{z} +
z\L(1-z\R)\R] \,.
\end{alignat}

We highlight that for the $0 \to 1$ splitting, connecting events with no extra partons to events with one extra parton, the PDFs are evaluated at the exact momentum
fractions $x_a (z)$ and $x_b (z)$ of the real emission phase space $\Phi_{1}$ rather than their infrared
limits. This has proven to be necessary to obtain an accurate
description also in the tail of the colour singlet
transverse momentum distribution.
We note that in this case we also reproduce the correct soft limit, as shown in \sec{0to1splitting}.
For the $1 \to 2$ splitting the true $x_a$ and $x_b$ also depend on $\phi$. In this case they are approximated by dropping this additional dependence, which still
represents an improvement with respect to the strict collinear limit.

\subsubsection{Soft limit of the $0 \to 1$ splitting}
\label{sec:0to1splitting}

In the following we show that the expression of $f_{kj}$ introduced in
\eq{Altarelli-Parisi_splitting_functions} correctly reproduces both
the singular soft and collinear limits at $\ord{\alphaS}$ in the $0
\to 1$ splitting.

In the case of colour singlet production in hadron-hadron collisions,
let us consider the $k \to i + j$ splitting connecting the Born matrix element $\mathcal{B}_0$ and the real matrix element $\mathcal{B}_1$ (in both cases excluding parton densities).
This can be expressed in terms of the FKS
variables $\xi = 2 \, E / \sqrt{s} = 1 - Q^2 / s$ and $y = \cos \theta$. Here,
$s$ is the partonic centre-of-mass energy squared, and $E$ and $\theta$ are the energy of the emitted parton and the angle between the emitted and the right-moving incoming parton in the partonic centre-of-mass frame, respectively.

In the soft limit of the emitted particle $i$, we have
\begin{equation}
  \label{eq:0to1_soft_limit}
  \lim_{\xi \to 0} \mathcal{B}_1 = \frac{64\pi\alphaS\<\L(\muR\R)}{Q^2}
  \frac{C_{k}}{\xi^2\L(1-y^2\R)} \> \mathcal{B}_0,
\end{equation}
where $C_{k} = \CF$ for the quark-initiated processes and $C_{k} =\CA$
for the gluon-initiated.

In the azimuthally averaged collinear limit between particles $i$ and $j$, we have
\begin{equation}
  \label{eq:0to1_collinear_limit}
  \lim_{y \to \pm 1} \mathcal{B}_1 = \frac{16\pi\alphaS\<\L(\muR\R)}{Q^2}
  \frac{1-\xi}{\xi\L(1\mp y\R)} \> \hat{P}_{jk}\<\L(1-\xi\R) \mathcal{B}_0 \,,
\end{equation}
where $y \to 1$ and $y \to -1$ represent the collinear limits with respect to incoming parton $a$ and $b$ respectively.
If the colour singlet production process is quark-initiated or has only
scalar particles in the final state, the
above expressions also hold prior to averaging over the azimuthal
angle.

We consider a configuration with one final-state parton with momentum $p$,
where
\begin{align}
\label{eq:tau0_z_definition}
  \Tau_0 &= \h{p}^\pm \,,  \nonumber \\
  z &= Q / (Q + \h{p}^\mp) \,.
\end{align}
Here $\h{p}$ is obtained by longitudinally boosting $p$ from the
laboratory frame to the frame where the colour singlet has zero
rapidity, and $\h{p}^\pm = \h{p}_0 \mp \h{p}_3$.
We have chosen $z$ such that in the collinear limit it reduces to the energy fraction of the emitter with respect to the sister, while providing the correct scaling also for the single soft limit.

In order to show that the singular limits in
\eq{Altarelli-Parisi_splitting_functions} reproduce the above
results, we rewrite $\Tau_0$ and $z$ in terms of the FKS variables $\xi$ and $y$,
and then compare the ensuing expression to \eqs{0to1_soft_limit}{0to1_collinear_limit}. They read
\begin{align}
  \label{mCS_yCS_Tau0_z}
    \DS \Tau_0 &= \frac{Q\, \xi}{2 \sqrt{1-\xi}} \> \L(1\mp y\R)
    \sqrt{\frac{2-\xi\L(1\pm y\R)}{2-\xi\L(1\mp y\R)}} \,,
    \nonumber \\
    \DS z &= \L( \DS 1 + \frac{\xi\L(1\pm y\R)}{2\sqrt{1-\xi}}
      \sqrt{\frac{2-\xi\L(1\mp y\R)}{2-\xi\L(1\pm y\R)}}\, \R)^{-1} \, .
\end{align}
Therefore in the infrared singular limit one obtains
\begin{alignat}{2}
  \Tau_0 &\to Q\, \frac{\xi}{2}\L(1\mp y\R) \,, &&  \\
  z &\to 1-\frac{\xi}{2}\L(1\pm y\R) \qquad && \text{in the soft limit,} \\
  z &\to 1-\xi \qquad && \text{in the collinear limit.}
\end{alignat}

Multiplying the NLL singular $\Tau_0$ spectrum expanded at $\mathcal{O} (\alpha_s)$ by the splitting functions, in the infrared limit we find
\begin{equation}
\label{eq:dsigmaNLOIRlimits}
\mathcal{P}_{kj} (\Phi_0, \Tau_0, z) \, \left. \frac{\df \sigma^\text{NLL}}{\df \Phi_0 \df \Tau_0} \right|_{\mathcal{O}(\alpha_s)}
\to
  \frac{8\pi\alphaS\<\L(\muR\R)}{Q\,\Tau_0} \>
  z \> \hat{P}_{jk}\<\L(z\R) \>
  f_a (x_a, \mu_F) \, f_b (x_b, \mu_F) \>
  \mathcal{B}_0 (\Phi_0) \,,
\end{equation}
up to power corrections.
By using the above expressions for
$\Tau_0$ and $z$, it can be shown that this
reproduces both the soft and collinear limits given in
\eqs{0to1_soft_limit}{0to1_collinear_limit}.
We remark that with the choice of $z$ given in \eq{tau0_z_definition} the soft limit can be entirely captured by using the Altarelli-Parisi splitting collinear kernels. The validity of \eq{dsigmaNLOIRlimits} can be understood to be a consequence of the fact that the noncusp soft anomalous dimension is zero at one loop order, resulting in the lack of a single logarithmic contribution to the $\Tau_0$ spectrum coming from the soft function.

\subsubsection{Numerical validation}

%%%%%
\begin{figure}[t]
   \centering
   \includegraphics[width=\rescaletwoplots]{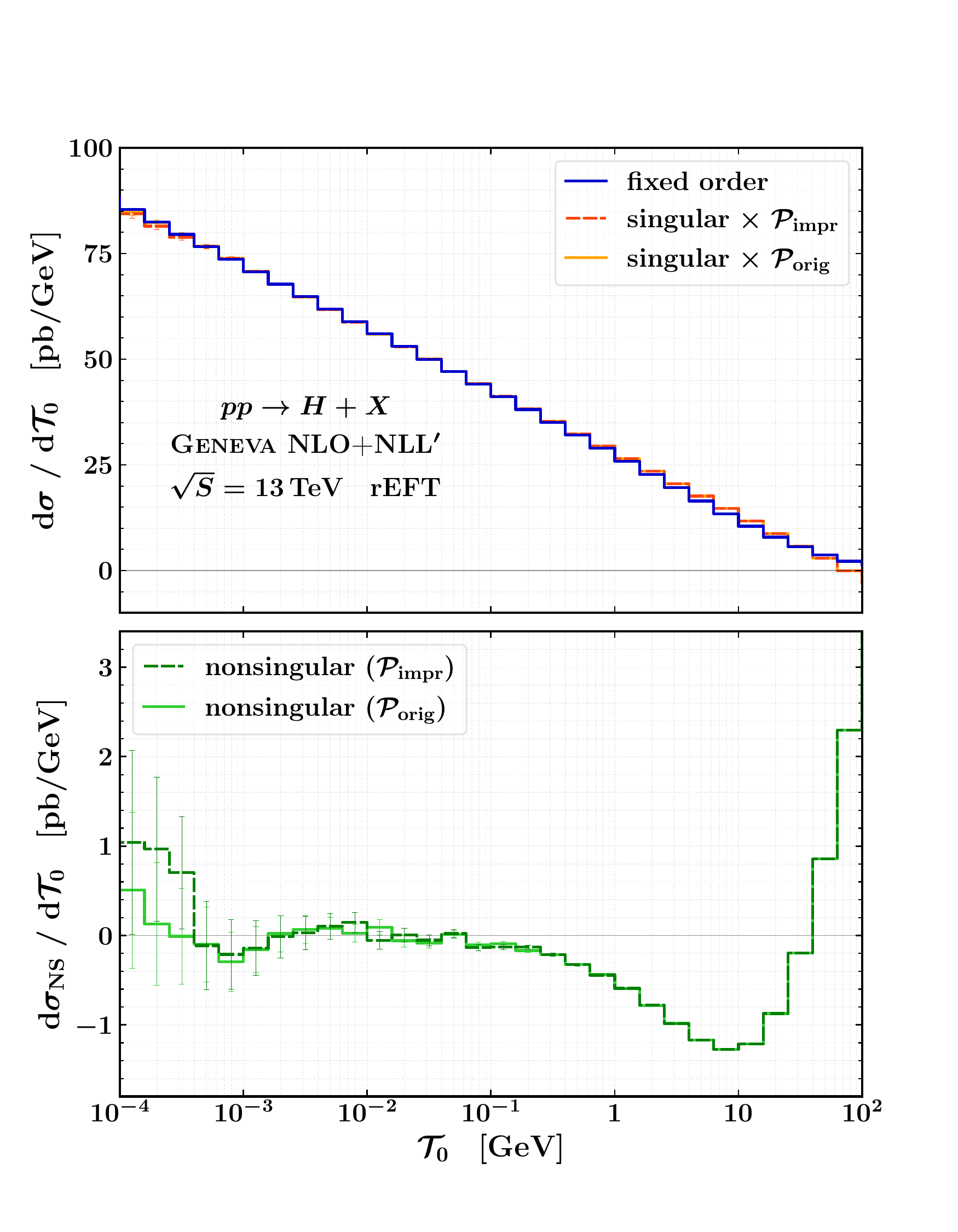}%
   \includegraphics[width=\rescaletwoplots]{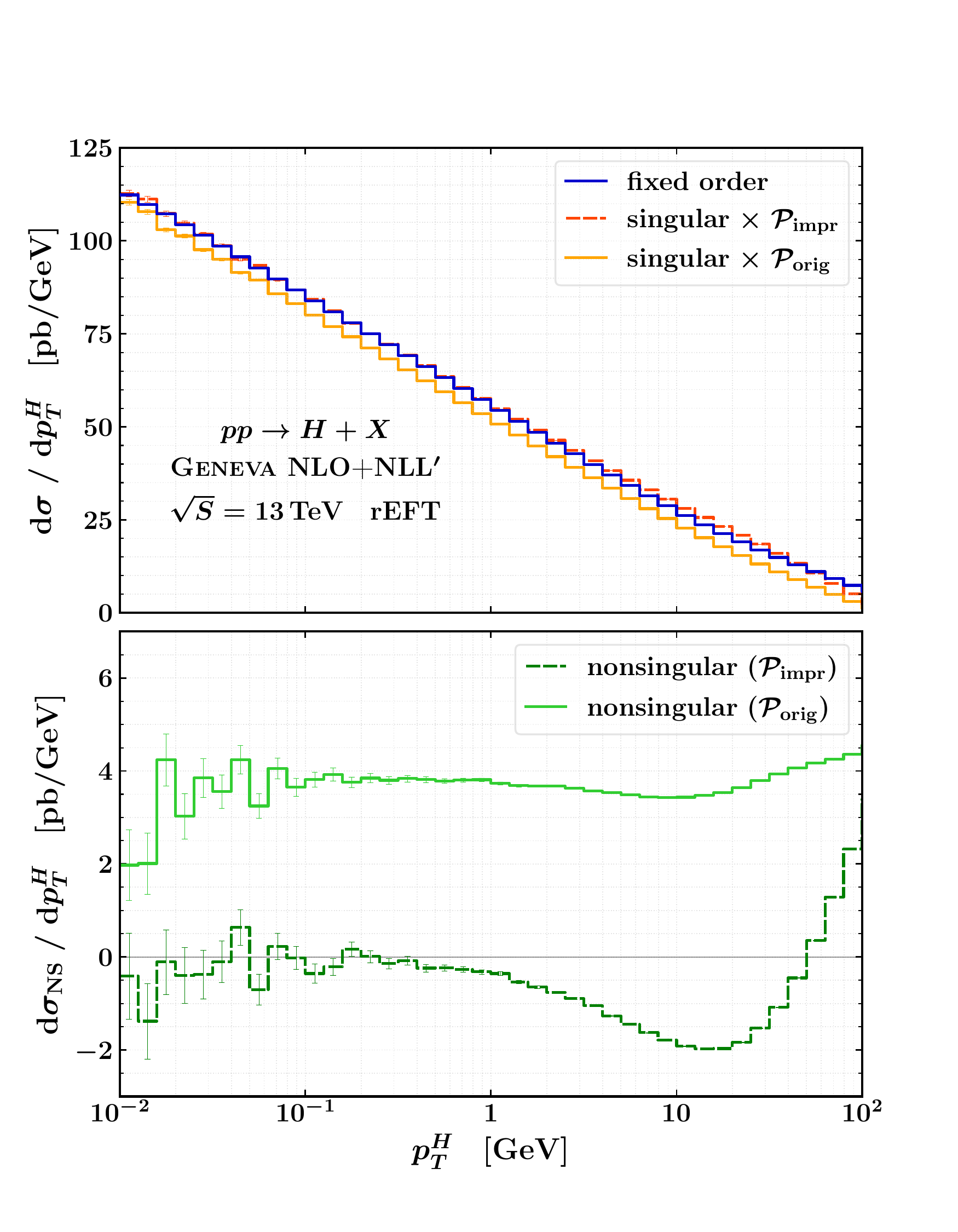}%
   \caption{Comparison of the fixed-order, singular, and nonsingular distributions at NLO+NLL$^\prime$, both for $\Tau_0$ (left) and $p_T^H$ (right). We show the singular and nonsingular distributions both for the original and improved versions of the splitting function implementation in \geneva.}
   \label{fig:splitting_nonsing_nll}
\end{figure}
%%%%%

%%%%%
\begin{figure}[t]
   \centering
   \includegraphics[width=\rescaletwoplots]{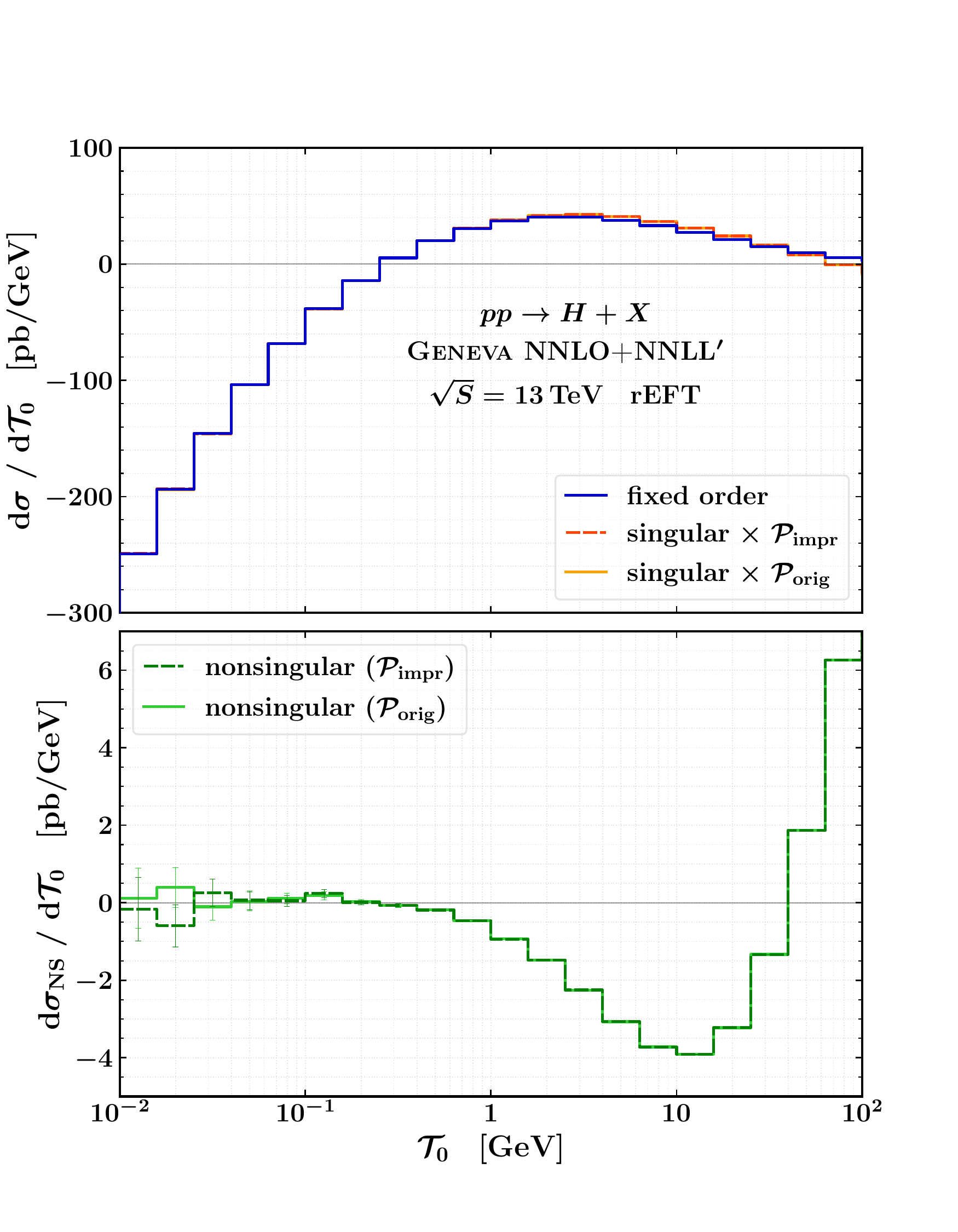}%
   \includegraphics[width=\rescaletwoplots]{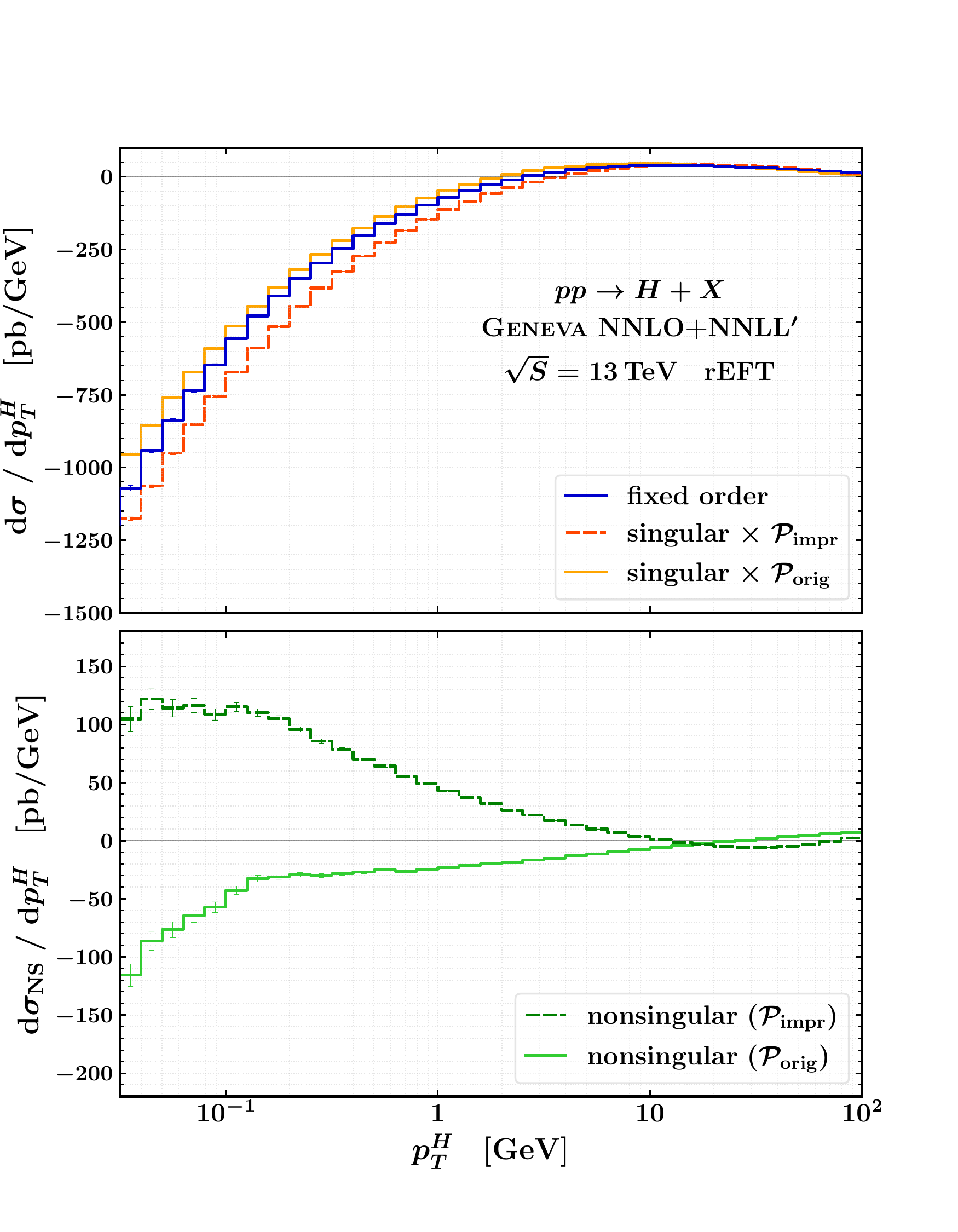}%
   \caption{Comparison of the fixed-order, singular, and nonsingular distributions at NNLO+NNLL$^\prime$, both for $\Tau_0$ (left) and $p_T^H$ (right). We show the singular and nonsingular distributions both for the original and improved versions of the splitting function implementation in \geneva.}
   \label{fig:splitting_nonsing_nnll}
\end{figure}
%%%%%

In this section we present the effects of the improved splitting function $\mathcal{P}_\mathrm{impr}$ implementation
described above
in the case of Higgs boson production via gluon fusion, setting $Q = m_H$;
we focus on the $p_T^H$ and the $\Tau_0$ spectrum.
We compare the results of a fixed-order calculation with those obtained by truncating the resummation formula in \eq{standardresum} multiplied by the splitting function to the same order.
We do so for the results at LO$_1$ compared with the NLL$^\prime$ resummed-expanded in \fig{splitting_nonsing_nll}, and for those at NLO$_1$ compared with the NNLL$^\prime$ resummed-expanded in \fig{splitting_nonsing_nnll}.
We also show the nonsingular contribution, defined as the difference between these fixed-order and resummed-expanded pieces. In all plots we also show the results obtained with the original implementation $\mathcal{P}_\mathrm{orig}$ of the splitting function in \eq{splitting_function}, which was based on a hit-or-miss method using upper bounds tabulated on a grid.

We begin the discussion with the results for the $\Tau_0$ distribution. As
expected, the improved implementation gives identical results to the original,
both at LO$_1$ and NLO$_1$. This is a consequence of the fact that $\Tau_0$ is
preserved by the splitting, by construction.  We observe that at extremely
low values of $\Tau_0$ the presence of technical cuts in the fixed-order calculation affects
the convergence to the singular predictions in both approaches.
When instead considering the LO$_1$ results for the $p_T^H$ distribution, we  notice how the improved implementation of the splitting functions correctly captures the logarithmic behaviour of the matrix element at fixed order. This can be seen by the fact that the improved nonsingular distribution converges to zero, contrary to the original case which converges to a finite value.
Similarly, an improvement is also visible for the NLO$_1$ case. Here, however, the new splitting function $\mathcal{P}_\mathrm{impr}$ is not able to exactly reproduce the complete logarithmic behaviour of the NLO$_1$ result, as it appears to miss a single logarithmic contribution $\sim 1/p_T^H$. This is implied by the fact that the improved nonsingular contribution converges to a nonzero constant at low values of $p_T^H$. This must however be compared with the original approach, $\mathcal{P}_\mathrm{orig}$,
where the divergent behaviour of the nonsingular plot suggests that that implementation also fails to capture
the logarithmic structure up to $\sim \ln^2 (p_T^H)/p_T^H$.

We examine the effects of the $\mathcal{P}_\text{impr}$ implementation on the Drell-Yan process in \app{splitting-dy}, where we compare different \geneva results with the ATLAS experimental data.

%%%%%%%%%%%%%%%%%%%%%%%%%%%%%%%%%%%%%%%%%%%%%%%%%%%%%%%%%%%%%%%%%%%%%%%%%%%%%%%%
\subsection{Independent scale variations}
\label{sec:independent_scale_vars}
%%%%%%%%%%%%%%%%%%%%%%%%%%%%%%%%%%%%%%%%%%%%%%%%%%%%%%%%%%%%%%%%%%%%%%%%%%%%%%%%

In traditional implementations of fixed-order QCD calculations,
a differentiation is made between the factorisation scale $\mu_F$ and the renormalisation scale $\mu_R$. The former is associated with the scale of collinear factorisation, while the latter is introduced in dimensional regularisation in order to render the strong coupling dimensionless.

To date, implementations of \geneva have assumed these scales to be equal.
Doing so facilitated the matching to the resummed calculation,
where a sole ``nonsingular'' scale $\mu_\NS$ appears as the endpoint of the RGE running, typically taken to be a hard scale $Q$ of the problem.
The two scales were then varied in a correlated fashion (``diagonal'' in the $\{ \mu_R,\, \mu_F \}$ space) when probing the higher order uncertainties.
This approach, however, can hinder a complete and thorough uncertainty estimation
as it neglects those variations which are off-diagonal, i.e.~where
$\mu_R$ and $\mu_F$ are varied independently.
In this section we provide an improved and robust uncertainty estimation within the \geneva framework
by exposing the $\mu_F$ dependence of the singular cross section that eventually allows for off-diagonal scale variations, and discuss the choice
of $\mu_F$ in the infrared region.

%===============================================================================
\subsubsection{Exposing the $\mu_F$ dependence of the singular cross section}
\label{sec:muF_dependence_singular_xsec}
%===============================================================================

The collinear beam functions $B_i$ entering the $\Tau_0$ factorisation in \eq{tau0_factorisation_def} satisfy the OPE in \eq{beam_func_ope_def}.
In resummed predictions, they are evaluated at a scale $\mu = \mu_B$ where $\mu_B$ minimises the singular
logarithmic structure of $B_i$, whereas at fixed order $\mu = \mu_R = \mu_F = Q$, where for example $Q=m_H$ for on-shell Higgs boson production.

In order to expose the $\mu_F$ dependence of the beam functions, we rewrite \eq{beam_func_ope_def} as
%%%
\begin{align} \label{eq:beam_func_muF_all_orders}
B_i(t, x, \mu)
&= \sum_j \cI_{ij}(t, x, \mu)  \otimes_x f_j(x, \mu)
\nn \\
&= \sum_{j,k} \cI_{ik}(t, x, \mu) \otimes_x \mathcal{U}_{kj}(x, \mu, \mu_F) \otimes_x f_j(x, \mu_F)
\nn \\
&\equiv \sum_j \hat{\cI}_{ij} (t, x, \mu, \mu_F) \otimes_x f_j(x, \mu_F)
\,,\end{align}
%%%
where we dropped the power corrections.
Here we evolve the PDFs from $\mu_F$ to $\mu$ using the evolution kernel $\mathcal{U}_{ij}(x, \mu, \mu_F)$ that results from the solution
of the DGLAP equations,
%%%
\begin{align} \label{eq:dglap_eq_def}
\mu \frac{\df}{\df \mu} f_i(x, \mu)
&= 2 \sum_j P_{ij}(x, \mu) \otimes_x f_j(x, \mu)
\,,\end{align}
%%%
and we follow the conventions of \refcite{Billis:2019vxg} for the perturbative expansion of the splitting kernels $P_{ij}(x, \mu)$.%
\footnote{The splitting functions used here are multiplied by a factor of 2 with respect to the unregulated ones defined in \eq{unregulated_dglap_P}.}
Although the $\mu_F$ dependence in \eq{beam_func_muF_all_orders} cancels exactly between the PDFs and the evolution kernel,
as soon as $\hat{\cI}_{ij}$ is truncated at a given order, a residual $\mu_F$
dependence appears in the beam function,
%%%
\begin{align}
B_i(t, x, \mu) \mapsto B_i(t, x, \mu, \mu_F)
\,.\end{align}
%%%

In order to manifest this dependence explicitly, we note that
the matching coefficients and the evolution kernel in \eq{beam_func_muF_all_orders} admit perturbative expansions in the strong coupling constant,
%%%
\begin{align}
\hat{\cI}_{ij}(t, x, \mu, \mu_F)
&= \delta(t) \id(x) +
\sum_{n=1}^\infty \hat{\cI}^{(n)}_{ij}(t, x, \mu, \mu_F) \Bigl(\frac{\alpha_s(\mu)}{4\pi}\Bigr)^n
\,, \\
\mathcal{U}_{ij}(x, \mu, \mu_F)
&= \id(x) + \sum_{n=1}^\infty \mathcal{U}^{(n)}_{ij}(x, \mu, \mu_F) \Bigl(\frac{\alpha_s(\mu)}{4\pi}\Bigr)^n
\end{align}
%%%
where $\id(x) \equiv \delta_{ij} \delta(1-x)$.
We first obtain closed-form expressions for the $\mathcal{U}_{ij}$, which can be achieved either by directly solving \eq{dglap_eq_def}
or by using the solutions of the RGE satisfied by $\cI_{ij}$  (see eq.~(2.17) of \refcite{Billis:2019vxg}).
We have%
%%%
\begin{align} \label{eq:rge_pdf_kern}
\mu \frac{\df}{\df \mu} \mathcal{U}^{-1}_{ij}(x, \mu, \mu_F) \otimes_x f_j(x, \mu)
&= - \mathcal{U}^{-1}_{ij}(x, \mu, \mu_F) \otimes_x \mu \frac{\df}{\df \mu} f_j(x, \mu)
\nn \\
&= -2 \,\, \mathcal{U}^{-1}_{ik}(x, \mu, \mu_F) \otimes_x P_{kj}(x, \mu) \otimes_x f_j(x, \mu)
\nn \\
\Rightarrow \quad
\mu \frac{\df}{\df \mu} \mathcal{U}^{-1}_{ij}(x, \mu, \mu_F)
&= -2 \,\, \mathcal{U}^{-1}_{ik}(x, \mu, \mu_F) \otimes_x P_{kj}(x, \mu)
\,,\end{align}
%%%
where $\mathcal{U}_{ij}^{-1}$ denotes the inverse of $\mathcal{U}_{ij}$.
Here and in the following, repeated flavour indices are implicitly summed over.
We note that \eq{rge_pdf_kern} is exactly the same as eq.~(2.17) of \refcite{Billis:2019vxg} if we set $\gamma_B = \gamma_\nu = 0$.
It is therefore straightforward to use its solution, which up to $\ord{\as^2}$ reads
%%%
\begin{align}
\label{eq:pdf_EK_inv}
\mathcal{U}^{-1\, (0)}_{ij} (x, \mu, \mu_F)
&= \id(x)
\,, \\
\label{eq:pdf_EK_inv_NLO}
\mathcal{U}^{-1\, (1)}_{ij} (x, \mu, \mu_F)
&= - 2\, L\, P_{ij}^{(0)}(x)
\,, \\
\label{eq:pdf_EK_inv_NNLO}
\mathcal{U}^{-1\, (2)}_{ij} (x, \mu, \mu_F)
&= 2\, L^2 \Bigl[P^{(0)}_{ik}(x) \otimes_x P^{(0)}_{kj}(x) -\beta_0 P_{ij}^{(0)}(x)\Bigr] -2 \, L \, P_{ij}^{(1)}(x)
\,,\end{align}
%%%
where we have abbreviated $L= \ln(\mu/\mu_F)$. Solving the closure equation of the evolution kernels
$\mathcal{U}_{ik}^{-1}(x, \mu, \mu_F) \otimes_x \mathcal{U}_{kj}(x, \mu, \mu_F) = \id(x)$ at each order in perturbation theory yields
%%%
\begin{align}
\label{eq:dglap_evol_kern_sol_nlo}
\mathcal{U}_{ij}^{(1)}(x, \mu, \mu_F)
&= 2 \, L \, P_{ij}^{(0)}(x)
\,, \\
\label{eq:dglap_evol_kern_sol_nnlo}
\mathcal{U}_{ij}^{(2)}(x, \mu, \mu_F)
&= L^2 \bigl[2\beta_0 P_{ij}^{(0)}(x) + 2P^{(0)}_{ik}(x) \otimes_x P^{(0)}_{kj}(x) \bigr] + 2 \, L\, P_{ij}^{(1)}(x)
\,.\end{align}
%%%
Substituting \eqscs{dglap_evol_kern_sol_nlo}{dglap_evol_kern_sol_nnlo} in \eq{beam_func_muF_all_orders}, we arrive at the explicit results for the $\mu_F$-dependent
matching coefficients. They read
%%%
\begin{align} \label{eq:cI_nlo_muf_dependent}
\hat{\cI}_{ij}^{(1)}(t, x, \mu, \mu_F)
&=\cI_{ij}^{(1)}(t, x, \mu) + 2  \, \delta(t) \, L \, P^{(0)}_{ij}(x)
\,,\\
\label{eq:cI_nnlo_muf_dependent}
\hat{\cI}_{ij}^{(2)}(t, x, \mu, \mu_F)
&=\cI_{ij}^{(2)}(t, x, \mu)
    + \cL_1(t, \mu^2) \Bigl[ 2\,L\,\Gamma_0^i \, P^{(0)}_{ij}(x) \Bigr]
    \nn \\ &\quad
    + \cL_0(t, \mu^2) \Bigl[ -\gamma_{B\, 0}^i \, L \, P^{(0)}_{ij}(x) + 2\, L \, P^{(0)}_{ik}(x) \otimes_x P^{(0)}_{kj}(x) \Bigr]
    \nn \\ &\quad
    + \delta(t) \bigl\{ L^2 \bigl[2\beta_0 P_{ij}^{(0)}(x) + 2 P^{(0)}_{ik}(x) \otimes_x P^{(0)}_{kj}(x) \bigr]
      \nn \\ & \quad \quad \quad \quad
      + 2 \, L\, \bigl[ P_{ij}^{(1)}(x) + I^{(1)}_{ik}(x) \otimes_x P^{(0)}_{kj}(x) \bigr] \bigr\}
\,,
\end{align}
%%%
where the expressions for the ($\mu_F$-independent) matching coefficients $\cI_{ij}^{(n)}(t, x, \mu)$, the anomalous dimensions $\gamma_{B \,0}^i$ and $\Gamma_0^i$, and the plus distributions $\mathcal{L}_n$ can be found in \refcite{Billis:2019vxg}.

%===============================================================================
\subsubsection{Choice of the factorisation scale}
\label{sec:choice_muF}
%===============================================================================

The choice of the factorisation scale $\mu_F$ is in principle subject to different requirements in the fixed-order and in the resummation region.
In order to minimise the size of the logarithms $L = \ln (\mu/\mu_F)$ in \eqs{cI_nlo_muf_dependent}{cI_nnlo_muf_dependent}, in the resummation region we demand that $\mu_F \sim \mu_B$, i.e.~the beam scale.
On the other hand,
in the fixed-order region, a natural scale setting is
$\mu_F \sim \mu_R \sim \mu_\NS$ so that the fixed-order perturbative convergence is not jeopardised.
However, given that the beam function profile scale $\mu_B (\Tau_0)$ flows to $\mu_\NS$ in the fixed-order region,
%%%
\begin{align}
\mu_B \xrightarrow[]{\Tau_0 \to Q} Q \equiv \mu_\NS
\,,\end{align}
%%%
choosing $\mu_F = \mu_B$ satisfies both conditions.

When considering the scale variations for the estimation of the theoretical uncertainty,
the definition of the profile scales is extended to
%%%
\begin{align}
\label{eq:muS_profile_scale_def}
\mu_S(\Tau_0/Q, \alpha) &= \kappa_R \ \mu_\NS \ f_{\mathrm{run}}(\Tau_0/Q) \, f^{\alpha}_{\mathrm{vary}}(\Tau_0/Q) \, ,\\
\label{eq:muB_profile_scale_def}
\mu_B(\Tau_0/Q, \alpha, \beta) &= \kappa_R \ \mu_\NS \bigl[ f_{\mathrm{run}}(\Tau_0/Q) \, f^{\alpha}_{\mathrm{vary}}(\Tau_0/Q) \bigr]^{1/2 - \beta} \, ,\\
\label{eq:muF_profile_scale_def}
\mu_F(\Tau_0/Q, \alpha_f, \beta)
&= \kappa_F \ \mu_\NS \bigl[ f_{\mathrm{run}}(\Tau_0/Q) \, f^{\alpha_f}_{\mathrm{vary}}(\Tau_0/Q) \bigr]^{1/2 - \beta}
\,,\end{align}
%%%
where the central predictions are obtained setting $\kappa_R=\kappa_F=1$, $\alpha \!=\! \alpha_f \!=\! \beta \!=\! 0$.
The function $f_\mathrm{run}$ is defined in \refcite{Alioli:2019qzz} and the function $f_\mathrm{vary}$ in \refcite{Gangal:2014qda}.
We note that a different parameter is used in the exponent of the function $f_{\mathrm{vary}}$ for $\mu_B$ and $\mu_F$
in \eqs{muB_profile_scale_def}{muF_profile_scale_def} above, while we use the same $\beta$ parameter for both.
This is justified because the $\beta$-variations are introduced in order to disentangle the variations of $\mu_B$ and $\mu_S$;
no ratios of $\mu_F/\mu_S$ appear in the singular cross section, therefore there is no need for an independent $\beta$ parameter in $\mu_F$.

We also extend the fixed-order uncertainty $\Delta_\FO$ to include the off-diagonal $\mu_F$ and $\mu_R = \kappa_R\ \mu_\NS $ scale variations, resulting from the envelope of
a 7-point scale variation
%%%
\begin{align}
\label{eq:fixed_order_variations_kappa}
(\kappa_R, \, \kappa_F)
= \{
(1,\, 1)
,\,
(2,\, 2)
,\,
(1/2,\, 1/2)
,\,
(1,\, 2)
,\,
(1,\, 1/2)
,\, (2,\, 1)
,\, (1/2,\, 1)
\}
\,,\end{align}
where we have excluded the cases when $\mu_F$ and $\mu_R$ are varied in opposite directions.
The resummation uncertainty $\Delta_\text{res}$ estimated by the usual profile scale and transition point variations~\cite{Gangal:2014qda}
(already present in \geneva) is extended by considering in addition the variations of the parameters $\alpha$, $\alpha_f$, and $\beta$ appearing in \eq{muF_profile_scale_def}.
Explicitly, we extend the original procedure to include the additional parameter variations:
%%%
\begin{align}
(\beta, \alpha, \alpha_f) =\{
&(0,\, 0,\, 0)
,\,
(1/6,\, 0,\, 0)
,\,
(-1/6,\, 0,\, 0)
,\, \\
&(0, \, 1, \, 1)
,\,
(0, \, -1, \, -1)
,\,
(0, \, 1, \, 0)
,\,
(0, \, -1, \, 0)
,\,
(0, \, 0, \, 1)
,\,
(0, \, 0, \, -1)
\}
\,. \nonumber
\end{align}
%%%
These variations are enveloped with the variations of the transition points $x_i$, and are summed in quadrature with the fixed-order variations in \eq{fixed_order_variations_kappa} to obtain the final theoretical uncertainty.

%%%%%%%%%%%%%%%%%%%%%%%%%%%%%%%%%%%%%%%%%%%%%%%%%%%%%%%%%%%%%%%%%%%%%%%%%%%%%%%%
\subsection{Treatment of timelike logarithms}
\label{sec:timelike}
%%%%%%%%%%%%%%%%%%%%%%%%%%%%%%%%%%%%%%%%%%%%%%%%%%%%%%%%%%%%%%%%%%%%%%%%%%%%%%%%

Radiative corrections in colour singlet production processes
such as Drell-Yan or gluon-initiated Higgs production contain Sudakov logarithms
of the form $\as^n \ln^m(-q^2/\mu_R^2)\,,\, m\leq 2n$, where $q^\mu$ is the momentum of the
colour singlet system. They primarily arise in the
calculation of the corresponding form factor, and their coefficients are linked
to the structure of its infrared singularities~\cite{Altarelli:1979ub}. For such processes $q^\mu$ is a timelike
vector, i.e.~$q^2>0$, and the scale choice $\mu_R^2 = q^2$ results in the Sudakov logarithms developing an imaginary part, since
$\ln^{m} (-1) = (\pm \mathrm{i} \pi)^m$.
The presence of such `timelike' logarithms at each order
might negatively affect the perturbative convergence of the cross section, where they result in additional terms proportional to powers of $\pi^2$.
The severity of this effect is process specific; for Drell-Yan or gluon-initiated Higgs production it has been explicitly studied for both
exclusive~\cite{Stewart:2010pd, Berger:2010xi, Becher:2012yn, Becher:2013xia, Stewart:2013faa, Jaiswal:2014yba, Gangal:2014qda, Neill:2015roa, Ebert:2016idf}
and inclusive~\cite{Ebert:2017uel, Billis:2021ecs} observables.
A way to mitigate the impact of said contributions is to choose $\mu_R$ such that timelike logarithms are eliminated,
i.e.~to evaluate the form factor at the complex scale $\mu_R = -\mathrm{i} \lvert q \rvert = - \mathrm{i}\, Q$~\cite{Parisi:1979xd, Sterman:1986aj, Magnea:1990zb, Eynck:2003fn}.

In factorised singular cross sections, the square of the form factor naturally appears as the hard function $H (Q^2, \mu_H)$ of the process.
Following the above discussion, the hard function can be evaluated at the complex scale $\mu_H = -\mathrm{i}\, Q$. This choice implies a nontrivial
renormalisation group evolution between $\mu_H$ and the real scale $Q$, in the form of a rotation in the complex $\mu$ plane. This procedure is also referred to as `timelike resummation'. It has been applied to a multitude of exclusive and inclusive processes,
including the case of Higgs boson production, see e.g.~\refscite{Stewart:2010pd, Berger:2010xi, Ahrens:2008qu, Ahrens:2009cxz, Billis:2021ecs, Ebert:2017uel}.

In addition, it has been shown that not only the singular, but also the
nonsingular~\cite{Stewart:2013faa, Berger:2010xi} and the total~\cite{Ebert:2017uel} cross section
can benefit from this prescription, in terms of
an improved perturbative convergence and a reduced ensuing scale uncertainty.
This can be better understood when one considers that by integrating the timelike resummed exclusive cross section, one obtains the corresponding resummed inclusive prediction.
For the nonsingular contribution, a factorisation formula remains in general unknown, meaning that it cannot be directly evaluated at the complex scale $\mu_H$ -- it still, however, contains the form factor plagued by timelike logarithms.
In this case, the procedure for the treatment of these logarithms involves re-expanding the nonsingular contribution, extracting from it the hard function evaluated at $Q$, and replacing it with that evaluated at the scale $\mu_H$, as detailed in~\refcite{Ebert:2017uel}.

In our implementation we perform these steps at the same order as the $\Tau_0$ resummation, both in the singular and the nonsingular
terms.  This implies that the improved perturbative convergence following the choice of a complex-valued scale $\mu_H= -\mathrm{i}\, Q$ does not only apply to the singular $\Tau_0$ spectrum but also to the inclusive predictions.

In order to study the uncertainty associated with this choice of scale, we  follow the prescription introduced in~\refcite{Ebert:2017uel}, designed to probe the structure of the timelike logarithms.
The uncertainty $\Delta_{\varphi}$
is estimated by the envelope of the phase variations
%%%
\begin{align} \label{eq:timelike_vars}
\mu_H = Q \, e^{-\mathrm{i} \varphi}
\,, \qquad
\varphi \in [\pi/4, 3\pi/4]
\,,\end{align}
%%%
while the central value predictions correspond to $\varphi = \pi/2$.
Since there is no dynamical parameter governing the choice of the scale $\mu_H$, the timelike resummation is performed throughout the $\Tau_0$ spectrum,
i.e.~even when $\Tau_0$ resummation is off. We therefore consider the uncertainty resulting from variations in \eq{timelike_vars} as an independent source
and add it to the other uncertainties in quadrature.
Thus, for inclusive predictions we have
%%%
\begin{align}
\Delta_{\mathrm{incl}}^2
= \Delta_{\mathrm{FO}}^2 + \Delta_{\varphi}^2
\,,\end{align}
%%%
whereas for exclusive predictions we use
%%%
\begin{align}
\Delta_{\mathrm{excl}}^2
= \Delta_{\mathrm{FO}}^2 + \Delta_{\mathrm{res}}^2 + \Delta_{\varphi}^2
\,.\end{align}
%%%

\begin{figure}[t]
\centering
\includegraphics[width=\rescaletwoplots]{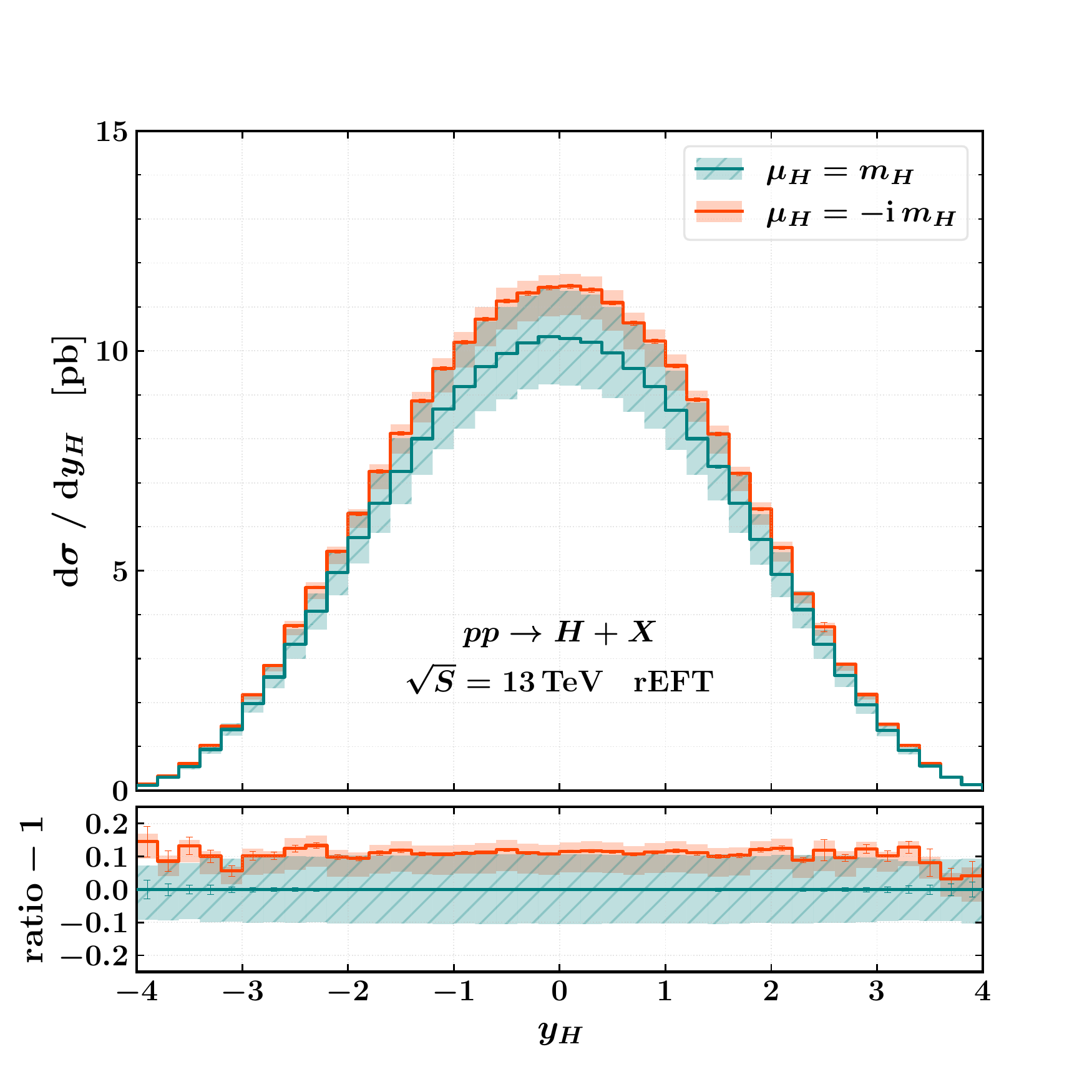}%
\includegraphics[width=\rescaletwoplots]{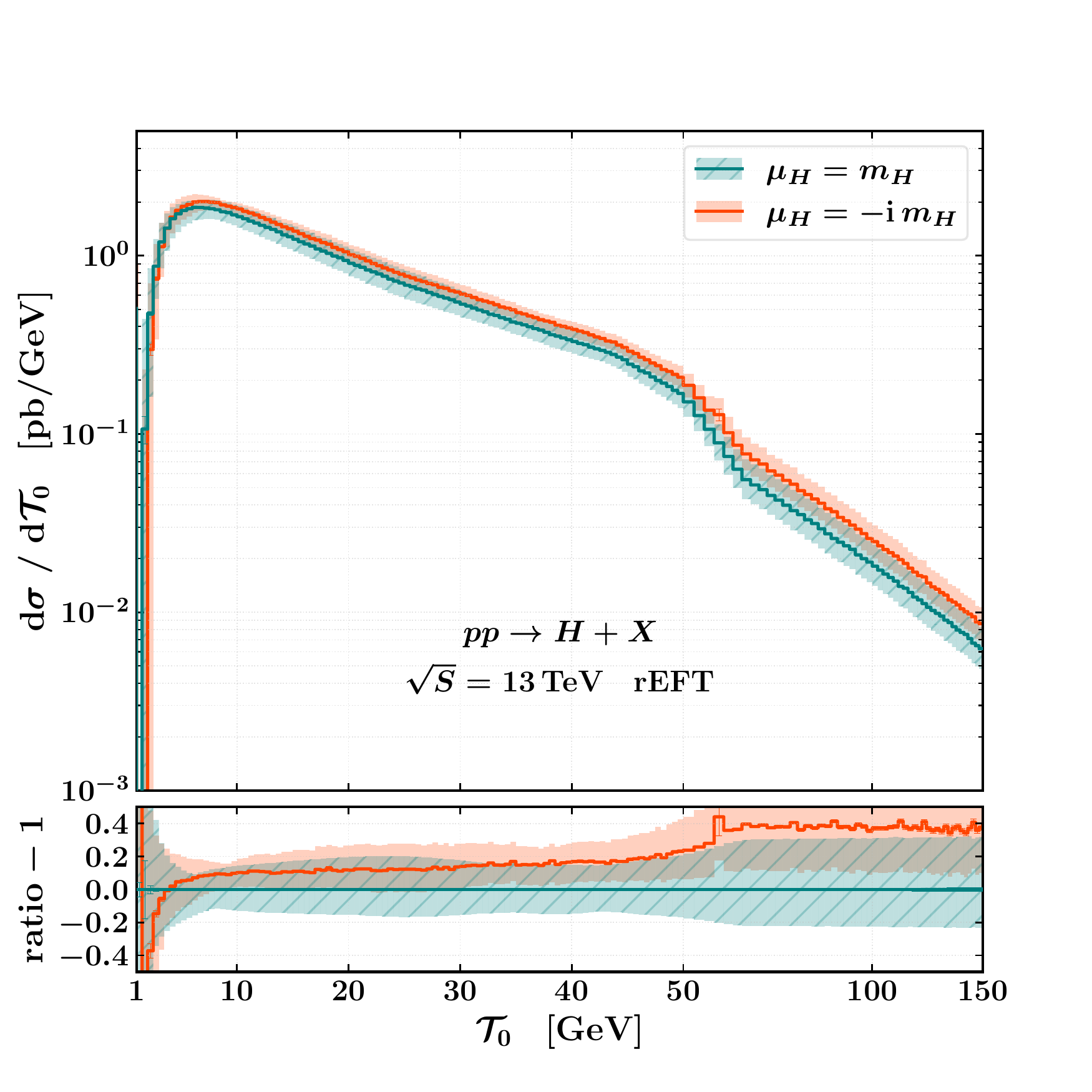}%
\caption{Comparison of the $y_H$ (left) and $\Tau_0$ (right) distributions with different choices of $\mu_H$.}
\label{fig:timelike_tau0}
\end{figure}

In the \geneva implementation of the $gg \to H$ process we use the \texttt{hardfunc} module from \scetlib~\cite{scetlib} for the hard function evaluation and evolution in the complex plane.
Since for this process we set $Q = m_H$, we pick
%%%
\begin{align}
\mu_H = -\mathrm{i}\, m_H
\,.
\end{align}
%%%
With this choice, we observe a difference in the total cross section result with respect to the $\mu_H = m_H$ case that can be substantial despite being formally of higher order.
The effects of the complex choice of scale $\mu_H$ on differential observables are illustrated in \fig{timelike_tau0}, where we compare predictions at NNLO+NNLL$^\prime$ for the $\Tau_0$ and $y_H$ distributions
with $\mu_H = m_H$ and $\mu_H = -\mathrm{i}\, m_H$.
In this and the following figures, the theoretical uncertainty is shown as a shaded band, while the Monte Carlo integration errors are shown as thin vertical bars.
For the Higgs boson rapidity distribution, we observe an increase of around 10\% that is almost independent of $y_H$, and a reduction in the uncertainty band as expected.
The $\Tau_0$ spectrum shows a larger effect, especially in the tail of the distribution, where our prediction is entirely driven by the fixed-order result. Nonetheless, we observe a reduction in the uncertainty band particularly in the peak and transition regions of the spectrum, between $5$ and $45 \GeV$.

%%%%%
%%%%%

%%%%%%%%%%%%%%%%%%%%%%%%%%%%%%%%%%%%%%%%%%%%%%%%%%%%%%%%%%%%%%%%%%%%%%%%%%%%%%%%
\section{Validation of the $gg \to H$ process}
\label{sec:validation_intro}
%%%%%%%%%%%%%%%%%%%%%%%%%%%%%%%%%%%%%%%%%%%%%%%%%%%%%%%%%%%%%%%%%%%%%%%%%%%%%%%%

In this section we validate our predictions.
We first compare our partonic NNLO results with two independent calculations,
and then discuss the interface to the \pythiaEight shower.

%%%%%%%%%%%%%%%%%%%%%%%%%%%%%%%%%%%%%%%%%%%%%%%%%%%%%%%%%%%%%%%%%%%%%%%%%%%%%%%%
\subsection{Partonic results at NNLO}
\label{sec:validation}
%%%%%%%%%%%%%%%%%%%%%%%%%%%%%%%%%%%%%%%%%%%%%%%%%%%%%%%%%%%%%%%%%%%%%%%%%%%%%%%%

Here we validate the NNLO accuracy of the total cross section obtained with \geneva
and that of the only differential inclusive quantity available, the Higgs boson rapidity.
We compare the total cross section with the independent calculations implemented in \texttt{ggHiggs}~\cite{Ball:2013bra, Bonvini:2014jma, Bonvini:2016frm, Bonvini:2018ixe, Bonvini:2018iwt} and \Matrix~\cite{Grazzini:2017mhc},
and the rapidity distribution with \Matrix only.
The \Matrix predictions are based on the $q_T$-subtraction approach and are extrapolated towards the zero $q_T$-cut value.
We set the input parameters of our calculations as described in \sec{ggh_definition},
and we choose the central factorisation and renormalisation scales equal to each other and to the Higgs boson mass, $\mu_R = \mu_F = m_H$.
We set our resolution cutoffs to $\Tau_0^\cut = \Tau_1^\cut = 1 \GeV$.
We employ the PDF set \texttt{PDF4LHC15\_nnlo\_100} from \textsc{LHAPDF}~\cite{Buckley:2014ana},
and take the value of $\alpha_s (m_Z)$ from the same set, so that $\alpha_s (m_H) = 0.11263$.

\begin{table}[t]
\centering
\begin{tabular}{|c|c|c|c|}
\hline
& \geneva & \texttt{ggHiggs} & \Matrix \\
\hline
$\sigma^{\text{NNLO, rEFT}}_{gg \to H}$ [pb] & $42.33^{+4.39}_{-4.34}$ & $42.35^{+4.55}_{-4.41}$ & $42.33^{+4.54}_{-4.40}$ \\
\hline
\end{tabular}
\caption{Comparison of the \geneva, \texttt{ggHiggs}, and \Matrix results for the $gg \to H$ inclusive cross section. The results are obtained at NNLO in the HTL approximation, and rescaled with the $r_\text{EFT}$ factor.}
\label{tab:nnlo_xs_validation}
\end{table}

In \tab{nnlo_xs_validation} we report the values of the inclusive $gg \to H$ cross section and the associated 7-point scale variations calculated at NNLO and rescaled with the rEFT factor using \geneva, \texttt{ggHiggs}, and \Matrix.%
\footnote{The impact of the 7-point scale variations on Higgs production via gluon
fusion is small to moderate. For this process, scale variations are largely driven by $\muR$ variations, and
therefore an independent variation of $\muR$ and $\muF$ leads to a theory uncertainty that is not extremely
different from the one obtained by varying those scales homogeneously. We find that the scale uncertainty
bands increase from roughly $\pm 9\%$ to $\pm 10\%$ for both the total cross-section and the Higgs rapidity distribution.}
We observe excellent agreement between the three predictions; by choosing $\Tau_0^{\rm cut} = 1\GeV$, the neglected power-suppressed
terms in \geneva are at the permille level and amount to an acceptable $\sim 0.02 \text{ pb}$ error for the central value.

\begin{figure}[t]
   \centering
   \includegraphics[width=\rescaletwoplots]{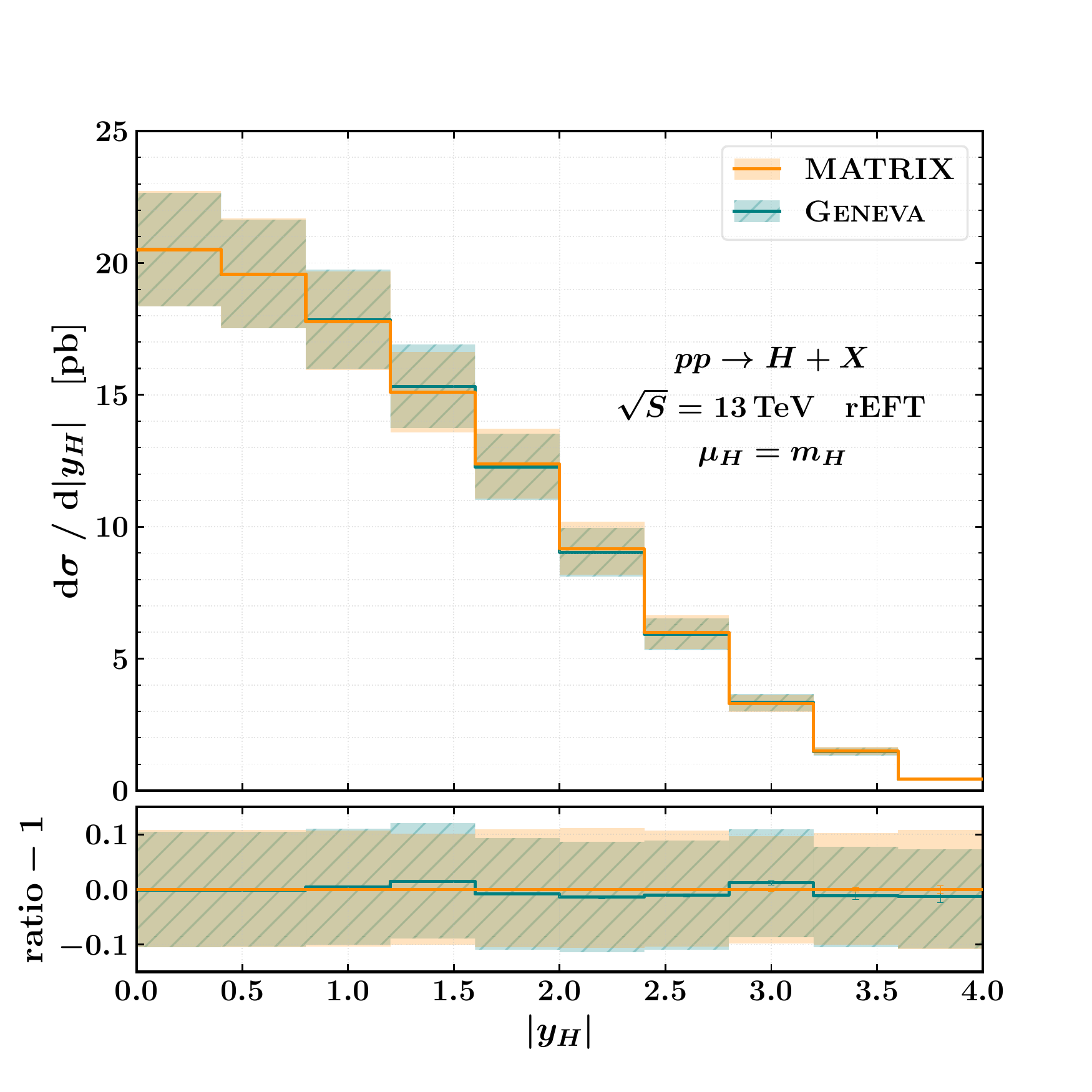}%
   \caption{Comparison of \geneva and \Matrix at NNLO for the $y_H$ distribution.}
   \label{fig:rapidity_matrix}
\end{figure}

In \fig{rapidity_matrix} we compare the Higgs rapidity spectrum obtained with \geneva with the NNLO result provided by \Matrix, including the $7$-point scale variations.
We observe very good agreement both in the central values and in the envelope of the scale variations, up to large values of $|y_H|$. The symmetry of the $pp$ collider allows us to show only the absolute value of $y_H$, and thus further reduce the Monte Carlo uncertainty.

%%%%%
%%%%%

%%%%%%%%%%%%%%%%%%%%%%%%%%%%%%%%%%%%%%%%%%%%%%%%%%%%%%%%%%%%%%%%%%%%%%%%%%%%%%%%
\subsection{Interface with P{\scriptsize YTHIA}8}
\label{sec:pythia}
%%%%%%%%%%%%%%%%%%%%%%%%%%%%%%%%%%%%%%%%%%%%%%%%%%%%%%%%%%%%%%%%%%%%%%%%%%%%%%%%

 % - recap what we do briefly

 % - add plot of Tau0 before /after the shower, also including with N3LL (?)

In this section we briefly recap the main features of the interface used in
\geneva to match the partonic results to the
\pythiaEight~\cite{Sjostrand:2014zea} parton shower. As this
is not the main focus of this work, however, we refer the interested reader to
\refcite{Alioli:2015toa} for a detailed discussion and \refcite{Alioli:2022dkj}
for additional details on the accuracy of the matched calculation.
Given that so far we have constructed partonic results with NNLL$^\prime$ accuracy in the resolution variable $\Tau_0$, we wish to preserve this resummed accuracy after the parton shower as far as is possible.
At the same time, for all other observables we need to guarantee
that the accuracy of the parton shower is preserved.
This is a nontrivial condition: since the ordering variable
of the \pythiaEight parton shower is the relative transverse momentum while the resolution
variable we use is the $N$-jettiness, the shower can in principle produce emissions which double-count
regions of the phase space.

To avoid this issue, we perform the matching employing the following
prescription. We set the starting scale of the parton shower by taking
the maximum relative $k_\perp$ determined by the lower scale of the
resummation. The latter is defined on an event-by-event basis and corresponds to
either $\Tau_{N}^{c} \equiv \Tau_0^{\mathrm{cut}}$, $\Tau_1^{\mathrm{cut}}$ or
$\Tau_1\left(\Phi_2\right)$, depending on whether the relative partonic
configuration has $N=0$, $1$ or $2$ jets, respectively. We then let the shower run down
to the internal minimum $p_\perp$, which produces a certain number of emissions $k$.
Lastly, we check that the resulting event fulfils the condition
\begin{equation}
  \Tau_N(\Phi_{N+k})\leq\Tau_N^{c}\, ,
\end{equation}
which ensures that both accuracies are correctly preserved.
For unshowered events with one jet in the final state, we perform the first
shower emission directly within \geneva, by implementing eqs.~(48) and (49) of \refcite{Alioli:2019qzz}.
Showered events will therefore almost exclusively originate from events with either zero or two final state partons.

%%%%%
\begin{figure}[t]
   \centering
   \includegraphics[width=\rescaletwoplots]{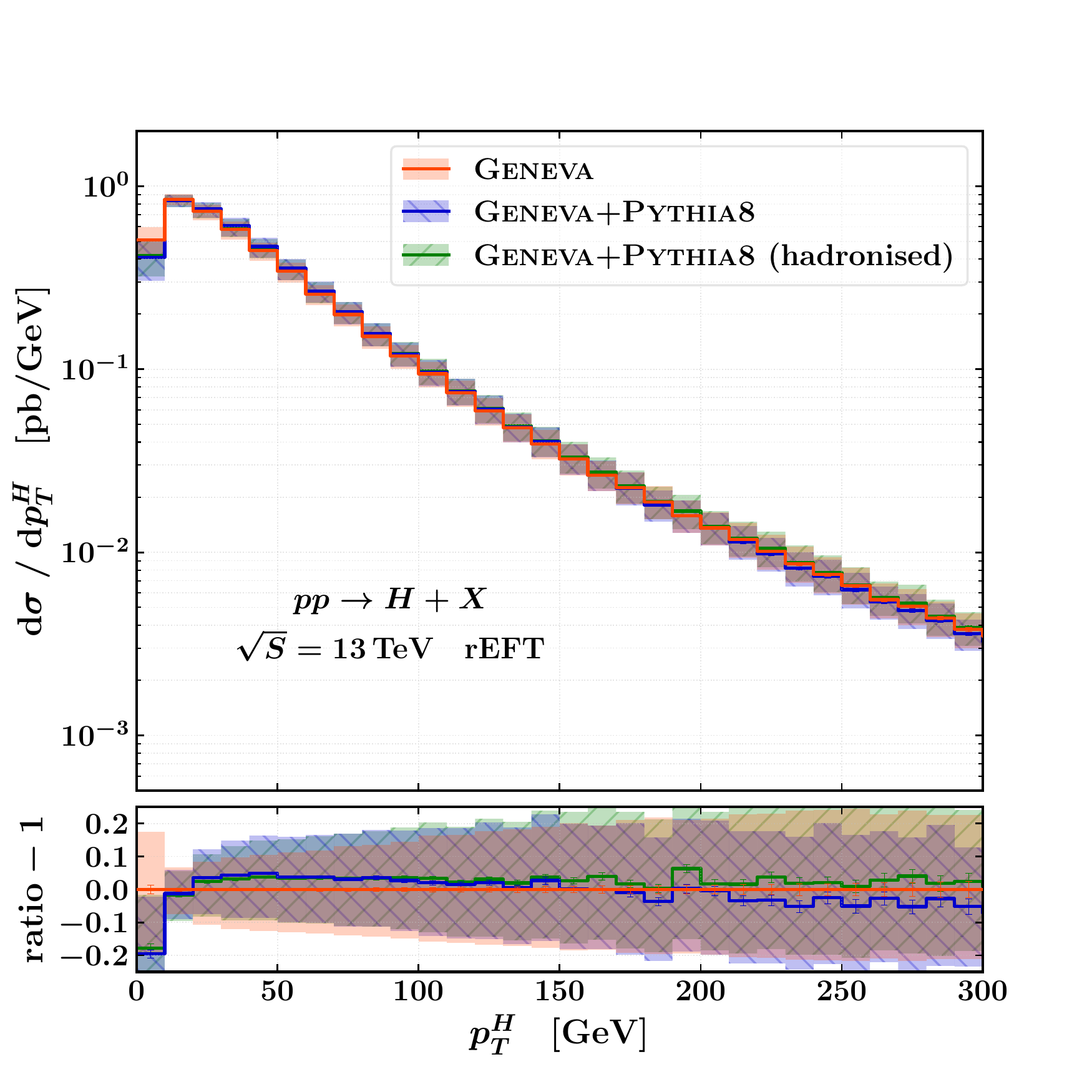}%
   \includegraphics[width=\rescaletwoplots]{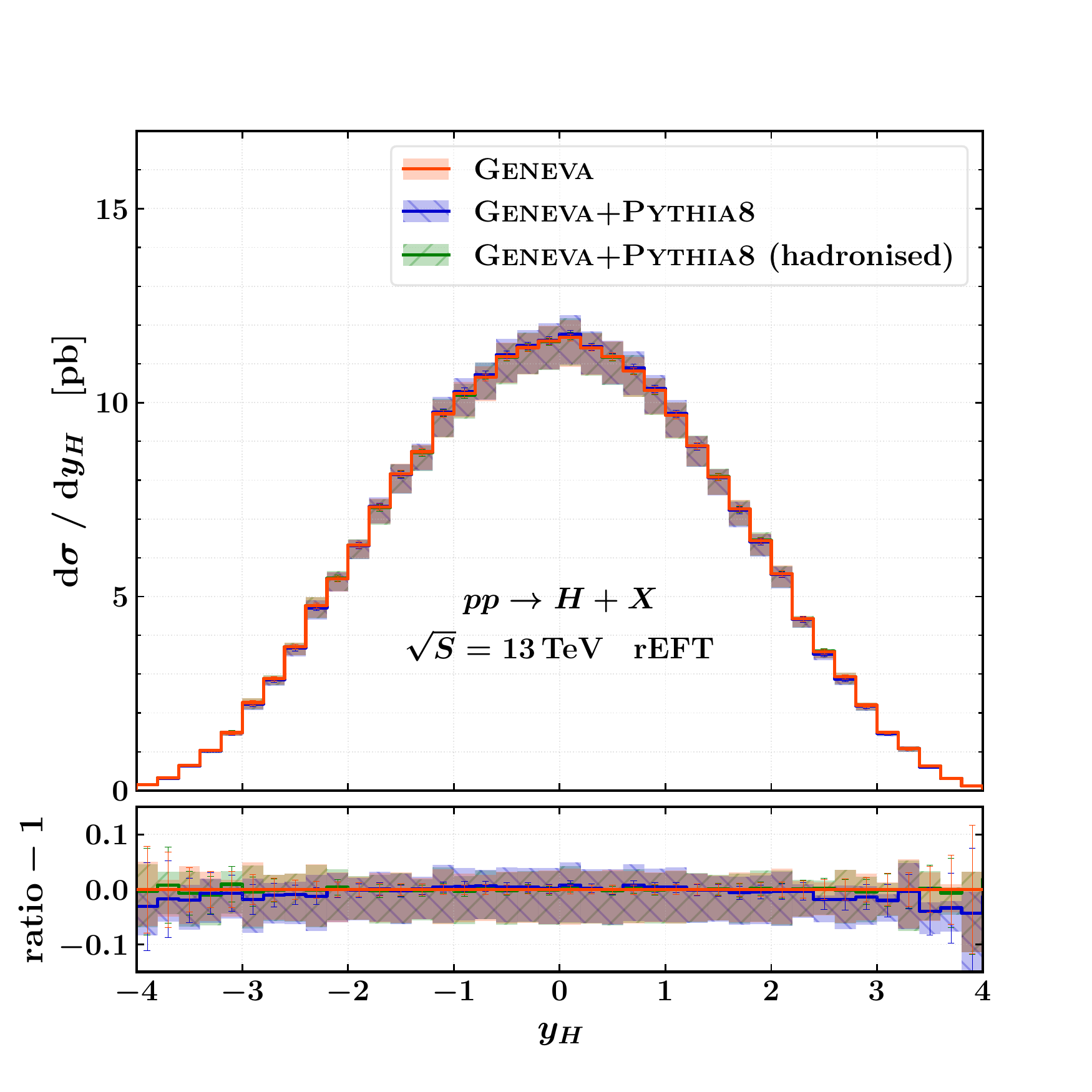}%
   \caption{Comparison of the partonic, showered, and hadronised results for the $p_H^T$ (left) and $y_H$ (right) distributions.}
   \label{fig:pythia_cmp_1}
\end{figure}
%%%%%

In \fig{pythia_cmp_1} we show the effect of the \pythiaEight shower
on the $p_T^H$ and $y_H$ partonic distributions.
For the results presented in this section we use the default \pythiaEight parameters
for the shower and the hadronisation model.
The rapidity distribution, being an inclusive observable, is exactly preserved by the shower, as expected.
The Higgs transverse momentum is an exclusive observable, and the shower can therefore have a significant impact on its shape:
in this case we see an effect of $\sim 15 \%$ in the $p_T^H < 15 \GeV$ bin, and smaller effects $\lesssim 5 \%$ in the rest of the spectrum, especially in the tail of the distribution.
After hadronisation, we find that most of these discrepancies are reduced.
%

%%%%%
\begin{figure}[t]
   \centering
   \includegraphics[width=\rescaletwoplots]{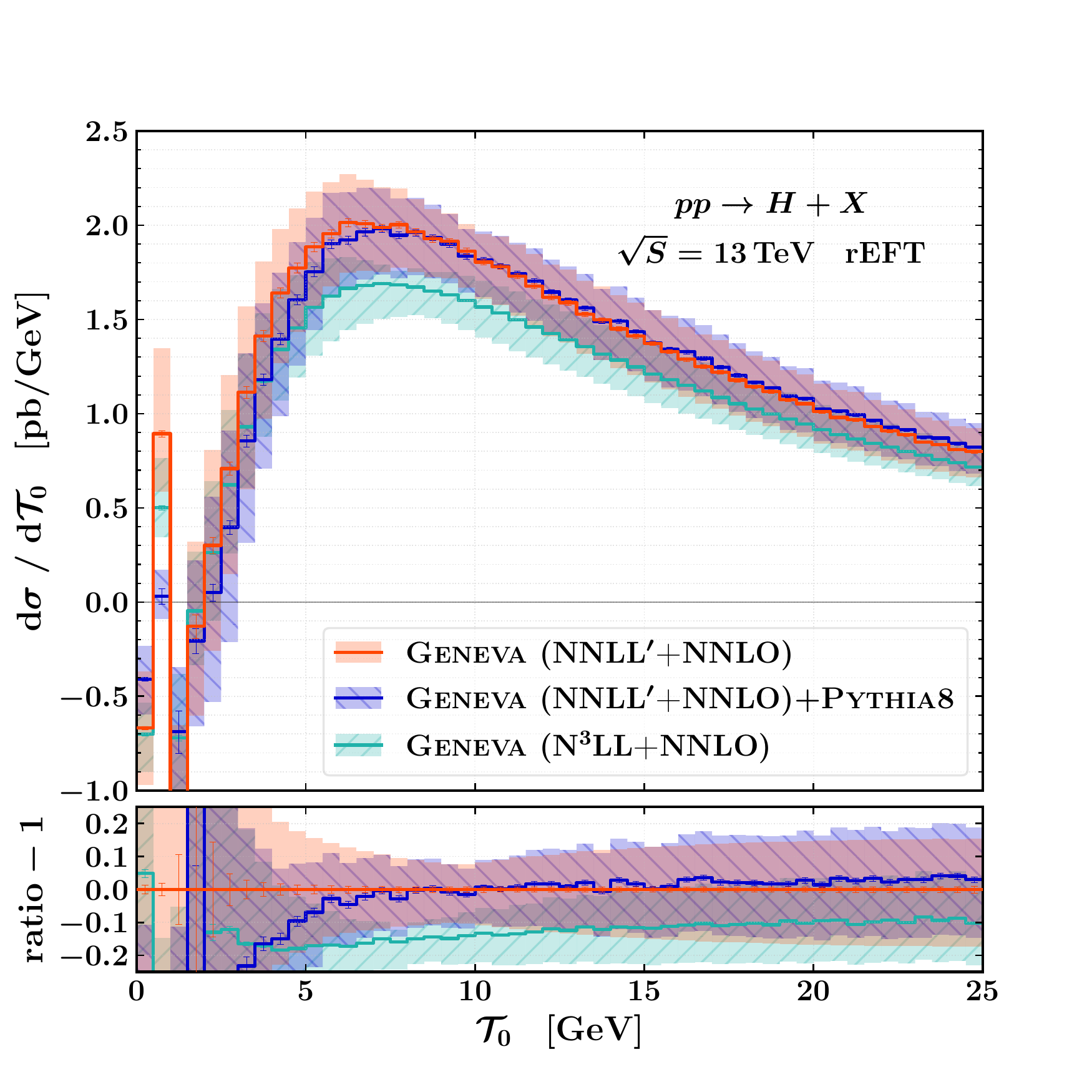}%
   \includegraphics[width=\rescaletwoplots]{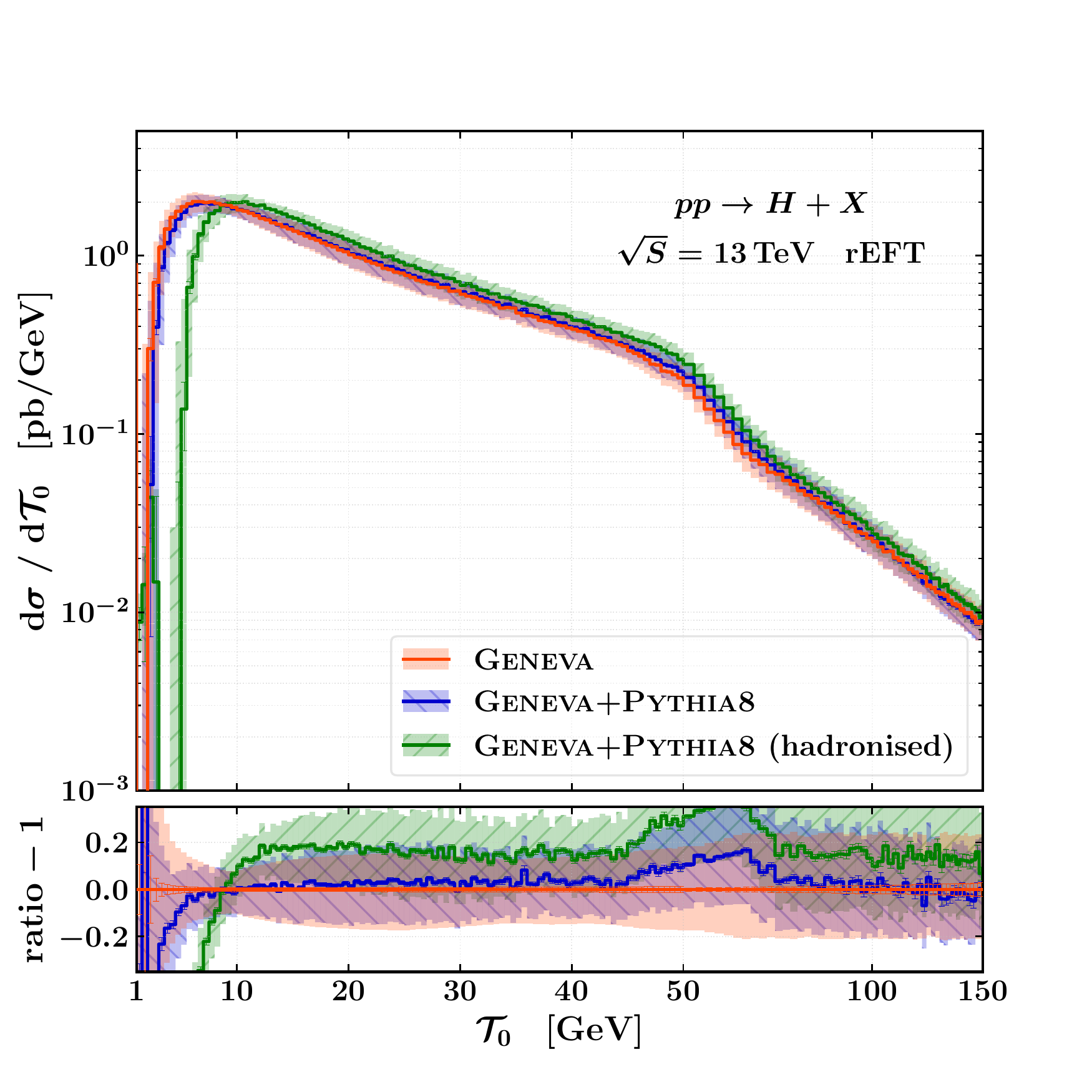}%
   \caption{Effects of the parton shower on the $\Tau_0$ spectrum: comparison of the partonic, showered, and N$^3$LL-resummed distribution (left), and comparison of the partonic, showered, and hadronised results (right).}
   \label{fig:pythia_cmp_2}
\end{figure}
%%%%%

The parton shower and hadronisation effects on the $\Tau_0$ distribution are displayed in \fig{pythia_cmp_2}.
As mentioned above, our matching procedure to the \pythiaEight shower is designed with the aim that the $\Tau_0$ logarithmic accuracy is not spoiled.
We explicitly check this in the left panel, where we compare the $\Tau_0$ distribution at NNLL$^\prime$ before and after the parton shower matching with the partonic prediction at N$^3$LL.
Note that for this process, which is gluon-initiated, one expects that the parton shower effects are larger than for quark-initiated processes, e.g.~because of the larger Casimir factors.
Nonetheless, in the peak region $\Tau_0 < 25 \GeV$, we find that the showered distribution lies in between the central NNLL$^\prime$ and N$^3$LL curves, and within the overlap of the two uncertainty bands.
We therefore conclude that the quantitative effects of the shower are on par with (or smaller than) the effects of the next logarithmic order in the resummation.

In principle it is possible to directly interface the N$^3$LL \geneva\ results to the parton shower.
In this work, however, we refrain from doing so --- in particular when comparing to data ---
because, due to the lack of the N$^4$LL prediction for the $\Tau_0$ spectrum, we cannot verify
that the large distortions induced by the shower are compatible with the next logarithmic correction.

The hadronisation effects on the $\Tau_0$ distribution are displayed in the right panel of \fig{pythia_cmp_2}. As expected for this observable, we observe $\mathcal{O}(1)$ effects in the peak region, which decrease for larger values of $\Tau_0$. In the region around $\Tau_0 \approx m_H / 2$, which corresponds to the point at which the $\Tau_0$ resummation is switched off, we find a more pronounced discrepancy between the \geneva partonic and showered results. We have verified that this is an artefact related to our choice of setting the $\Tau_0$ spectrum equal to the derivative of the cumulant as explained at the end of \sec{tau0}.

%

%%%%%%%%%%%%%%%%%%%%%%%%%%%%%%%%%%%%%%%%%%%%%%%%%%%%%%%%%%%%%%%%%%%%%%%%%%%%%%%%
\section{Comparison with LHC data}
\label{sec:data}
%%%%%%%%%%%%%%%%%%%%%%%%%%%%%%%%%%%%%%%%%%%%%%%%%%%%%%%%%%%%%%%%%%%%%%%%%%%%%%%%

%%%%%
\begin{figure}[t]
   \centering
   \includegraphics[width=\rescaletwoplots]{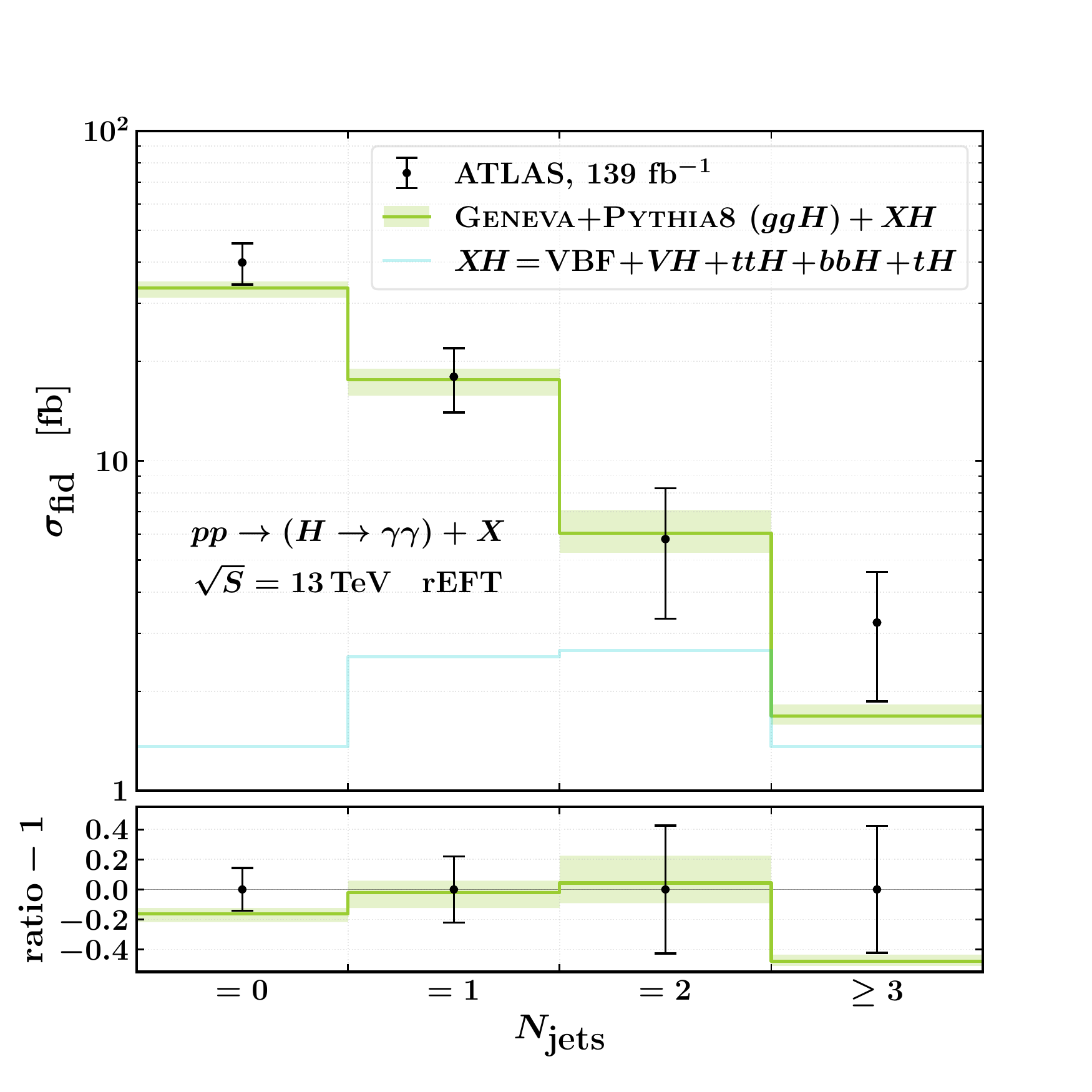}%
   \includegraphics[width=\rescaletwoplots]{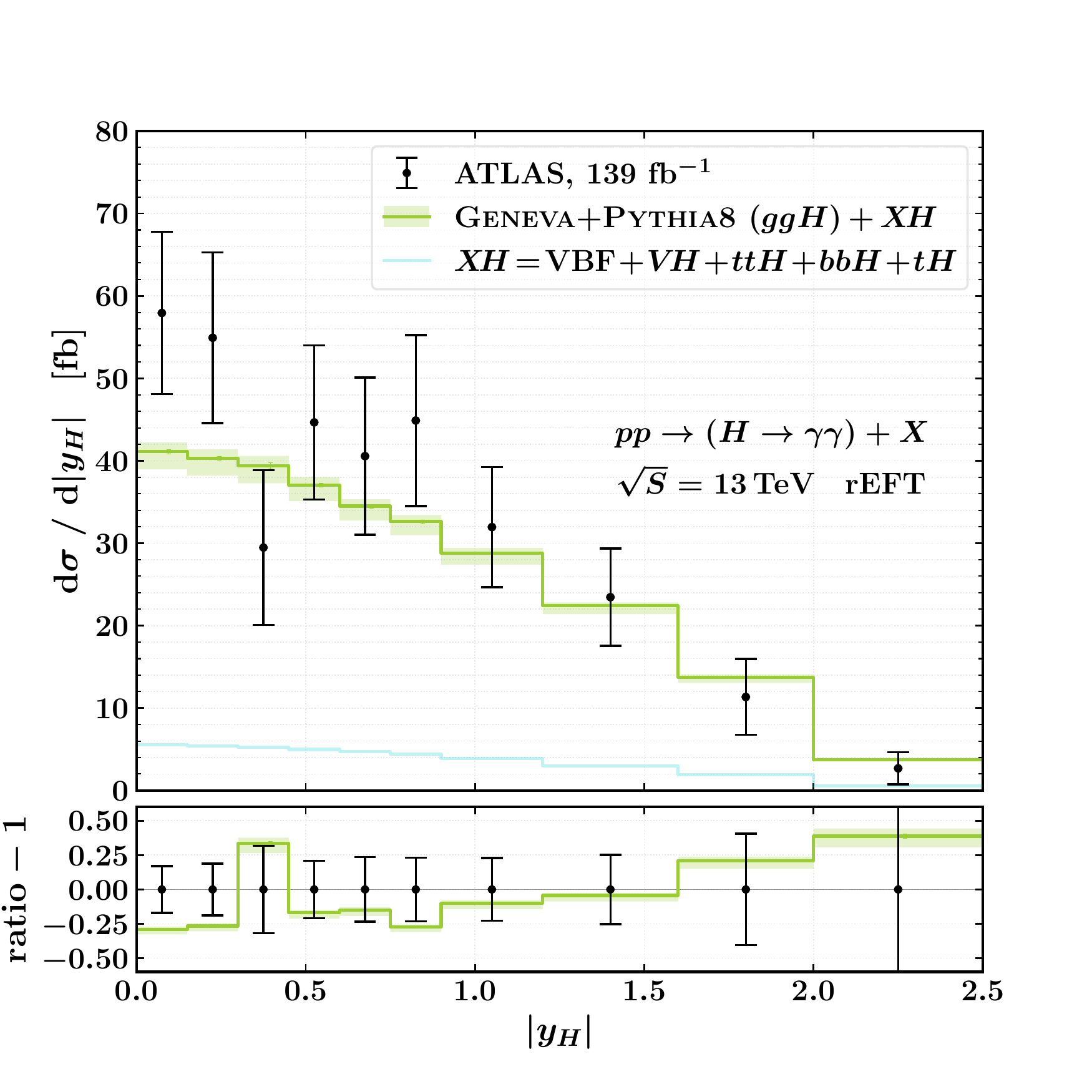}\\%
   \includegraphics[width=\rescaletwoplots]{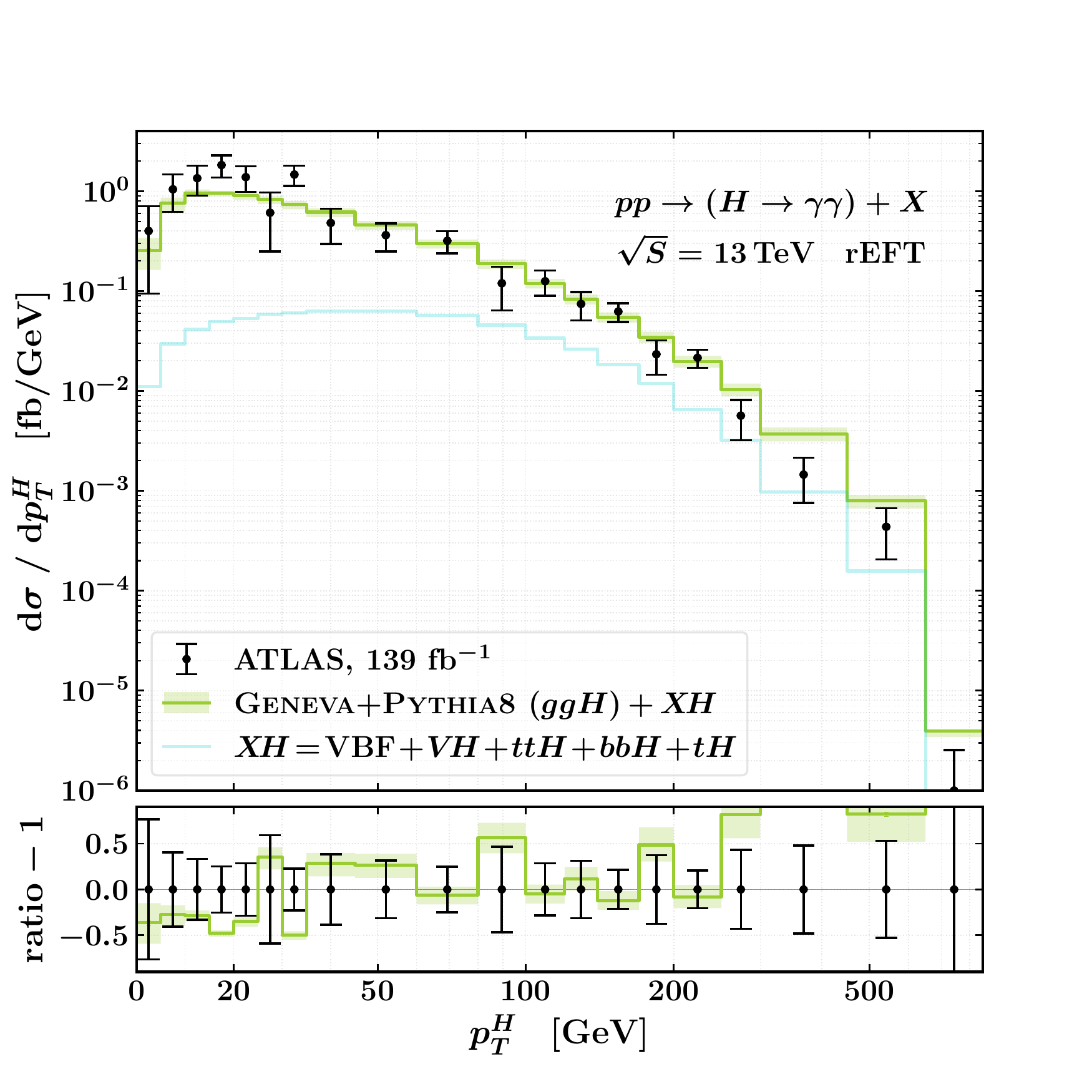}%
   \includegraphics[width=\rescaletwoplots]{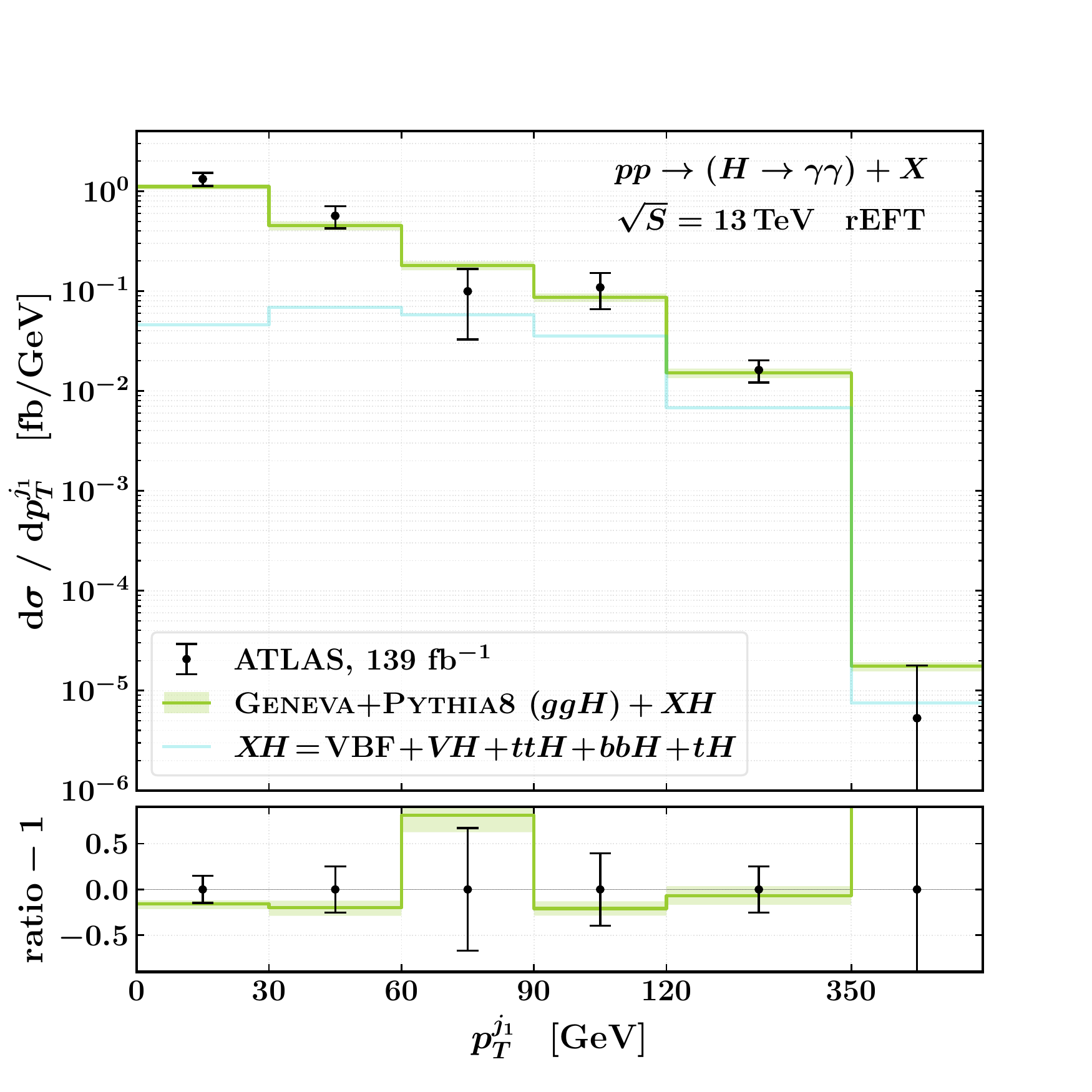}%
   \caption{Comparison of the ATLAS data \cite{ATLAS:2022fnp} with the \geneva{}+\pythiaEight results at $13 \TeV$. We show the fiducial cross sections for different values of $N_\text{jets}$ (top left), as well as the distributions of $|y_H|$ (top right), $p_T^H$ (bottom left), and $p_T^{j_1}$ (bottom right).}
   \label{fig:atlas_cmp_1}
\end{figure}
%%%%%

%%%%%
\begin{figure}[t]
   \centering
   \includegraphics[width=\rescaletwoplots]{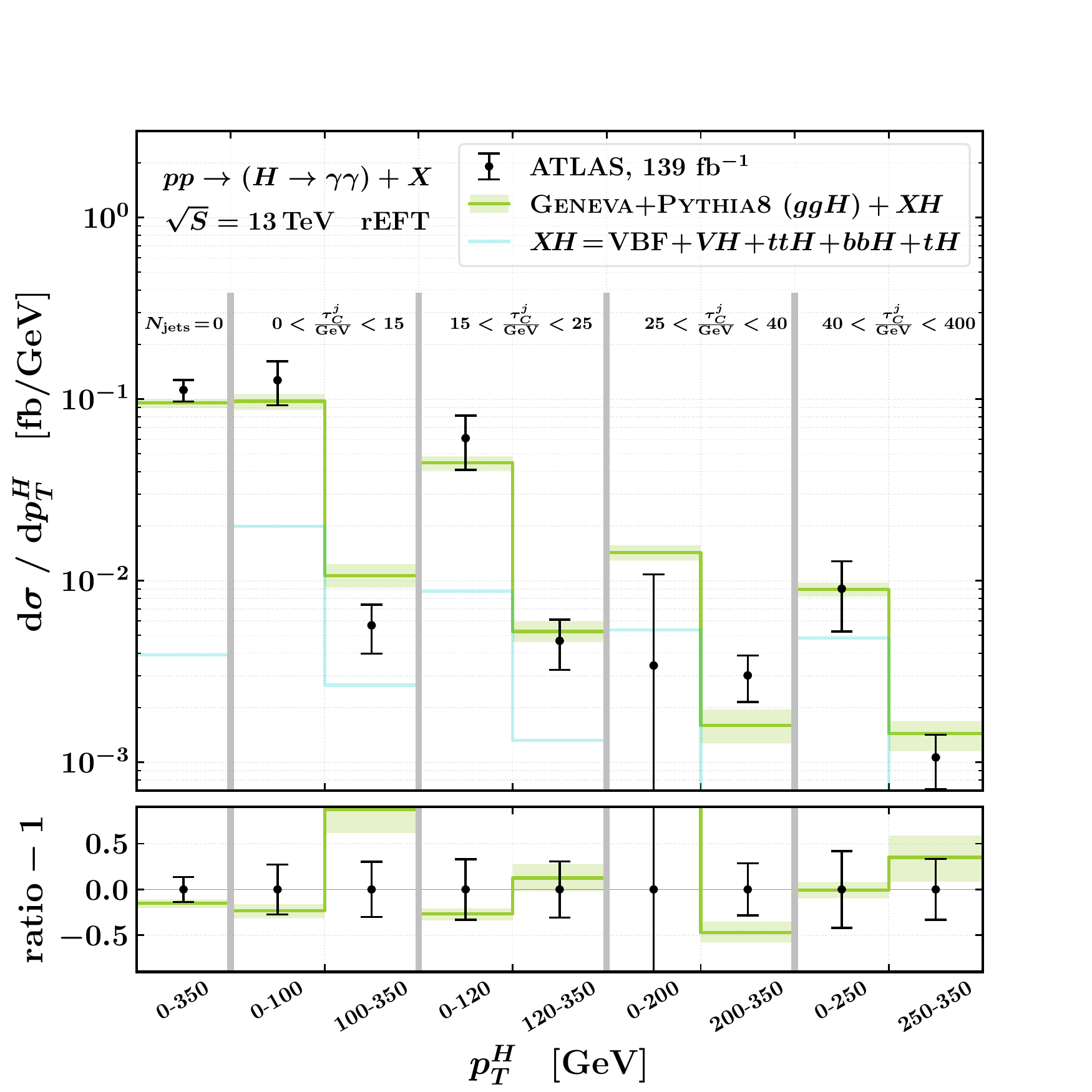}%
   \includegraphics[width=\rescaletwoplots]{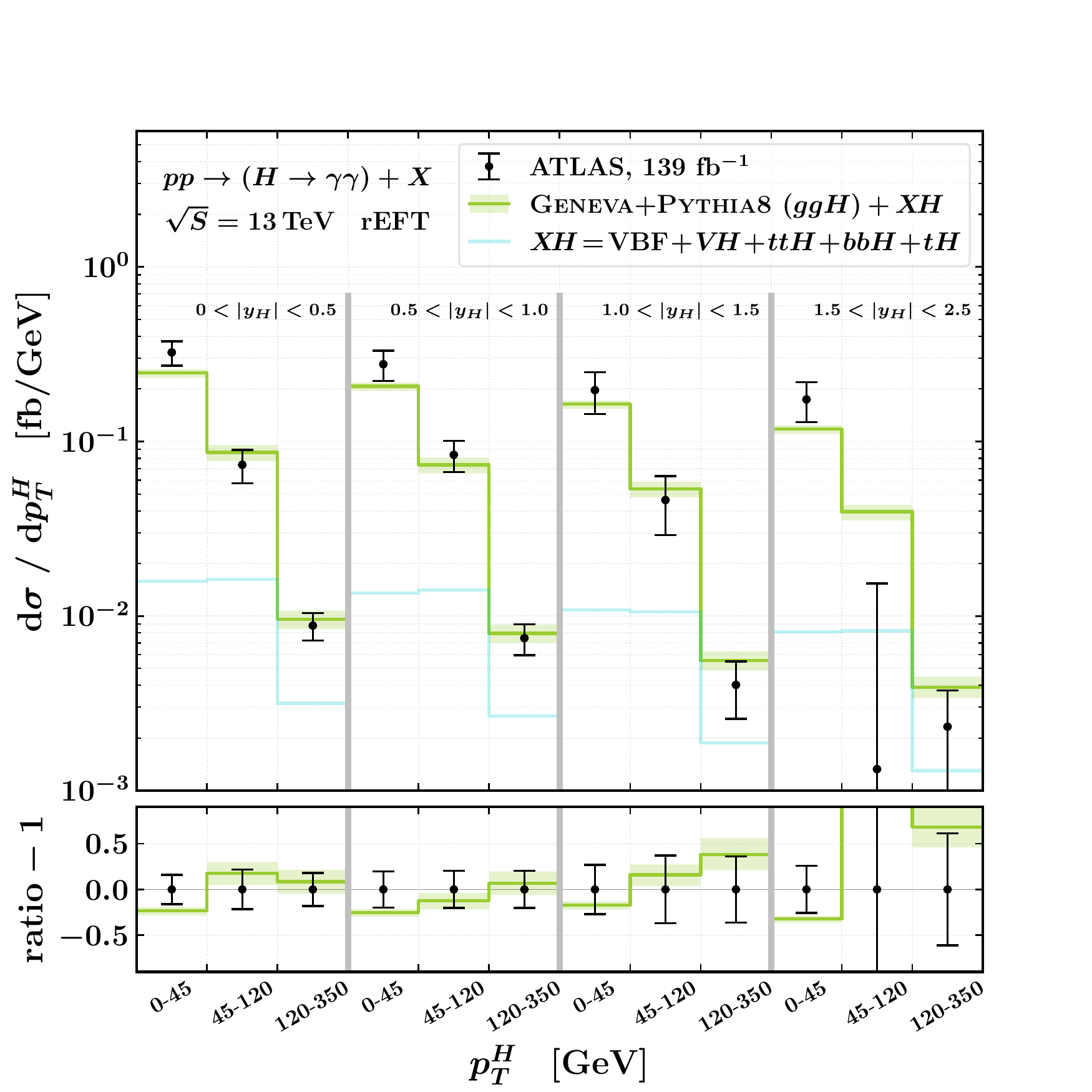}\\%
   \includegraphics[width=\rescaletwoplots]{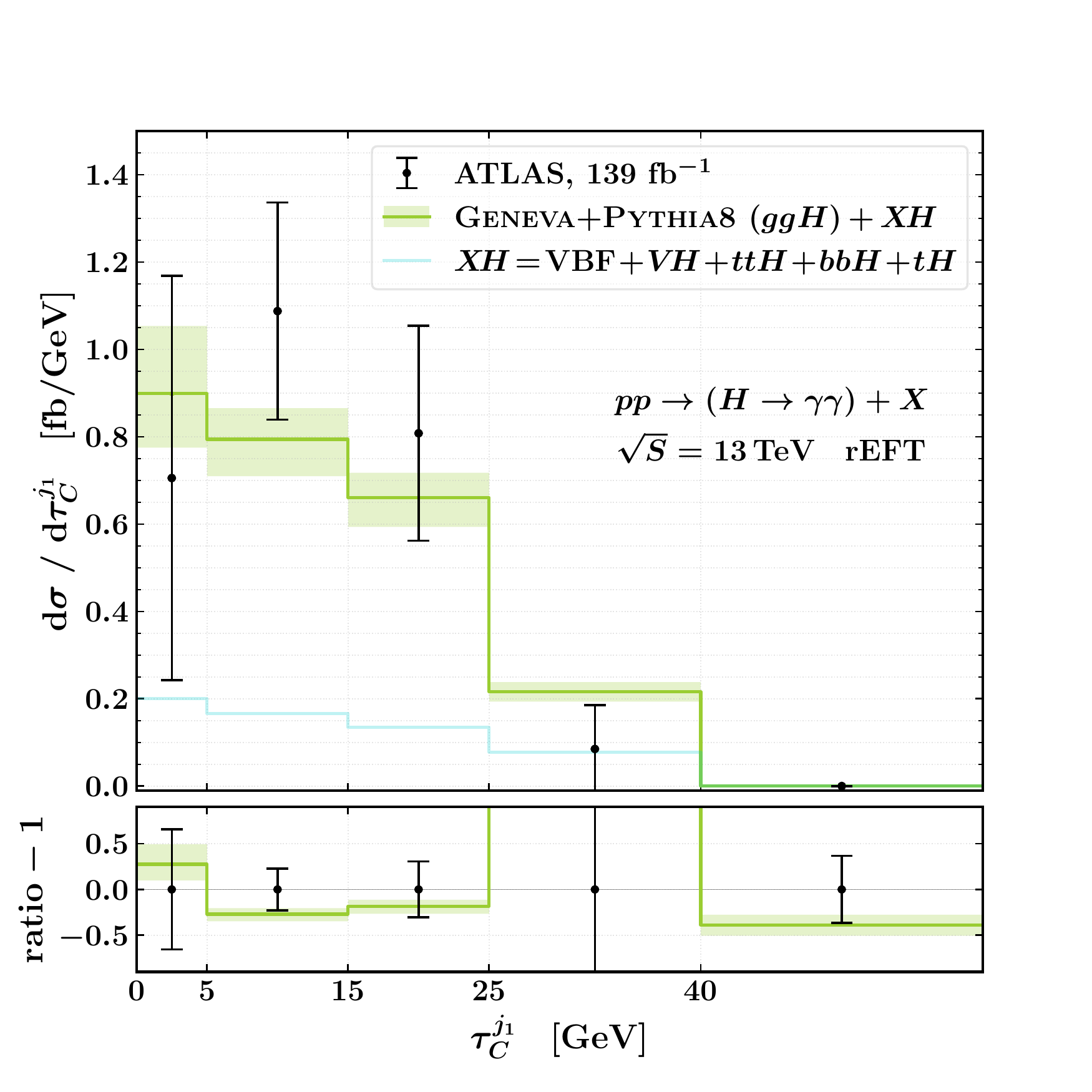}%
   \includegraphics[width=\rescaletwoplots]{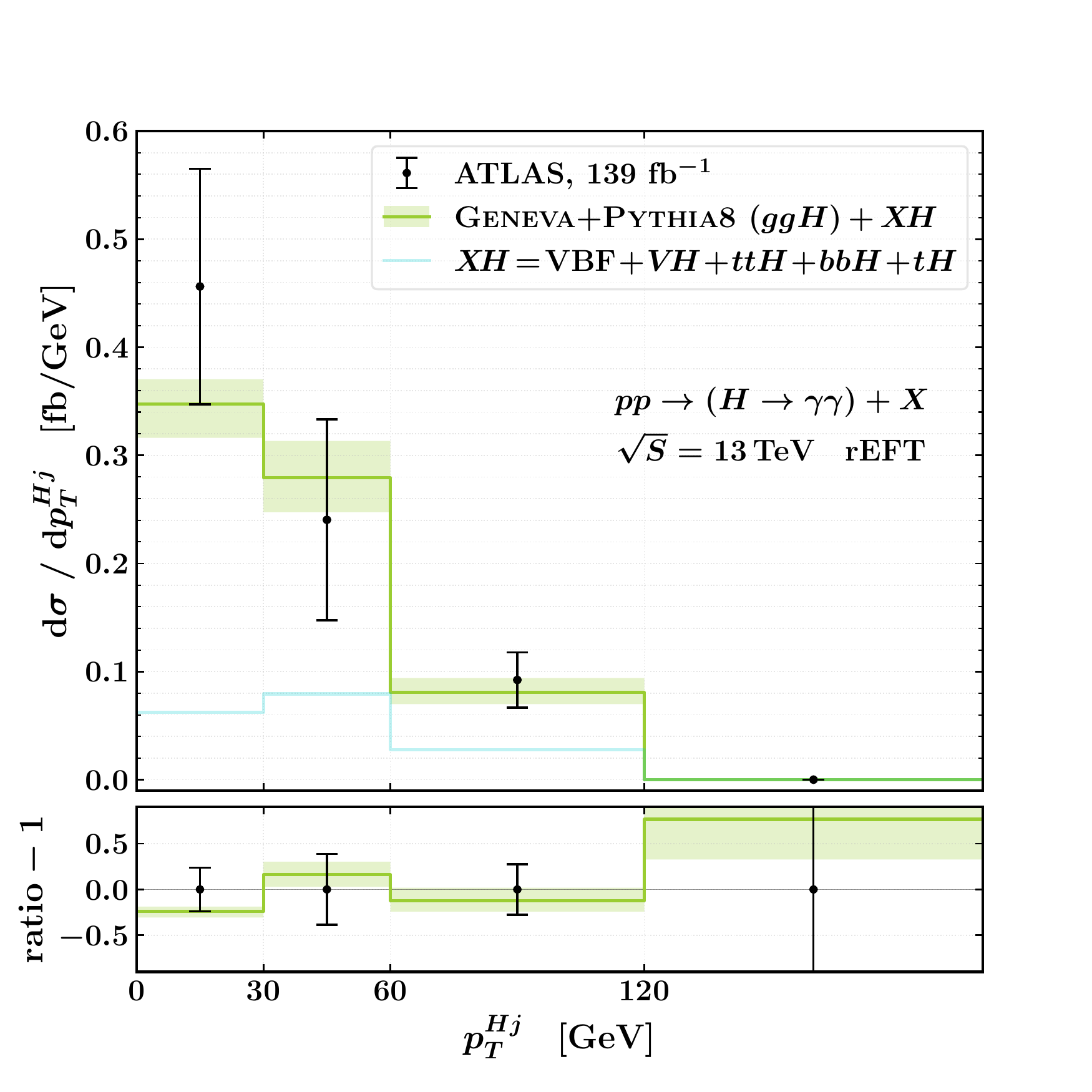}%
   \caption{Comparison of the ATLAS data \cite{ATLAS:2022fnp} with the \geneva{}+\pythiaEight results at $13 \TeV$. We show the $p_T^H$ distributions in bins of $\tau_C^{j_1}$ (top left) and of $|y_H|$ (top right), as well as the $\tau_C^{j_1}$ (bottom left) and the $p_T^{H j}$ (bottom right) distributions.}
   \label{fig:atlas_cmp_2}
\end{figure}
%%%%%

%%%%%
\begin{figure}[t]
   \centering
   \includegraphics[width=\rescaletwoplots]{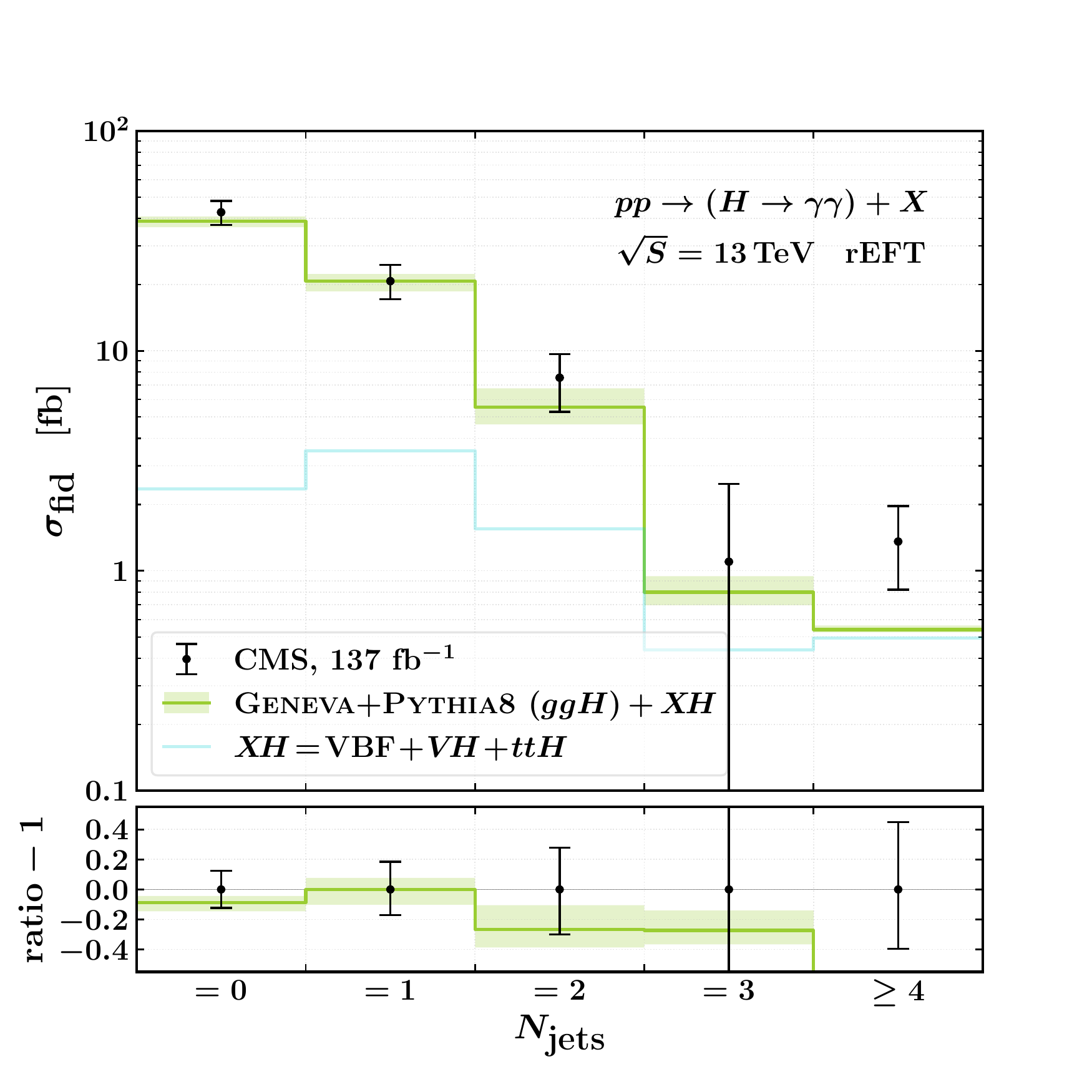}%
   \includegraphics[width=\rescaletwoplots]{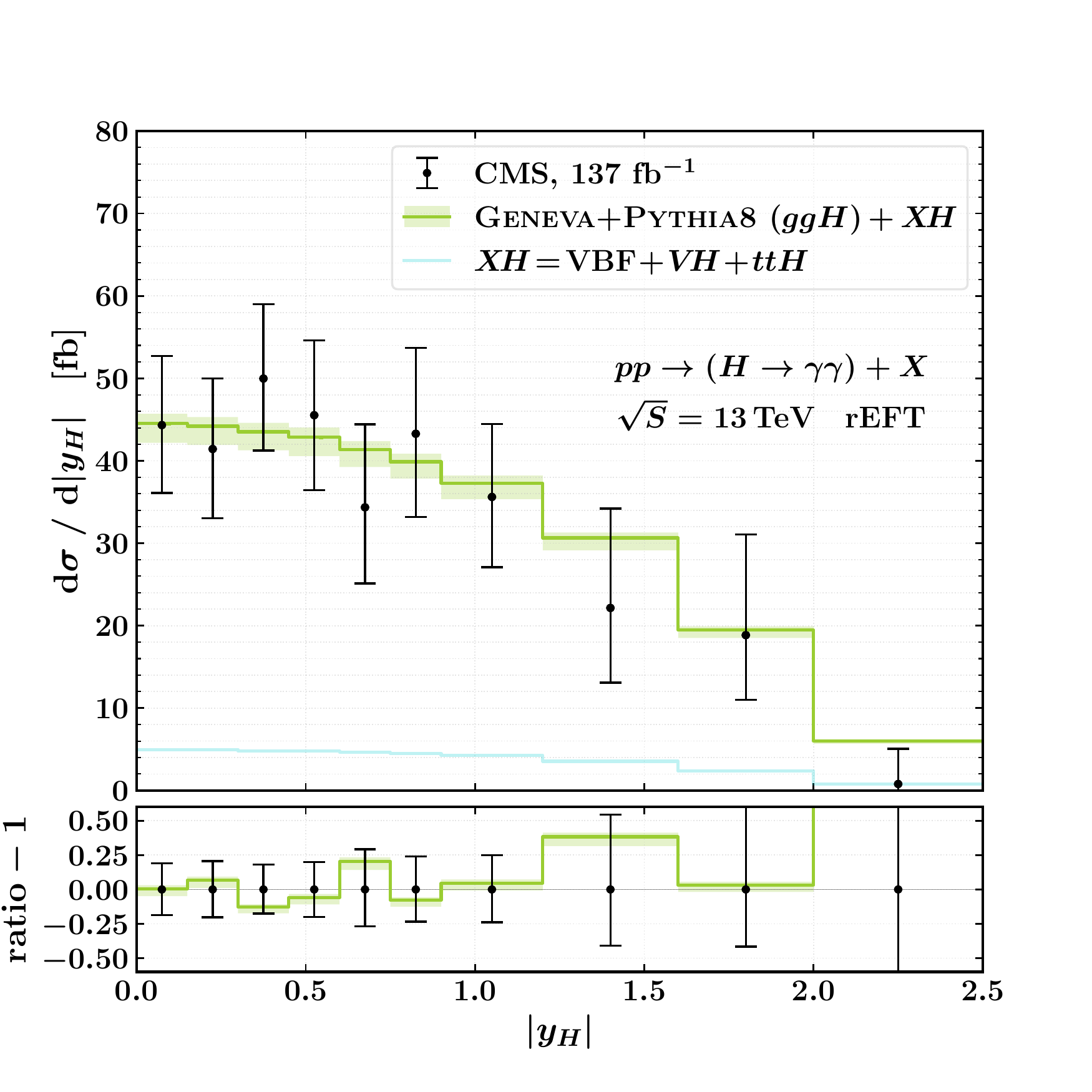}\\%
   \includegraphics[width=\rescaletwoplots]{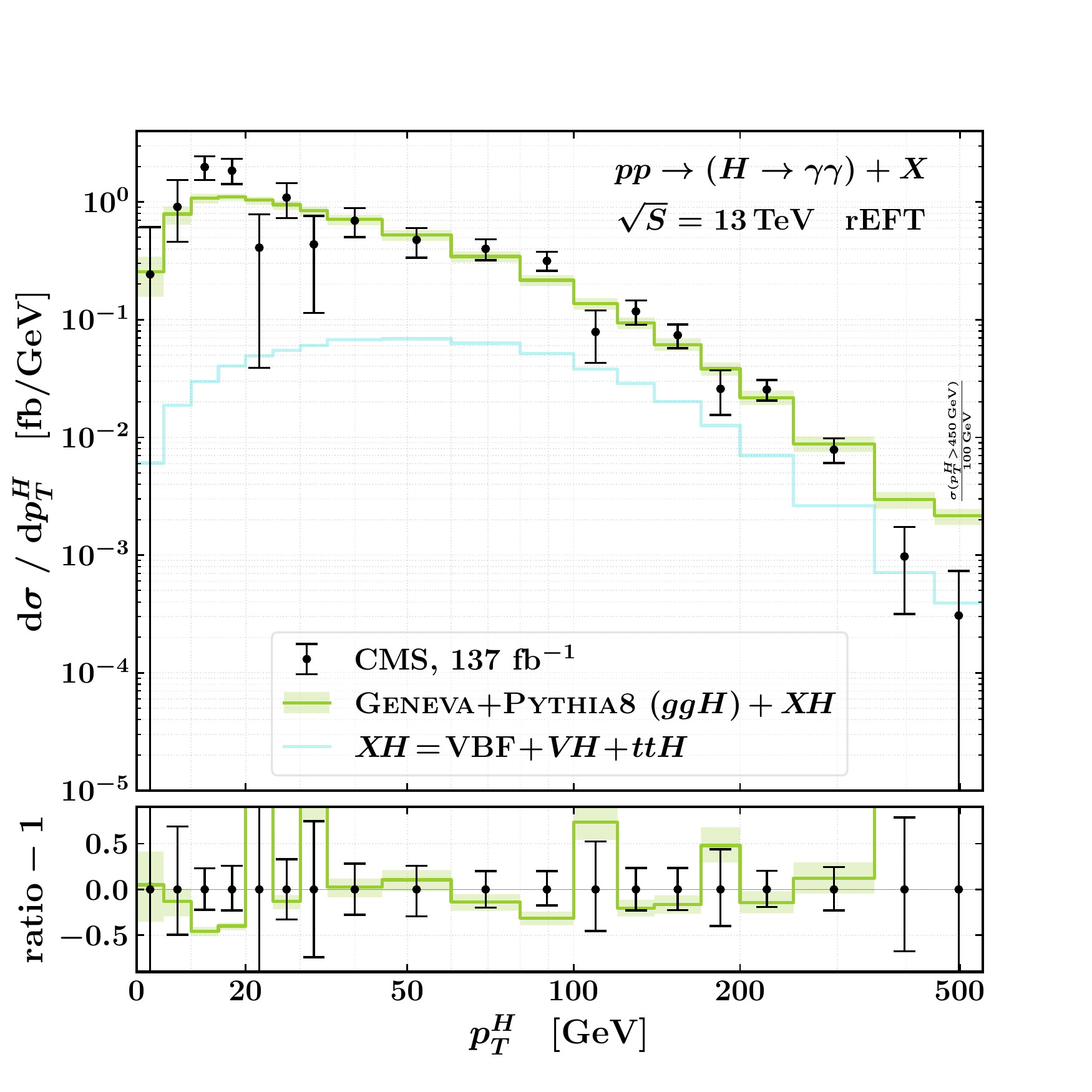}%
   \includegraphics[width=\rescaletwoplots]{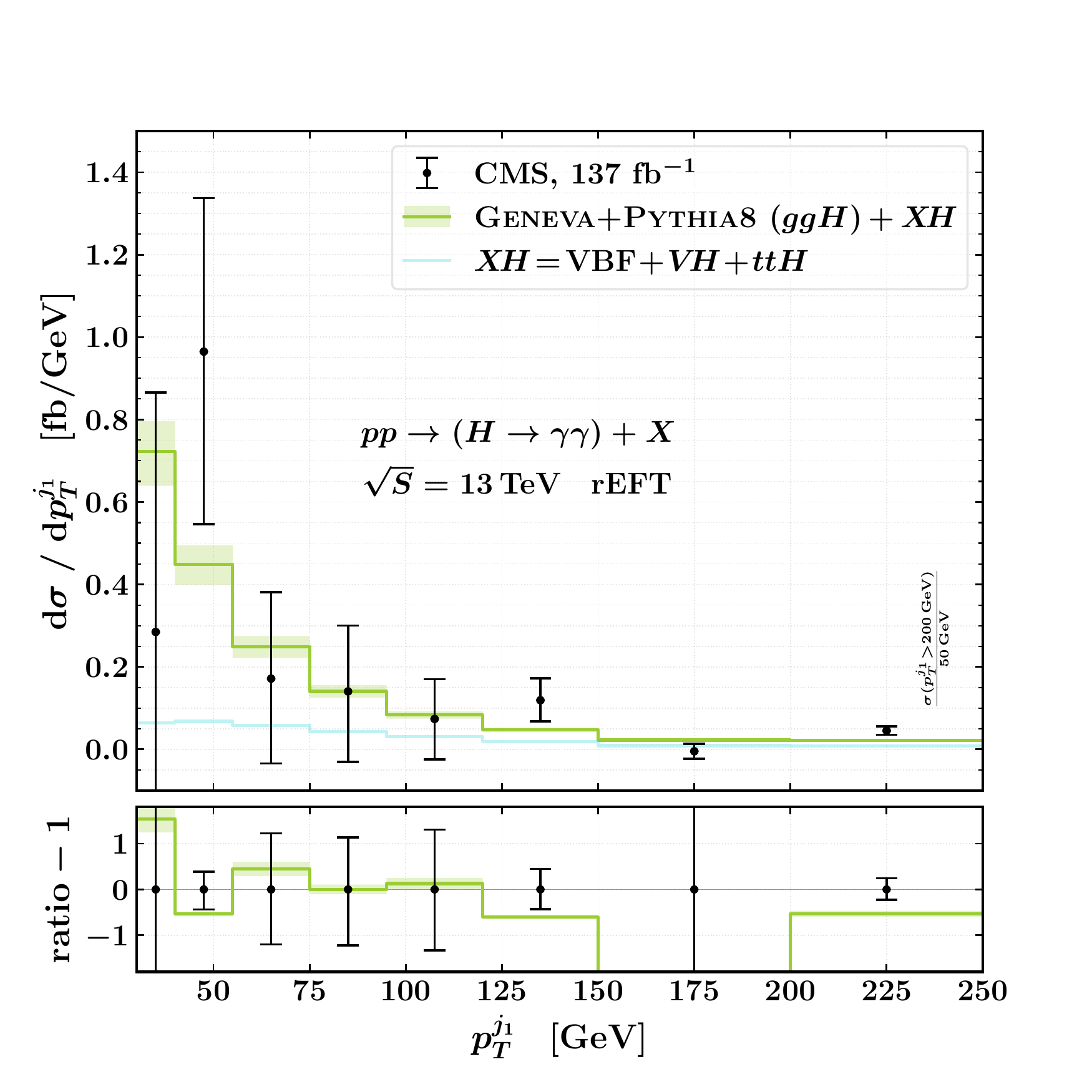}%
   \caption{Comparison of the CMS data \cite{CMS:2022wpo} with the \geneva{}+\pythiaEight results at $13 \TeV$. We show the fiducial cross sections for different values of $N_\text{jets}$ (top left), as well as the distributions of $|y_H|$ (top right), $p_T^H$ (bottom left), and $p_T^{j_1}$ (bottom right).}
   \label{fig:cms_cmp_1}
\end{figure}
%%%%%

%%%%%
\begin{figure}[t]
   \centering
   \includegraphics[width=\rescaletwoplots]{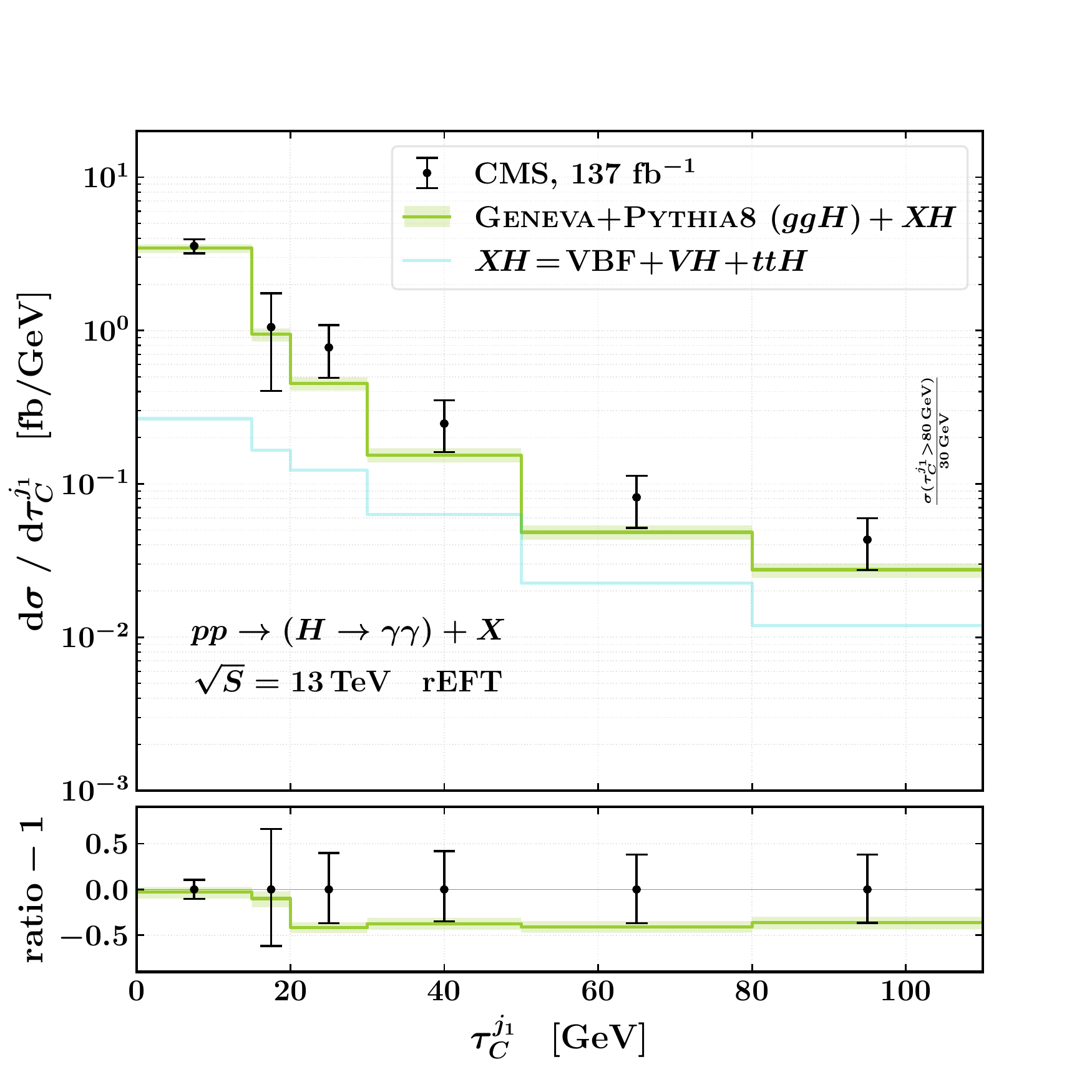}%
   \includegraphics[width=\rescaletwoplots]{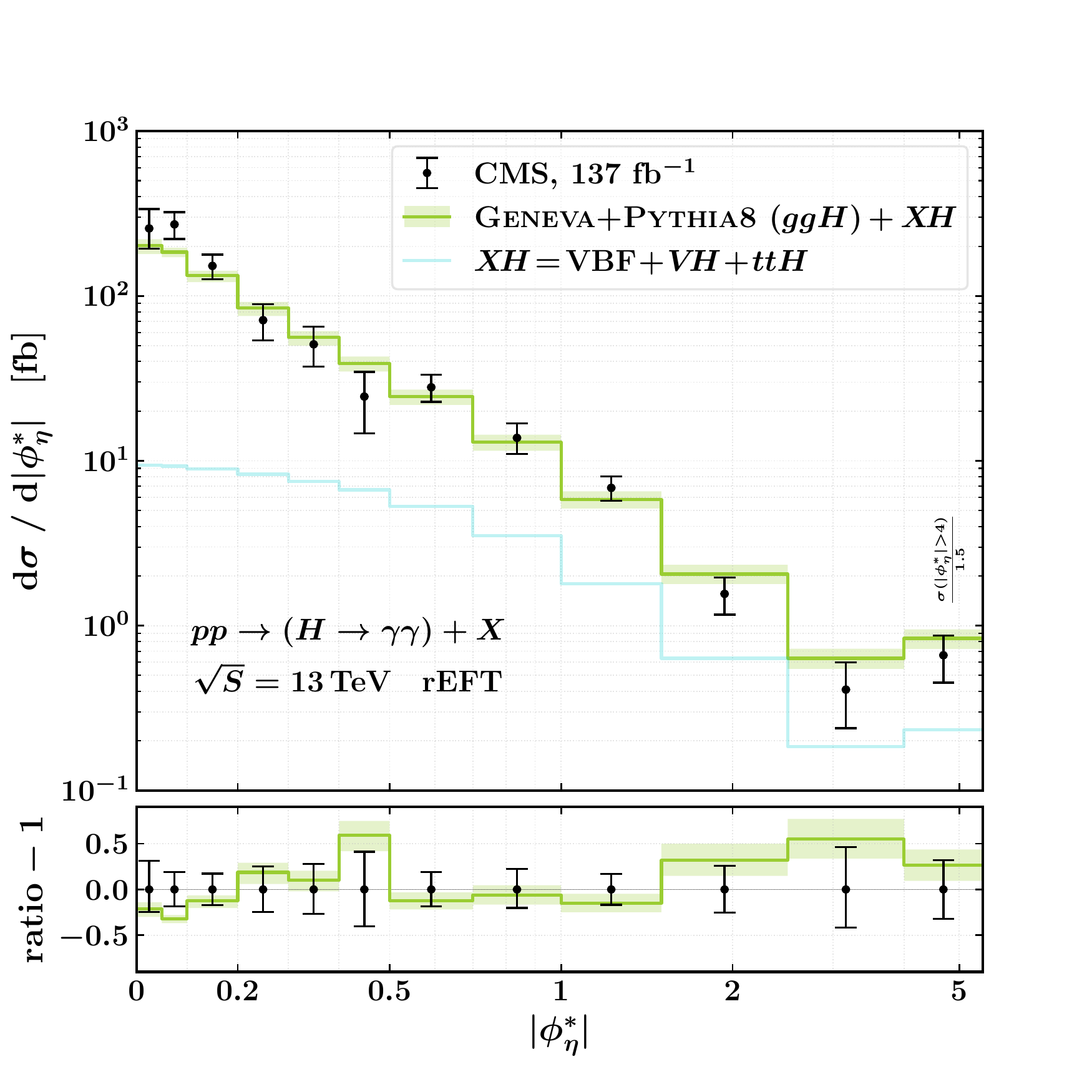}\\%
   \includegraphics[width=\rescaletwoplots]{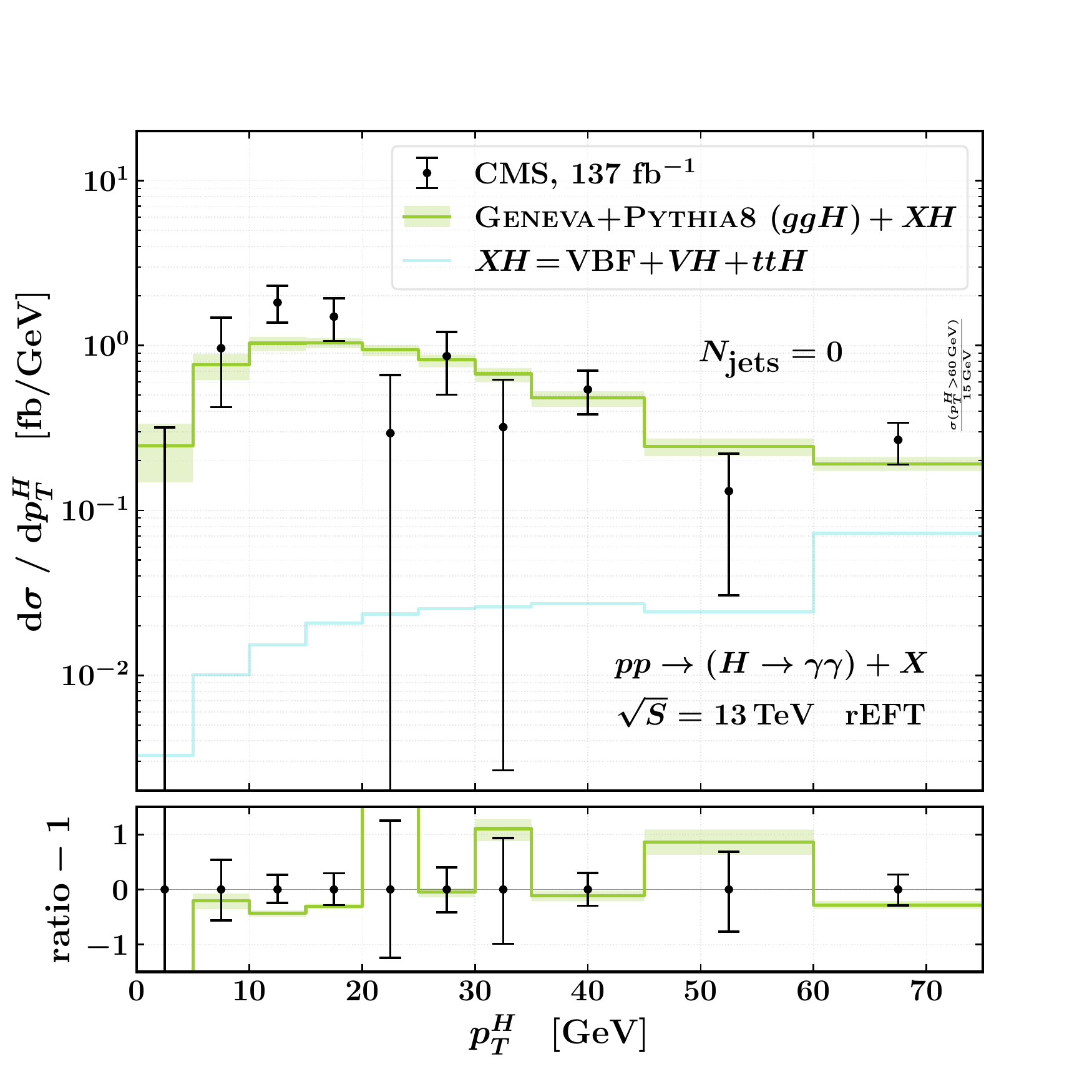}%
   \includegraphics[width=\rescaletwoplots]{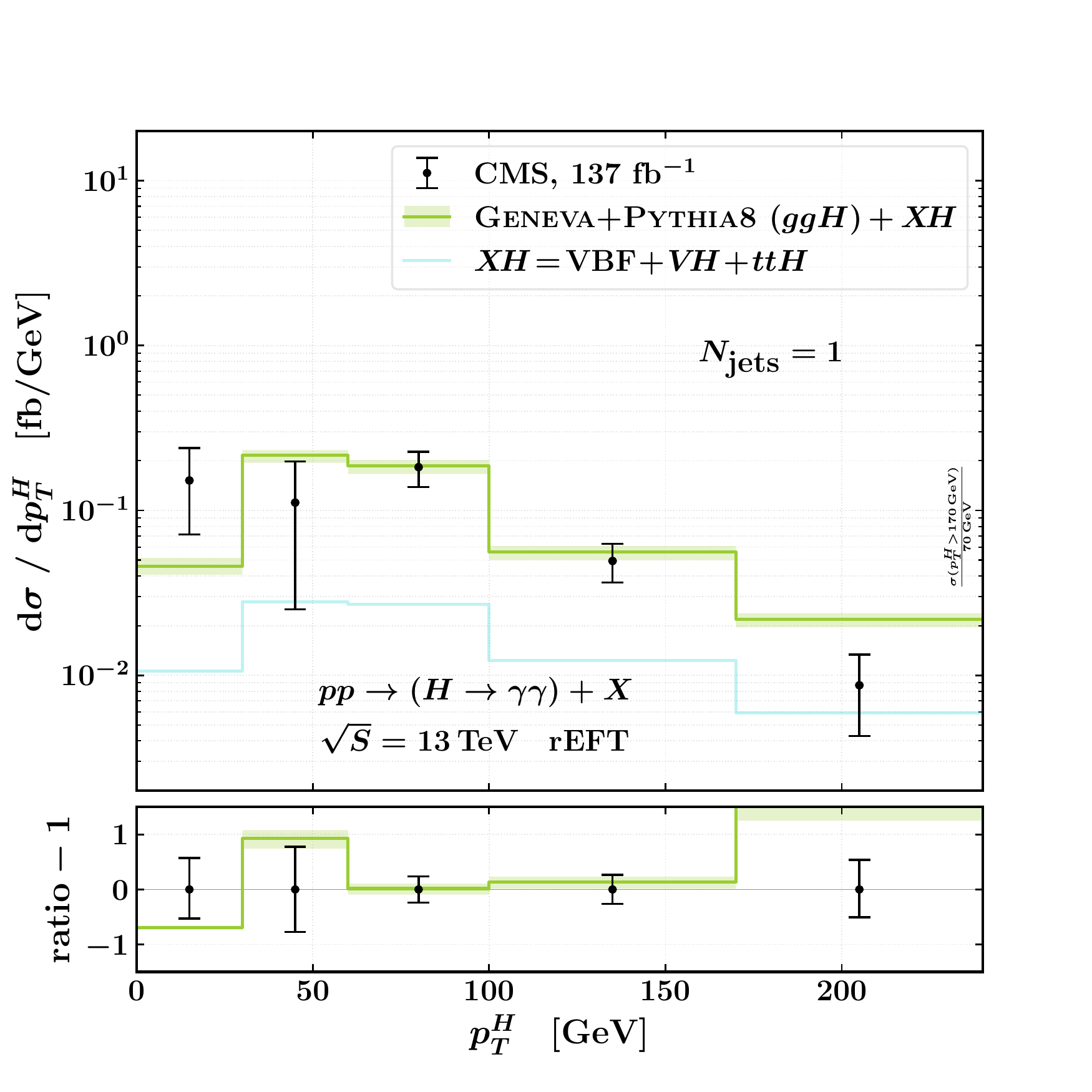}%
   \caption{Comparison of the CMS data \cite{CMS:2022wpo} with the \geneva{}+\pythiaEight results at $13 \TeV$. We show the $\tau_C^{j_1}$ (top left) and $|\phi^\ast_\eta|$ (top right) distributions, as well as the $p_T^H$ distributions for events with $N_\text{jets} = 0$ (bottom left) and $N_\text{jets} = 1$ (bottom right).}
   \label{fig:cms_cmp_2}
\end{figure}
%%%%%

We compare the predictions obtained with \geneva with the latest experimental results
for the Higgs boson inclusive and differential cross sections in the $H \to \gamma \gamma$ decay channel.
The results are provided both by the ATLAS \cite{ATLAS:2022fnp} and CMS \cite{CMS:2022wpo} experiments,
and are obtained from the LHC data at a centre-of-mass energy of $13 \TeV$
using $139\ \textrm{fb}^{-1}$ and $137\ \textrm{fb}^{-1}$ of proton-proton collision data, respectively.

In the ATLAS measurement \cite{ATLAS:2022fnp}, the fiducial phase space is identified
by requiring the existence of two isolated photons with $p_T^\gamma > 25 \GeV$ in the final state.
Photons are considered isolated if the transverse energy of charged particles with $p_T > 1 \GeV$
within a cone of radius $R_\text{iso} = 0.2$ around the photon direction
does not exceed 5\% of the photon's transverse momentum.
The two isolated photons must additionally have transverse momenta larger than 35\% and 25\% of the diphoton invariant mass,
for the leading and subleading photons respectively.
The invariant mass of the diphoton system must be in the range $105 \GeV < m_{\gamma \gamma} < 160 \GeV$.
Moreover, photons are required to have pseudorapidity $|\eta_\gamma| < 1.37$ or $1.52 < |\eta_\gamma| < 2.37$.
For this measurement, jets are defined using the anti-$k_T$ algorithm
with radius $R = 0.4$, and must have $p_T^{j} > 30 \GeV$ and $|y_j| < 4.4$.
Jets must also be separated from photons with $p_T^\gamma > 15 \GeV$ by a distance $\Delta R_{\gamma j} > 0.4$.

Similarly, in the CMS measurement \cite{CMS:2022wpo}, the fiducial region is defined
by having two isolated photons in the final state.
In this case, photons are isolated if the transverse energy of all particles
inside a cone of radius $R_\text{iso} = 0.3$ is less than $10 \GeV$.
The transverse momenta of the leading (subleading) isolated photon must satisfy $p_T^\gamma > 35\, (25) \GeV$, and amount to at least $1/3$ ($1/4$)
of the reconstructed Higgs invariant mass.
In turn, the Higgs invariant mass must lie between $100$ and $180 \GeV$.
Photons must also satisfy $|\eta_\gamma| < 2.5$.
Also in this case, jets are constructed using the anti-$k_T$ algorithm with $R = 0.4$,
and are required to have $p_T^j > 30 \GeV$. Jets with $|\eta_j| < 2.5$ are used
for observables with one extra jet or to count the number of jets,
while a looser cut $|\eta_j| < 4.7$ is applied
for observables requiring at least two jets in the final state.

Due to the lack of availability of these analyses in the \textsc{Rivet} \cite{Buckley:2010ar} framework, we have implemented the ATLAS and CMS analyses within the \geneva code.
The $H \to \gamma \gamma$ decay is inserted by the \pythiaEight particle decays handler on top of the events produced by \geneva.
Its kinematics are treated at leading order in QCD, and
we set the branching ratio to $\mathrm{BR} (H \to \gamma\gamma) = 2.27 \times 10^{-3}$, i.e.~the value reported in \refcite{LHCHiggsCrossSectionWorkingGroup:2016ypw} and calculated with HDECAY \cite{Djouadi:1997yw}.
The \geneva prediction for the gluon-fusion production channel is obtained
at NNLO+NNLL$^{\prime}_{\Tau_0}$+NLL$_{\Tau_1}$,
and setting the scale of the hard function $\mu_H = -\mathrm{i}\, m_H$.
We use matrix elements computed in the infinite-top-mass limit and rescaled in the rEFT scheme.
We set $\Tau_0^\cut = \Tau_1^\cut = 1 \GeV$.
We use the PDF set \texttt{PDF4LHC15\_NNLO}, and take the value of $\alpha_s(m_Z)$ from there.
The partonic prediction is matched to the \pythiaEight QCD+QED shower,
including multiparton interaction (MPI) contributions.
We use the AZNLO tune \cite{ATLAS:2014alx} for the ATLAS comparison,
and the CP5 tune \cite{CMS:2019csb} for CMS.
Showered events are then hadronised using the default \pythiaEight Lund string model \cite{Andersson:1983ia,Andersson:1997xwk}.
In order to obtain a meaningful comparison with the experimental data,
we include the contributions from other Higgs boson production modes
(labelled overall as $XH$) by summing them to the \geneva results for the gluon-fusion channel alone.\footnote{The values of the $XH$ distributions are taken from the plots in ATLAS and CMS publications.}
For ATLAS these include vector-boson fusion (VBF), Higgsstrahlung ($VH$),
and associated production with $t\bar{t}$, $b\bar{b}$, and $t$,
all computed at NLO accuracy in QCD.
For CMS these only include contributions from VBF, $VH$, and $t\bar{t}H$.

The outcome of the comparison with the experimental results is shown
in \figs{atlas_cmp_1}{atlas_cmp_2} for the ATLAS data,
and in \figs{cms_cmp_1}{cms_cmp_2} for the CMS data.
For the ATLAS data, we show the $p_T^H$, $N_\mathrm{jets}$, $|y_H|$, $p_T^{j_1}$, $p_T^{Hj}$, and $\tau_C^{j_1}$ distributions, as well as the $p_T^H$ spectra in bins of $\tau_C^{j_1}$ and in bins of $|y_H|$.
For the CMS data, we show the $p_T^H$, $N_\mathrm{jets}$, $|y_H|$, $p_T^{j_1}$, $\tau_C^{j_1}$, and $|\phi^\ast_\eta|$ distributions, as well as the $p_T^H$ spectra in different jet multiplicity bins ($N = 0$ and $N = 1$).
The definitions of $\tau_C^j$ and $\phi^\ast_\eta$ are given by
\begin{align}
\tau_C^j &= \frac{m_T^j}{2\,\cosh (y_j - y_H)} \,, \nonumber \\
\phi^\ast_\eta &=  \tan \left( \frac{\phi_\mathrm{acop}}{2} \right) \sin \theta^\ast_\eta\,,
\end{align}
where $\phi_\mathrm{acop} = \pi - |\Delta \phi_{\gamma \gamma}|$
and $\sin \theta^\ast_\eta  = \left[ \cosh (\Delta \eta_{\gamma \gamma} / 2) \right]^{-1}$ \cite{Banfi:2010cf}.

With the ATLAS fiducial cuts, we obtain a total fiducial cross section of
$58.8^{+1.5}_{-3.0}\ \rm fb$,
to be compared to the experimental finding of $67 \pm 6 \ \rm fb$.
In the CMS fiducial region, we obtain a total cross section of
$66.6^{+1.6}_{-3.3}\ \rm fb$,
which is compatible with the measurement of $73.4^{+6.1}_{-5.9}\ \rm fb$.
In both cases our predictions agree with the measured results within roughly one standard deviation.
We note that our results are not rescaled to the total N$^3$LO gluon-fusion cross section,
contrary to the theoretical predictions used in the ATLAS publication
for their comparison.

Regarding the distributions, we find overall good agreement between the \geneva predictions
and the measurements.
For the ATLAS data we find slight deviations in the $p_T^H$ peak and a more marked discrepancy in the tail of the distribution. The latter corresponds to the region where the HTL approximation is less accurate. We also find slight deviations in the $|y_H|$ spectrum.
The deviations in both spectra are consistent with those obtained using other calculations, as shown in \refcite{ATLAS:2022fnp}.
Similarly, for the CMS data our results underestimate the bins corresponding to the $p_T^H$ peak,
again in a similar fashion to other predictions \cite{CMS:2022wpo}.
Large deviations are also found in the first bin of the $p_T^H$ distribution
with $N_\text{jets} = 1$,
once again in agreement with other theoretical predictions.

%%%%%%%%%%%%%%%%%%%%%%%%%%%%%%%%%%%%%%%%%%%%%%%%%%%%%%%%%%%%%%%%%%%%%%%%%%%%%%%%
\section{Conclusions}
\label{sec:conclusions}
%%%%%%%%%%%%%%%%%%%%%%%%%%%%%%%%%%%%%%%%%%%%%%%%%%%%%%%%%%%%%%%%%%%%%%%%%%%%%%%%

We have described a number of improvements to the \geneva method,
which are particularly useful for all colour singlet production processes.
Specifically, we detailed a new implementation of the splitting
functions which serve to make the resummed
calculation fully differential in higher multiplicity phase
spaces. This results in an improved behaviour of the
  nonsingular cross section as a function of the colour singlet transverse momentum in
the infrared limit. In addition, following earlier
work~\cite{Billis:2021ecs} we have introduced a separation between the
beam scale in our SCET-based resummed calculation $\mu_B$ and the
scale associated with collinear factorisation $\mu_F$ which appears in
the fixed-order calculation. This allowed us to achieve a more robust
estimate of the theoretical uncertainties associated with our
calculation in the fixed-order region, including uncorrelated
variations of the renormalisation and factorisation scales. Finally,
we have addressed the issue of large contributions from $\pi^2$ terms
originating from timelike logarithms in $2 \to 1$ processes, by enabling the
choice of a complex-valued hard scale $\mu_H$. We
studied the associated resummation of said logarithms in our fully-differential calculation, and showed that, as previously noted in the literature, the perturbative convergence can thus be improved.

Throughout this work, we have used the gluon-initiated Higgs production process to study the effects of our
improvements. We have constructed an NNLO+PS event generator for the
process, including the resummation of the zero-jettiness variable up
to NNLL$^\prime$ accuracy.
We also studied the effects of the parton shower on the logarithmic accuracy
achieved at partonic level by comparing the showered results to the N$^3$LL partonic predictions.
The availability of recent experimental
results~\cite{ATLAS:2022fnp, CMS:2022wpo} for this process also
allowed us to make a detailed comparison of our final, showered events
with data. We stress, however, that the issues which we
addressed in this work have a more general
applicability, and we anticipate that the future implementations of processes in \geneva will make use of these developments.

In this study, we have consistently worked in a heavy-top limit in which the top-quark has been integrated out of the SM Lagrangian, resulting in an effective gluon-Higgs coupling.
We reweighted the results with the exact LO top-quark mass dependence using the so-called rEFT approximation.
Given the advancement in recent years towards including the exact quark mass dependence at NNLO~\cite{Czakon:2020vql,Czakon:2021yub}, it would be desirable to incorporate this progress into a \geneva event generator at NNLO+PS. We leave this issue to future work.

The code used for this study will be included in a future public release of \geneva,
and is available upon request to the authors together with the associated generated events.

%%%%%%%%%%%%%%%%%%%%%%%%%%%%%%%%%%%%%%%%%%%%%%%%%%%%%%%%%%%%%%%%%%%%%%%%%%%%%%%%
\section*{Acknowledgements}
\label{sec:Acknowledgements}
%%%%%%%%%%%%%%%%%%%%%%%%%%%%%%%%%%%%%%%%%%%%%%%%%%%%%%%%%%%%%%%%%%%%%%%%%%%%%%%%

We thank L.~Rottoli for his collaboration in the early stages of this project and for useful exchanges regarding the study presented in the appendix.
We are also grateful to A.~Cueto, M.~Donega, M.~Malberti, and S.~Pigazzini
for their help with the comparison of \geneva with the ATLAS and CMS results.
We thank F.~Tackmann for providing us with a preliminary version of \scetlib
and for useful comments to the manuscript.
This project has received funding from the European Research Council (ERC)
under the European Union's Horizon 2020 research and innovation programme
(Grant agreements No. 714788 REINVENT and 101002090 COLORFREE)
The work of SA and GB is supported by MIUR through the FARE grant R18ZRBEAFC.
SA also acknowledges funding from Fondazione Cariplo and Regione
Lombardia, grant 2017-2070. MAL is supported by the Deutsche
Forschungsgemeinschaft (DFG) under Germany's Excellence Strategy -- EXC 2121 ``Quantum Universe''
-- 390833306, and also by the UKRI guarantee scheme for the Marie Sk\l{}odowska-Curie postdoctoral
fellowship, grant ref. EP/X021416/1.
We acknowledge the CINECA and the National Energy Research Scientific Computing Center (NERSC), a U.S. Department of Energy Office of
Science User Facility operated under Contract No. DEAC02-05CH11231,
for the availability of the high performance computing resources
needed for this work.

\appendix

%%%%%%%%%%%%%%%%%%%%%%%%%%%%%%%%%%%%%%%%%%%%%%%%%%%%%%%%%%%%%%%%%%%%%%%%%%%%%%%%
\section{Effect of the improved splitting functions on Drell-Yan production}
\label{app:splitting-dy}
%%%%%%%%%%%%%%%%%%%%%%%%%%%%%%%%%%%%%%%%%%%%%%%%%%%%%%%%%%%%%%%%%%%%%%%%%%%%%%%%

We study the effect of the improved splitting functions introduced in \sec{splitting} for the Drell-Yan process $pp \to Z \to \ell^+ \ell^-$.
We compare the \geneva predictions obtained with the original and improved splitting functions ($\mathcal{P}_\text{orig}$ and $\mathcal{P}_\text{impr}$) to the $13 \TeV$ ATLAS data for the normalised $p_T^{\ell \ell}$ and $\phi^\ast_\eta$ spectra \cite{ATLAS:2019zci}.
We also compare these distributions to the \geneva{}+\radish result,
for which the resolution variable is $p_T^{\ell \ell}$ instead of $\Tau_0$
and is resummed to N$^3$LL accuracy \cite{Alioli:2021qbf}.

In the experimental analysis, the $Z$ boson is reconstructed from the two hardest final state leptons with same flavour and opposite sign.
Additionally, only events with $p_T^\ell > 27 \GeV$, $|\eta_\ell| < 2.47$, and $66 \GeV < m_{\ell \ell} < 116 \GeV$ are selected.
To produce our predictions, we take the settings used in \refcite{Alioli:2021qbf}. We set $\Tau_0^\cut = \Tau_1^\cut = p_T^{\ell \ell,\, \cut} = 1 \GeV$, $m_Z = 91.1876 \GeV$, $\Gamma_Z = 2.4952 \GeV$, and $\alpha_{e} (m_Z) = 7.55638 \times 10^{-3}$.
We use the \texttt{NNPDF31\_nnlo\_as\_0118} PDF set, and take the value of $\alpha_s$ from there.
All the \geneva results are matched to the \pythiaEight QCD shower
including MPI contributions.

\begin{figure}[t]
   \centering
   \includegraphics[width=\rescaletwoplots]{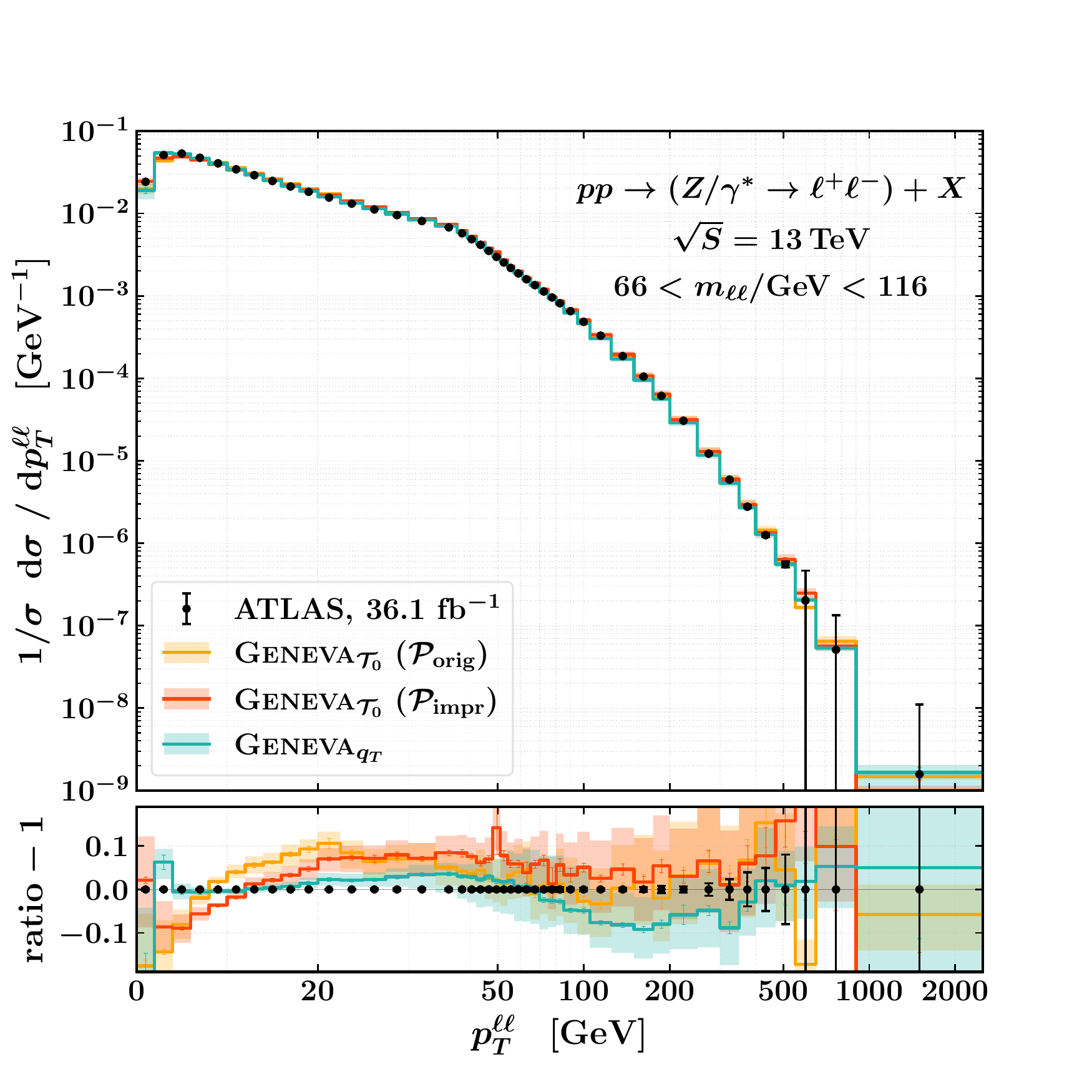}%
   \includegraphics[width=\rescaletwoplots]{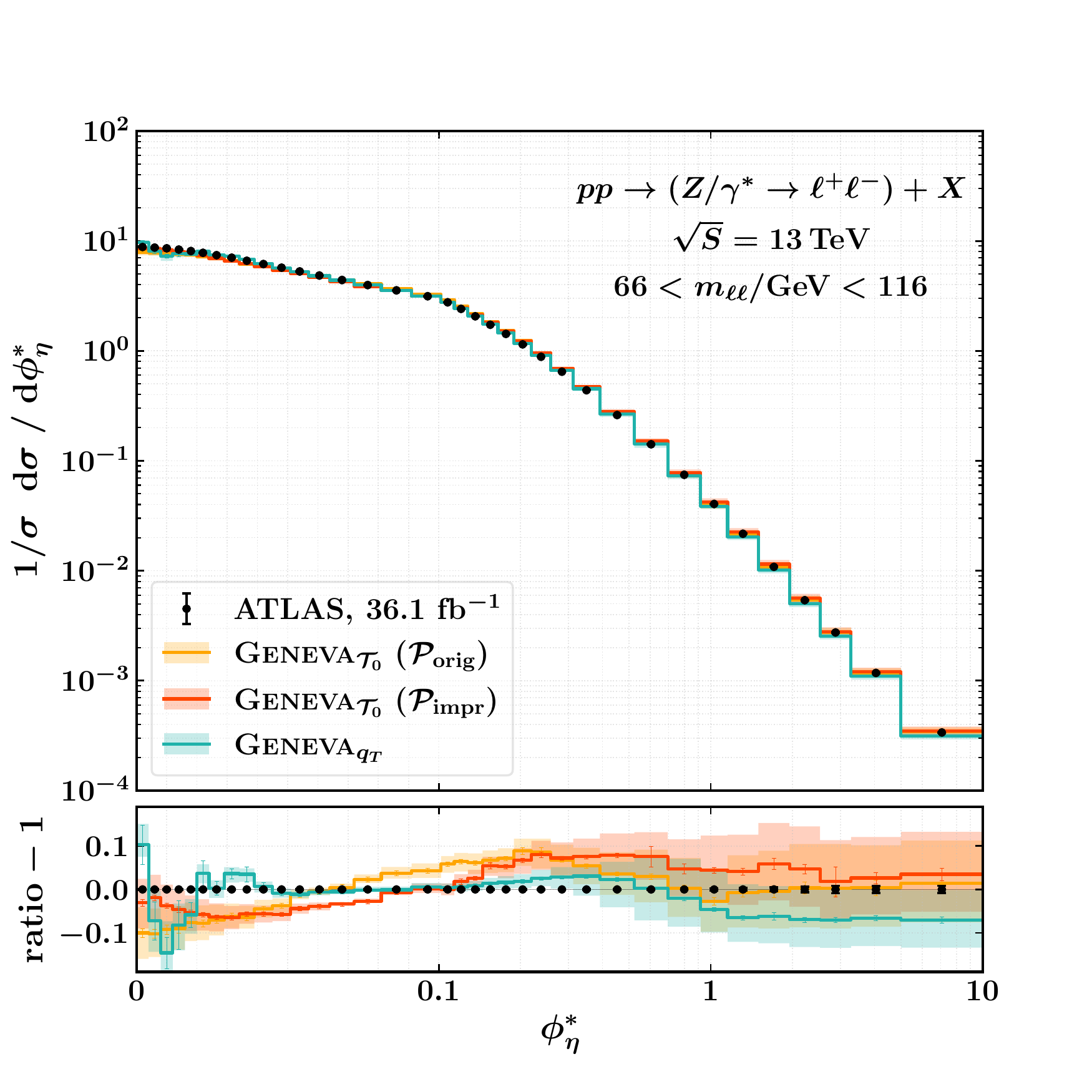}
   \caption{Comparison of the ATLAS normalised $p_T^{\ell \ell}$ and $\phi^\ast_\eta$ distributions \cite{ATLAS:2019zci} with the \geneva{}+\pythiaEight results at $13 \TeV$. We show the distributions obtained with \geneva using $\Tau_0$ as resolution variable and by using the original and improved splitting functions, and also the one obtained with \geneva{}+\radish using $p_T^{\ell \ell}$ as resolution variable.}
   \label{fig:atlas_dy_cmp}
\end{figure}

The results of this study are shown in \fig{atlas_dy_cmp}.
With the improved splitting functions we get a $p_T^{\ell \ell}$ distribution that is closer to data in the interval $[10, 100] \GeV$, although not as close as in the $p_T^{\ell \ell}$-resummed case.
This is an effect of the better physical description of the splitting behaviour encoded in the expressions defined in \eqs{splitting_function}{Altarelli-Parisi_splitting_functions}.
For the $\phi^\ast_\eta$ distribution, the results are affected by the parton shower to a greater extent,
and the different \geneva implementations produce mixed performances,
again with the $p_T^{\ell \ell}$-resummed case being the closest to data.
We notice, though, that in the small $\phi^\ast_\eta$ limit using the improved splitting functions in the $\Tau_0$-resummed case
gives a result that is closer to data than the one obtained with the original splitting functions.

%%%%%
%%%%%

%% Bibliography
\bibliographystyle{JHEP}
\bibliography{geneva}

\end{document}